%% file: paper.tex
\documentclass[iop]{emulateapj}

\usepackage{natbib}
\usepackage{amsmath, amssymb}
\usepackage{times}
\usepackage{apjfonts, graphicx, graphics}
\usepackage[breaklinks,colorlinks,urlcolor=blue,citecolor=blue]{hyperref}
\usepackage[all]{hypcap}
\usepackage{aas_macros}

\input{defn}
\providecommand{\e}[1]{\ensuremath{\times 10^{#1}}}

\newcommand{\beam}{\rm\; beam}
\newcommand{\Jykmps}{\hbox{$\Jy\km\s^{-1}\,$}}
\newcommand{\Jypbeamkmps}{\hbox{$\Jy\beam^{-1}\km\s^{-1}\,$}}

\shorttitle{Molecular Gas in 2A~0335+096}
\shortauthors{Vantyghem et al.}

\begin{document}

\title{Molecular Gas Along a Bright H$\alpha$ Filament in 2A~0335+096 Revealed by ALMA}
\author{A.~N. Vantyghem$^{1,\ast}$}
\author{B.~R. McNamara$^{1,2}$}
\author{H.~R. Russell$^{3}$}
\author{M.~T. Hogan$^{1,2}$}
\author{A.~C. Edge$^4$}
\author{P.~E.~J. Nulsen$^{5,6}$}
\author{A.~C. Fabian$^3$}
\author{F. Combes$^{7,8}$}
\author{P. Salom{\'e}$^7$}
\author{S.~A. Baum$^{9}$}
\author{M. Donahue$^{10}$}
\author{R.~A. Main$^{11}$}
\author{N.~W. Murray$^{11}$}
\author{R.~W. O'Connell$^{12}$}
\author{C.~P. O'Dea$^{9}$}
\author{J.~B.~R. Oonk$^{13,14}$}
\author{I.~J Parrish$^{11}$}
\author{J.~S. Sanders$^{15}$}
\author{G. Tremblay$^{16}$}
\author{G.~M. Voit$^{10}$}

\affil{
    $^1$ Department of Physics and Astronomy, University of Waterloo, Waterloo, ON N2L 3G1, Canada \\ 
    $^2$ Perimeter Institute for Theoretical Physics, Waterloo, Canada \\
    $^3$ Institute of Astronomy, Madingley Road, Cambridge CB3 0HA \\
    $^4$ Department of Physics, Durham University, Durham DH1 3LE \\
    $^5$ Harvard-Smithsonian Center for Astrophysics, 60 Garden Street, Cambridge, MA 02138, USA \\
    $^6$ ICRAR, University of Western Australia, 35 Stirling Hwy, Crawley, WA 6009, Australia \\
    $^7$ LERMA, Observatoire de Paris, CNRS, UPMC, PSL Univ., 61 avenue de l’Observatoire, 75014 Paris, France \\
    $^8$ Coll{\`e}ge de France, 11 place Marcelin Berthelot, 75005 Paris \\
    $^{9}$ Department of Physics and Astronomy, University of Manitoba, Winnipeg, MB R3T 2N2, Canada \\
    $^{10}$ Department of Physics and Astronomy, Michigan State University, 567 Wilson Road, East Lansing, MI 48824, USA \\
    $^{11}$ Canadian Institute for Theoretical Astrophysics, University of Toronto, 60 St. George Street, Toronto, ON M5S 3H8, Canada \\
    $^{12}$ Department of Astronomy, University of Virginia, PO Box 400235, Charlottesville, VA 22904, USA \\
    $^{13}$ ASTRON, the Netherlands Institute for Radio Astronomy, Postbus 2, NL-7990 AA Dwingeloo, the Netherlands \\
    $^{14}$ Leiden Observatory, Leiden University, PO Box 9513, NL-2300 RA Leiden, the Netherlands \\
    $^{15}$ Max-Planck-Institut f{\"u}r extraterrestrische Physik, Giessenbachstrasse 1, D-85748 Garching, Germany \\
    $^{16}$ Department of Physics and Yale Center for Astronomy \& Astrophysics, Yale University, 217 Prospect Street, New Haven, CT 06511, USA
}

\begin{abstract}

We present ALMA CO(1-0) and CO(3-2) observations of the brightest cluster galaxy 
(BCG) in the 2A~0335+096 galaxy cluster ($z=0.0346$). The total molecular gas mass of 
$1.13\pm0.15\e{9}\Msun$ is divided into two components: a nuclear region and a 
7~kpc long dusty filament.
The central molecular gas component accounts for $3.2\pm0.4\e{8}\Msun$ of the total 
supply of cold gas. Instead of forming a rotationally-supported ring or disk, it is 
composed of two distinct, blueshifted clumps south of the nucleus and a series of 
low-significance redshifted clumps extending toward a nearby companion galaxy. 
The velocity of the redshifted clouds increases with radius to a value consistent 
with the companion galaxy, suggesting that an interaction between these galaxies 
$<20\Myr$ ago disrupted a pre-existing molecular gas reservoir within the BCG.
Most of the molecular gas, $7.8\pm0.9\e{8}\Msun$, is located in the filament. 
The CO emission is co-spatial with a $10^4\K$ emission-line nebula and soft X-rays 
from 0.5~keV gas, indicating that the molecular gas has cooled out of the intracluster 
medium over a period of $25-100\Myr$. The filament trails an X-ray cavity, suggesting 
that the gas has cooled from low entropy gas that has been lifted out of the cluster 
core and become thermally unstable.
We are unable to distinguish between inflow and outflow along the filament with 
the present data. Cloud velocities along the filament are consistent with 
gravitational free-fall near the plane of the sky, although their increasing
blueshifts with radius are consistent with outflow.
 
\end{abstract}


\altaffiltext{*}{
    \href{mailto:a2vantyg@uwaterloo.ca}{a2vantyg@uwaterloo.ca}
}

\section{Introduction}

Located at the centres of galaxy clusters, brightest cluster galaxies (BCGs)
are the largest and most luminous galaxies in the universe. They are giant
elliptical galaxies with extended stellar envelopes and predominantly old, 
``red and dead'' stellar populations. 
However, BCGs situated in cooling flow clusters \citep{fabian94}, where the
cooling time of the hot gas is shorter than the age of the system, harbour 
upward of $10^{9}\Msun$ of molecular gas, approaching $10^{11}\Msun$ in the
most extreme systems \citep{edge01, salome03}. Star formation proceeding at
rates of several to tens of solar masses per year, which exceeds the star 
formation rates of many spiral galaxies, is also observed in these systems 
\citep{mcn04, odea08, mcdonald11, donahue15, tremblay15}.

Molecular clouds and stars in BCGs likely form from the cooling of the hot 
intracluster medium (ICM). Correlations between star formation rate and the 
rate of mass deposition from the ICM support this picture \citep{egami06, odea08}.
Furthermore, cold gas and star formation are observed almost exclusively in 
systems where the central cooling time is below a sharp threshold of $\sim5\e{8}\yr$
\citep{rafferty08}, or equivalently where the entropy is less than $30\keV\cmsq$ 
\citep{voit08, cavagnolo08}. This threshold has been attributed to the onset
of thermal instabilities in the ICM \citep{gaspari12, voit15}.
These systems also host diffuse emission-line nebulae, which are likely the 
ionized skins of molecular clouds \citep{heckman81, hu85, odea94, jaffe05, oonk10}.
Alternatively, the peculiar emission line ratios in BCGs \citep[e.g.][]{heckman89}
may originate from primarily neutral gas that is excited by collisions with energetic 
particles \citep{ferland09}.

Although the reservoirs of molecular gas observed in BCGs are quite massive,
they constitute only a few percent of the mass expected from unimpeded cooling 
\citep{peterson06}. Instead, feedback from the active galactic nucleus (AGN) 
is heating the ICM and regulating the rate of cooling \citep{mcn07, mcn12, fabian12}.
High resolution {\it Chandra} X-ray imaging of cool core clusters shows that
AGN launch jets that inflate cavities, drive shock fronts, and generate
sound waves, offsetting radiative losses from the ICM \citep[e.g.][]{mcn00, 
blanton01, fabian06}. The rate of heating is closely tied to the rate of cooling 
in a large sample of groups and clusters \citep{birzan04, dunn06, rafferty06}, 
and is sufficient to prevent the bulk of the hot gas from cooling.

Accretion of molecular gas potentially plays a key role in forming a feedback 
loop, as it connects residual cooling of the ICM with energetic outbursts from
the AGN \citep[e.g.][]{pizzolato05, gaspari13, li14a}. 
While the effects of AGN feedback on the hot atmosphere are clear, little is
known about its connection to the cold gas. 
Radio-jets are known to couple to emission-line nebulae, driving outflows of
ionized \citep{morganti05, nesvadba06, villarmartin06} and molecular gas 
\citep{alatalo11, morganti15} in radio galaxies.
NGC1275, at the centre of the Perseus cluster, 
hosts a filamentary H$\alpha$ nebula with two prominent filaments extending
toward an X-ray cavity \citep{fabian03}, suggesting that the filaments have been 
drawn out of a central reservoir. These filaments are well correlated
with soft X-ray emission as well as molecular hydrogen \citep{lim12} and CO
\citep{salome06, salome11}.
Infalling molecular gas toward the centre of the BCG suggests that the uplifted 
gas is returning in a molecular ``fountain'' \citep{lim08}. Narrow, redshifted 
absorption lines have been observed in NGC5044 \citep{david14} and A2597 
\citep{tremblay16}, indicating that molecular clouds are inflowing toward the
central black hole.
ALMA Early Science observations of A1835 revealed $10^{10}\Msun$ of molecular
gas being uplifted by the X-ray cavities in a bipolar outflow \citep{almaA1835}.
PKS0745-191 harbours an even more dramatic outflow -- virtually all of its 
molecular gas has been swept from the BCG in three low-velocity filaments 
\citep{almaPKS0745}.
These results have led \citet{mcn16} to postulate that molecular gas condenses
out of low entropy ICM that is lifted from the cluster centre by X-ray cavities.

Here we present ALMA Cycle 1 observations of the molecular gas in the BCG of 
2A~0335+096 (also known as RXJ0338+09), traced by CO(1-0) and CO(3-2) line emission. 
2A~0335+096 is one of the brightest X-ray objects on the sky \citep{edge90, 
reiprich02}, and has a centre with a short radiative cooling time
\citep{schwartz80, singh86, singh88, white91}. Its X-ray atmosphere is
complex, containing a series of cool clumps \citep{mazz03, werner06}, a cold
front that was likely induced by sloshing motions \citep{mazz03, sanders09}, and
several cavities corresponding to multiple generations of AGN feedback
with a total enthalpy of $5\e{59}\erg$ \citep{sanders09}. Multiphase gas in 
the ICM traces an extended H$\alpha$ filament within the BCG with a total
luminosity of $L_{\rm H\alpha}=8\e{41}\ergps$ \citep{RH88, donahue07, sanders09}.
\citet{farage12} argued that the H$\alpha$ filament consists of a 2~kpc, 
counterrotating disk within the 17~kpc filament.
Single dish IRAM-30m observations detected the 2A~0335+096 BCG at CO(1-0), measuring 
a total molecular gas mass of $2.7\pm0.3\e{9}\Msun$ \citep[][corrected for cosmology]{edge03}.
Optical, UV, and IR observations of the BCG show ongoing star formation at a rate of 
several solar masses per year \citep{RH88, donahue07, odea08}.
X-ray spectroscopy from {\it Chandra} and {\it XMM-Newton} indicate that the
0.5~keV gas within the ICM is cooling out of the hot atmosphere and depositing mass 
onto the BCG at $<30\Msunpyr$ \citep{sanders09}.
The ALMA observations presented here resolve the spatial and velocity structure of
the molecular gas within the BCG, revealing a striking correlation between 
molecular gas and the H$\alpha$ filament.

Throughout this paper we assume a standard $\Lambda$-CDM cosmology with $H_0=70\kmpspMpc$,
$\Omega_{\rm m, 0}=0.3$, and $\Omega_{\Lambda, 0}=0.7$. 
At the redshift of 2A~0335+096 \citep[$z=0.0346$;][]{mcn90}, the angular scale is 
$1\arcsec = 700\pc$ and the luminosity distance is $150\Mpc$.
This paper is organized as follows. Details of the ALMA observations and data reduction 
are given in Section \ref{sec:obs}. Our results pertaining to the morphological and kinematic 
distribution of the molecular gas and its relation to other wavelengths are described 
in Section \ref{sec:results}. The origin of the molecular gas is discussed in Section 
\ref{sec:discussion}, and the main results are summarized in Section \ref{sec:summary}.

\section{Observations and Data Reduction}
\label{sec:obs}

The brightest cluster galaxy (BCG) in the 2A~0335+096 galaxy cluster 
(RA: 03:38:40.50, Dec: +09:58:12.3) was observed by ALMA Cycle 1 
(Program ID 2012.1.00837.S, PI McNamara) centred
at $111.394\GHz$ and $334.169\GHz$ to cover the CO(1-0) and CO(3-2) lines.
The CO(1-0) observation was divided into two blocks, which were
observed in band 3 on July 22, 2014 and March 08, 2015. Both observations 
had an on-source integration time of 35 minutes. 
For the July 2014 (March 2015) observation the array was configured with 33 
(30) antennas with baselines of $17-716\m$ ($12-280\m$), each with a
primary beam diameter of $56''$.
Our observations employed the frequency division correlator mode, so 
had a frequency resolution of $488.281$ kHz ($1.31\kmps$) over a 
$1.875\GHz$ bandwidth.
The CO(3-2) line was observed in band 7 with one 30 minute on-source integration
on August 12, 2014. The observation used 34 antennas with baselines of $19-915\m$
and a primary beam of $18.5''$. At CO(3-2) the velocity resolution
in frequency division correlator mode is $0.44\kmps$.
Velocity channels are binned together during imaging to improve sensitivity.
An additional baseband with $2\GHz$ bandwidth was included in order to image
the continuum.

The observations were primarily calibrated in CASA version 4.4.0 \citep{casa}
using the automatic pipeline scripts. Additional phase self-calibration 
on the nuclear continuum improved the signal-to-noise of the CO(1-0) 
observation by a factor of $1.2$. The nuclear
continuum flux at CO(3-2) was too faint to perform successful self-calibration.

The phase calibrator chosen automatically at the time of the March 2015
CO(1-0) observation was located $10^{\circ}$ away from 2A~0335+096 and has a
flux 10 times fainter than the calibrator used in the July 2014 CO(1-0) 
observation. The resulting phase solutions determined by the pipeline were poor, 
and a manual recalibration of the data did not improve these solutions to
an acceptable level. The phase self-calibration also did not rectify the
problem. We therefore do not include this observation in our analysis.

Images of the continuum were created for each band by imaging the line-free 
channels. An unresolved point source was detected in each image.
At $109.84\GHz$ the nuclear continuum flux is $6.854\pm0.044\mJy$ in the
July 2014 observation and $6.36\pm0.15\mJy$ in the March 2015 observation.
This implies a variability of $7.2\pm2.3\%$ over an 8 month period, which
is consistent with the variability seen in other BCGs with active cores \citep{hoganb}.
At $335.12\GHz$ (August 2014) the nuclear continuum flux is $1.14\pm0.15\mJy$.
The corresponding spectral index, following the convention $S_{\nu}\propto \nu^{-\alpha}$ 
and considering only the 2014 observations, is $\alpha=1.61\pm0.27$.
The location of the continuum source is consistent with $5\GHz$ VLBI \citep{sanders09}. 
No continuum emission is associated with the nearby companion galaxy.
Imaging the continuum in narrow velocity channels ($3\kmps$) shows no 
evidence of line absorption against the continuum emission.

Images of line emission were reconstructed using \textsc{clean} with a
Briggs weighting of 2. No continuum subtraction was applied to the CO(3-2)
observation, as the continuum flux is small compared to the line emission.
The resulting images had a synthesized beam of $1.3''\times 0.92''$ 
(PA $-34.5^{\circ}$) at CO(1-0) and $0.39''\times 0.23''$ (PA $-50.0^{\circ}$)
at CO(3-2). The rms noise in $20\kmps$ line-free channels was 
$0.56\mJy~{\rm beam}^{-1}$ and $0.61\mJy~{\rm beam}^{-1}$ at CO(1-0) and CO(3-2), 
respectively. In regions of low signal-to-noise we further bin the spectra to 
either $40\kmps$ or $60\kmps$ channels, as needed.
Missing short spacings will filter out emission on spatial scales larger than
$\sim11''$ at CO(1-0) and $\sim3''$ at CO(3-2).

A reference velocity of $10368\kmps$ ($z=0.0346$) was adopted for this analysis. 
This is a stellar absorption measurement of the BCG that was measured by Huchra. 
The result was first published in \citet[][quoting private communication with Huchra]{mcn90}
and subsequently in the 2MASS catalog \citep{lavaux11,huchra12}, who quote an 
uncertainty of $10\kmps$. 
The centre of the CO(1-0) and CO(3-2) emission is better estimated by the adopted 
systemic velocity than by the redshifts adopted by previous studies. For reference, 
a redshift of $z=0.0349$, which is used by \citet{donahue07} and \citet{farage12}, 
corresponds to a velocity of $+92\kmps$ in our adopted frame.

\section{Results}
\label{sec:results}

\begin{figure}
  \centering
  \hspace*{0.0cm} \includegraphics[width=0.45\textwidth]{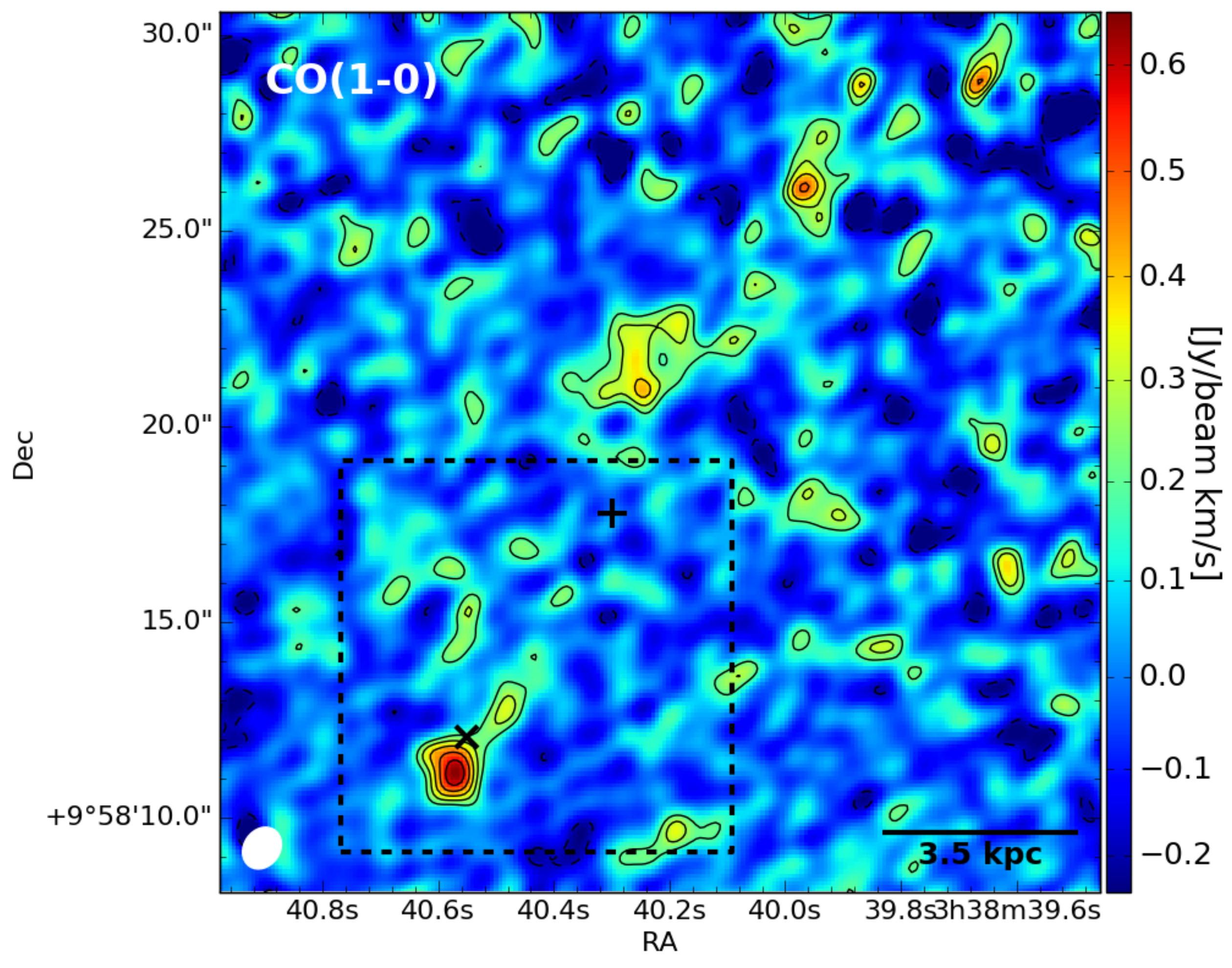}
  \hspace*{-1.0cm} \includegraphics[width=0.4\textwidth]{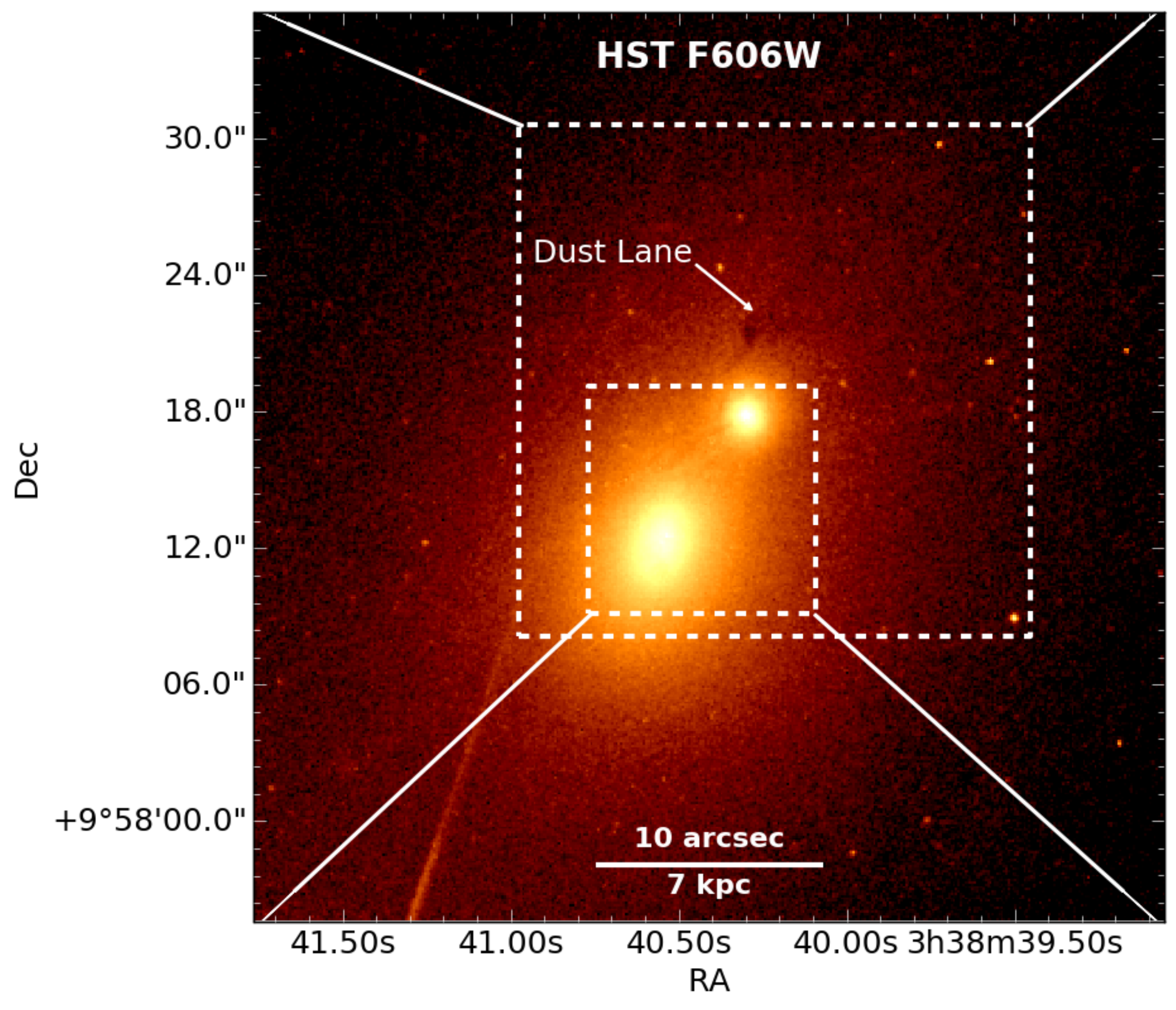}
  \hspace*{0.0cm} \includegraphics[width=0.45\textwidth]{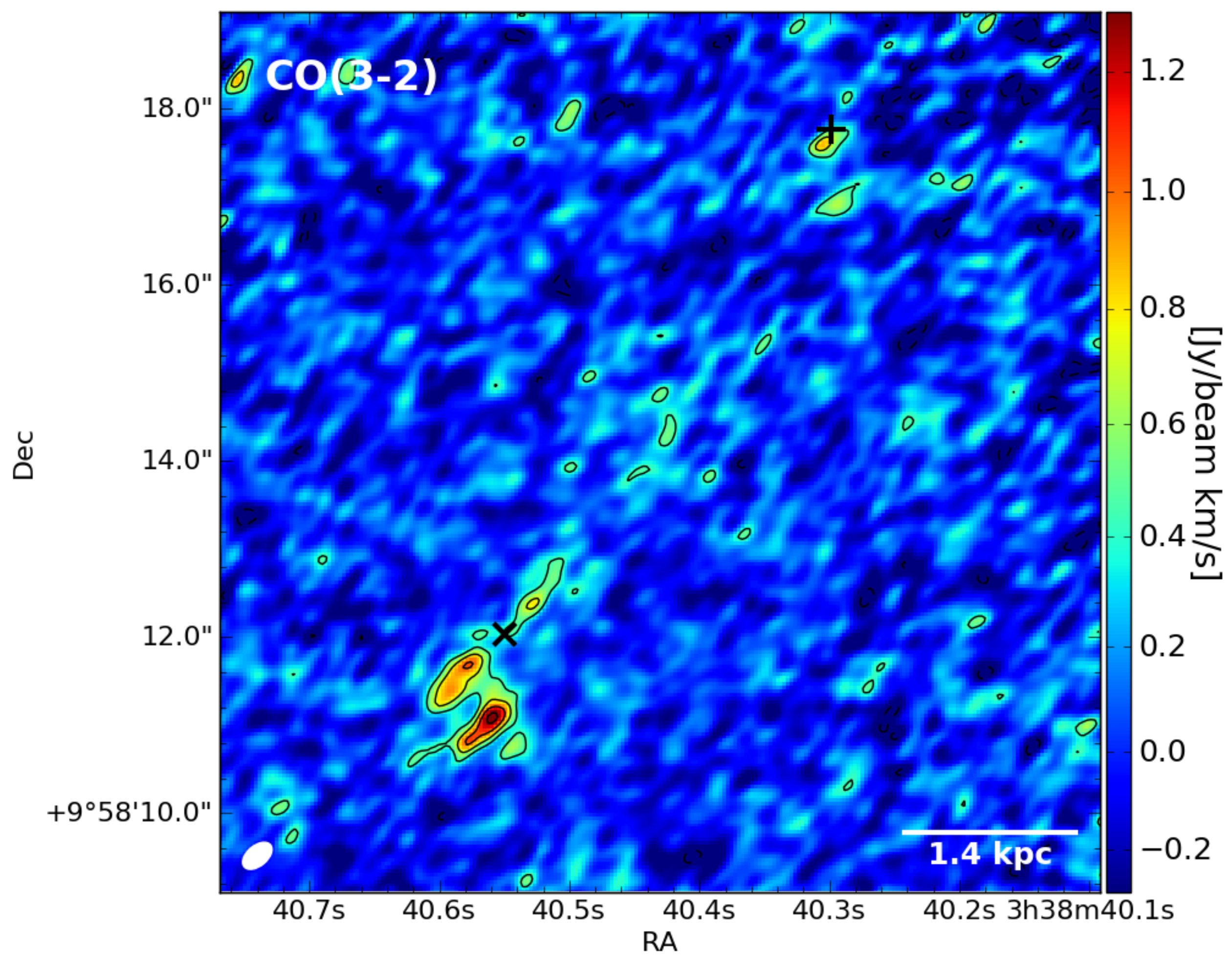}
  \caption{
    {\it Top:} Integrated flux of the CO(1-0) ALMA image. The contours are 
    $-2$, $2$, $3$, $4$, $5\sigma$, ... , where $\sigma=0.095\Jypbeamkmps$. 
    The $\times$ indicates the position of radio AGN determined from VLBI 
    \citep{sanders09}, and the $+$ indicates the flux centroid of the companion 
    galaxy determined from the HST F606W image.
    The $1.13\times 0.92\asec$ (PA $-35^{\circ}$) synthesized beam is shown in white
    in the lower left corner. The box indicates the CO(3-2) frame shown in the bottom
    panel.
    \newline
    {\it Middle:} {\it HST} WFPC2 F606W optical image of the 2A~0335+096 BCG and the 
    nearby companion galaxy.
    \newline
    {\it Bottom:} Integrated flux of the CO(3-2) ALMA image, with contours set at 
    $-3$, $3$, $5$, $7$, and $9\sigma$, where $\sigma=0.145\Jypbeamkmps$. The 
    $0.32\times 0.20\asec$ beam is shown in white in the lower left corner.
  }
  \label{fig:COmaps}
\end{figure}

\subsection{Distribution of Molecular Gas}
\label{sec:maps}

The {\it HST} WFPC2 F606W image presented in the middle frame of Fig. 
\ref{fig:COmaps} shows the central $28\times28\kpc$ of the galaxy cluster, 
encompassing the BCG and a nearby companion galaxy that is situated $5\kpc$
in projection from the BCG and well within its light profile. 
Stellar absorption lines from the two galaxies indicate that the companion is 
offset from the BCG by $212\pm58\kmps$ in velocity space \citep{gelderman96}. 
This low velocity suggests that the radial distance between the galaxies is
small, so the companion is not just a chance projection. However, this information
is not sufficient to determine the true 3D distance or relative velocity
between the BCG and companion.
Outside of the shown field of view, the closest galaxies are $25\kpc$ and $28\kpc$
from the BCG, respectively. These both lie well beyond the molecular gas, so are
unlikely to be relevant in our analysis.

\begin{figure}
  \includegraphics[trim=0.0cm 0.0cm 0.0cm 2.0cm, clip=true, width=0.5\textwidth]{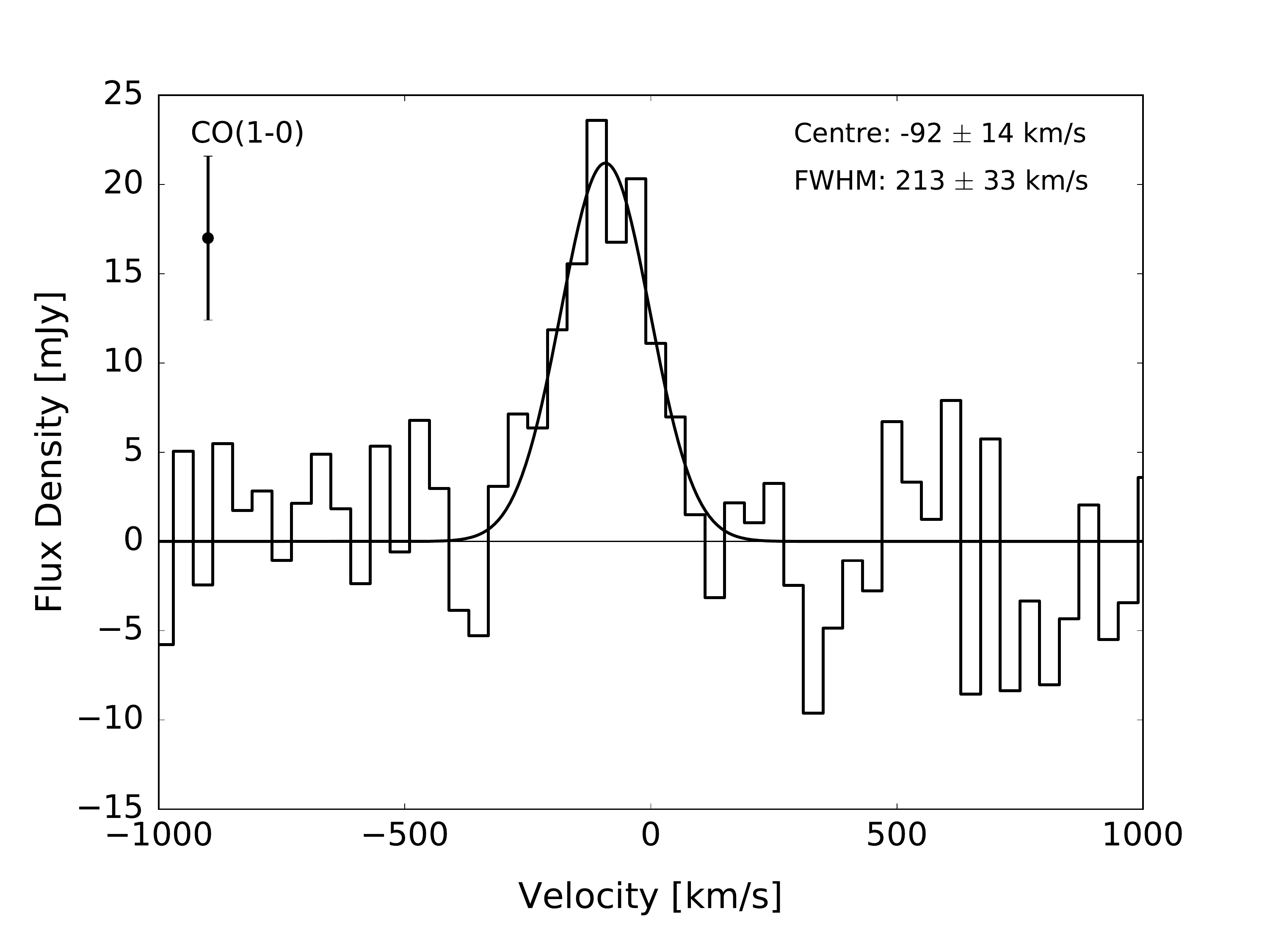}
  \caption{
    CO(1-0) spectrum extracted from a $17.5''\times 22''$ box that encompasses
    both the BCG and the extended filament. The best-fit parameters are given
    in Table \ref{tab:fitparameters}. The error bar indicates the rms variation
    in the line-free channels.
  }
  \label{fig:fullspec}
\end{figure}

Maps of the integrated CO(1-0) and CO(3-2) flux are presented alongside the 
{\it HST} image in Fig. \ref{fig:COmaps}. These maps were created by 
summing the flux in each pixel over the velocity range $-500$ to $500\kmps$, 
which encompasses all of the observed flux (see Fig. \ref{fig:fullspec}).
The maps have been corrected for the response of the primary beam, with contour
levels determined from the line-free regions of an uncorrected map.
Significant CO emission is observed in two primary locations: near the nucleus of
the BCG and in a filament situated beyond a nearby companion galaxy. The filament
falls outside of the CO(3-2) field of view.

Fig. \ref{fig:fullspec} shows the CO(1-0) spectrum extracted from a $17.5''\times 22''$
($12.3\times 15.4\kpc$) box encompassing all of the observed flux. This region is 
similar in size to the $21.5''$ IRAM 30m beam. 
The spectrum was well-fit by a single Gaussian component centred at
$-92\pm14\kmps$ with a linewidth of $213\pm33\kmps$ (full width at half maximum; FWHM). 
Spectral fitting throughout this work used the \textsc{lmfit} 
\footnote{https://lmfit.github.io/lmfit-py/} package with one to three Gaussian
components, as necessary. Each spectrum has been corrected for the response of the
primary beam as well as instrumental broadening introduced by the velocity binning.
Table \ref{tab:fitparameters} lists the best-fitting parameters for all spectra.

The total CO(1-0) flux recovered by these observations is $4.8\pm0.6\Jykmps$, which is
consistent with the OVRO measurement of $7.1\pm2.4\Jykmps$ within 2$\sigma$ \citep{edge03}.
This flux constitutes $43\pm7~\%$ of the IRAM 30m flux ($11.4\pm1.1\Jykmps$; \citealt{edge01}),
implying that over half of the molecular gas is undetected in our observations. 
Similar recovered fractions were noted in ALMA observations of NGC5044 \citep{david14}
and A1664 \citep{almaA1664}. Missing short spacings filter out emission
on scales larger than 11~arcsec at CO(1-0) or 3~arcsec at CO(3-2).

\begin{table*}
\begin{minipage}{\textwidth}
\caption{Parameters of Molecular Features}
\begin{center}
\begin{tabular}{l l c c c c c}
\hline
CO   & Region & $\chi^2/$dof & Velocity centre & FWHM       & Integrated intensity & Gas Mass      \\
line &        &              & ($\kmps$)       & ($\kmps$)  & ($\Jykmps$)          & ($10^8\Msun$) \\
\hline
J=1-0 & Total & $101/107$    & $-92\pm14$      & $213\pm33$  & $4.8\pm0.6$         & $11.3\pm1.5$ \\
      & BCG   & $154/67$     & $-219\pm21$     & $266\pm57$  & $0.73\pm0.13$       & $1.7\pm0.3$ \\
      &       &              & $109\pm12$      & $164\pm32$  & $0.63\pm0.11$       & $1.5\pm0.3$ \\
      & Filament & $132/107$ & $-96\pm11$      & $196\pm25$  & $3.3\pm0.4$         & $7.8\pm0.9$ \\
      & South BCG & $136/107$ & $-172\pm16$ & $297\pm38$ & $0.99\pm0.11$ & $2.33\pm0.26$ \\
      & North BCG & $131/107$ & $134\pm18$ & $238\pm43$ & $0.560\pm0.088$ & $1.31\pm0.21$ \\
      & Inner filament & $291/214$ & $-28.1\pm3.6$ & $100.9\pm8.8$ & $1.22\pm0.09$ & $2.86\pm0.21$ \\
      &                &           & $-164.7\pm3.0$ & $34.5\pm8.1$ & $0.32\pm0.06$ & $0.74\pm0.14$ \\
      & Outer filament & $240/211$ & $-132.0\pm3.4$ & $55.6\pm8.6$ & $0.75\pm0.10$ & $1.77\pm0.23$ \\
      &                &           & $-258.3\pm8.1$ & $51\pm21$    & $0.27\pm0.10$ & $0.64\pm0.22$ \\
      &                &           & $-31.6\pm6.9$  & $38\pm19$    & $0.22\pm0.09$ & $0.51\pm0.21$ \\
J=3-2 & BCG   & $54/34$      & $-223\pm15$    & $188\pm36$  & $4.8\pm0.8$ & $1.60\pm0.28$ \\
      &       &              & $121\pm27$     & $282\pm71$  & $4.7\pm1.0$ & $1.58\pm0.33$ \\
      & South BCG & $108/74$ & $-160\pm10$    & $279\pm17$  & $6.37\pm0.45$ & $2.14\pm0.15$ \\
      &           &          & $-281.1\pm3.5$ & $73\pm11$   & $1.5\pm0.3$ & $0.50\pm0.10$ \\
      & South BCG: NE Clump & $80/77$ & $-201.4\pm7.6$ & $275\pm18$ & $2.46\pm0.14$ & $0.83\pm0.05$ \\
      & South BCG: SW Clump & $65/74$ & $-147\pm19$    & $300\pm32$ & $1.92\pm0.23$ & $0.64\pm0.08$ \\
      &                     &         & $-278.6\pm3.2$ & $77\pm10$  & $0.89\pm0.16$ & $0.30\pm0.05$ \\
      & BCG Spur  & $83/74$ & $127\pm24$ & $368\pm67$  & $3.1\pm0.5$ & $1.03\pm0.16$ \\
      &           &         & $93.5\pm7.4$ & $79\pm23$ & $0.95\pm0.36$ & $0.32\pm0.12$ \\
      & Diffuse Gas  & $32/34$ & $188.3\pm4.9$ & $117\pm13$ & $2.56\pm0.23$ & $0.86\pm0.08$ \\
      &              &         & $22.2\pm8.2$ & $46\pm30$ & $0.45\pm0.22$ & $0.15\pm0.08$ \\
      & Companion Galaxy  & $36/37$ & $224\pm19$ & $256\pm46$ & $1.65\pm0.25$ & $0.56\pm0.09$ \\
\hline
\end{tabular}
\end{center}
Notes: All spectra have been corrected for the response of the primary beam and
instrumental broadening. Masses determined from the CO(3-2) line have been calculated 
assuming ${\rm CO(3-2)/CO(1-0)} = 7$.
\label{tab:fitparameters}
\end{minipage}
\end{table*}

\begin{figure}
  \includegraphics[trim=0.0cm 0.0cm 0.0cm 2.0cm, clip=true, width=0.5\textwidth]{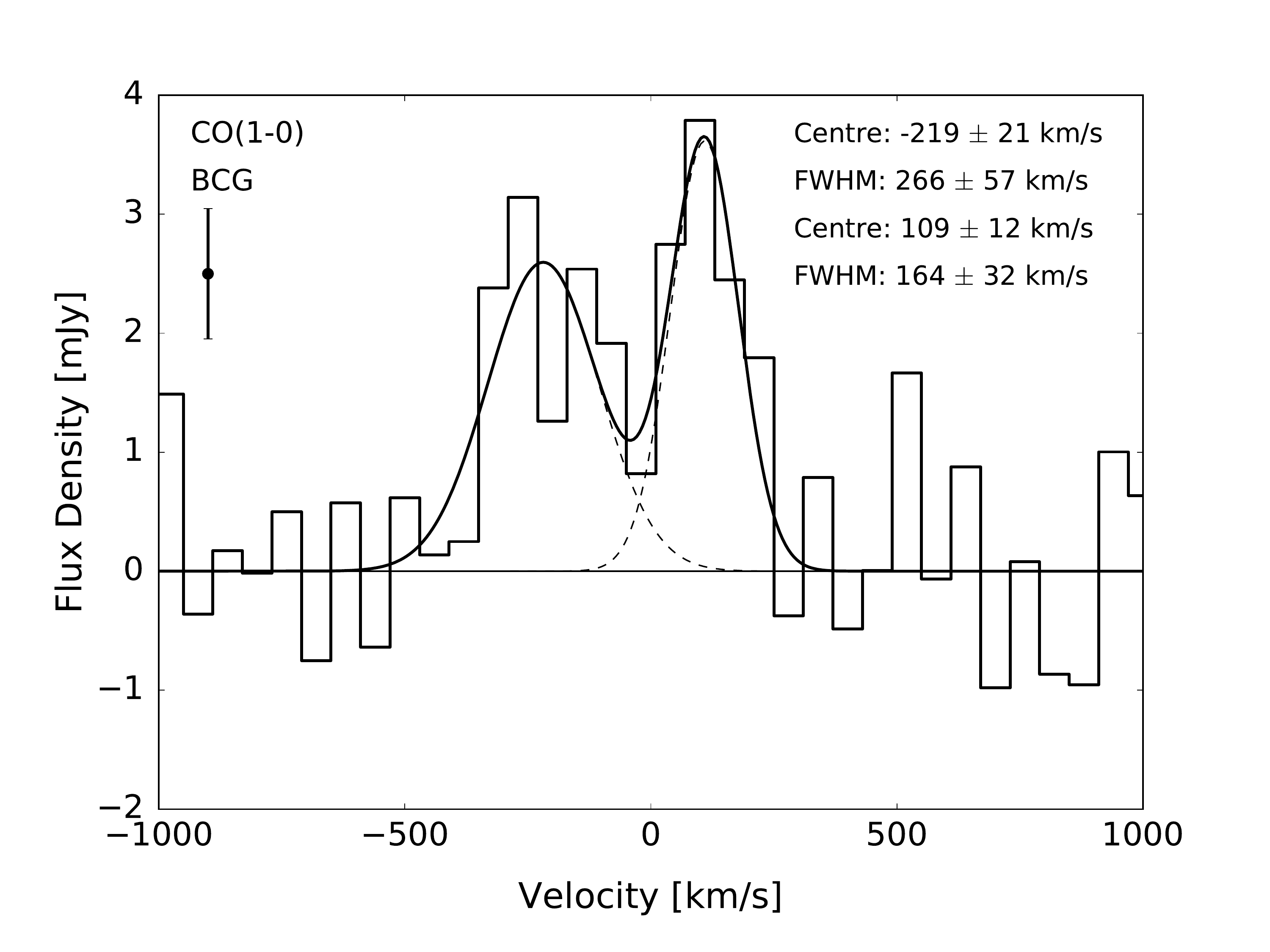}
  \includegraphics[trim=0.0cm 0.0cm 0.0cm 1.9cm, clip=true, width=0.5\textwidth]{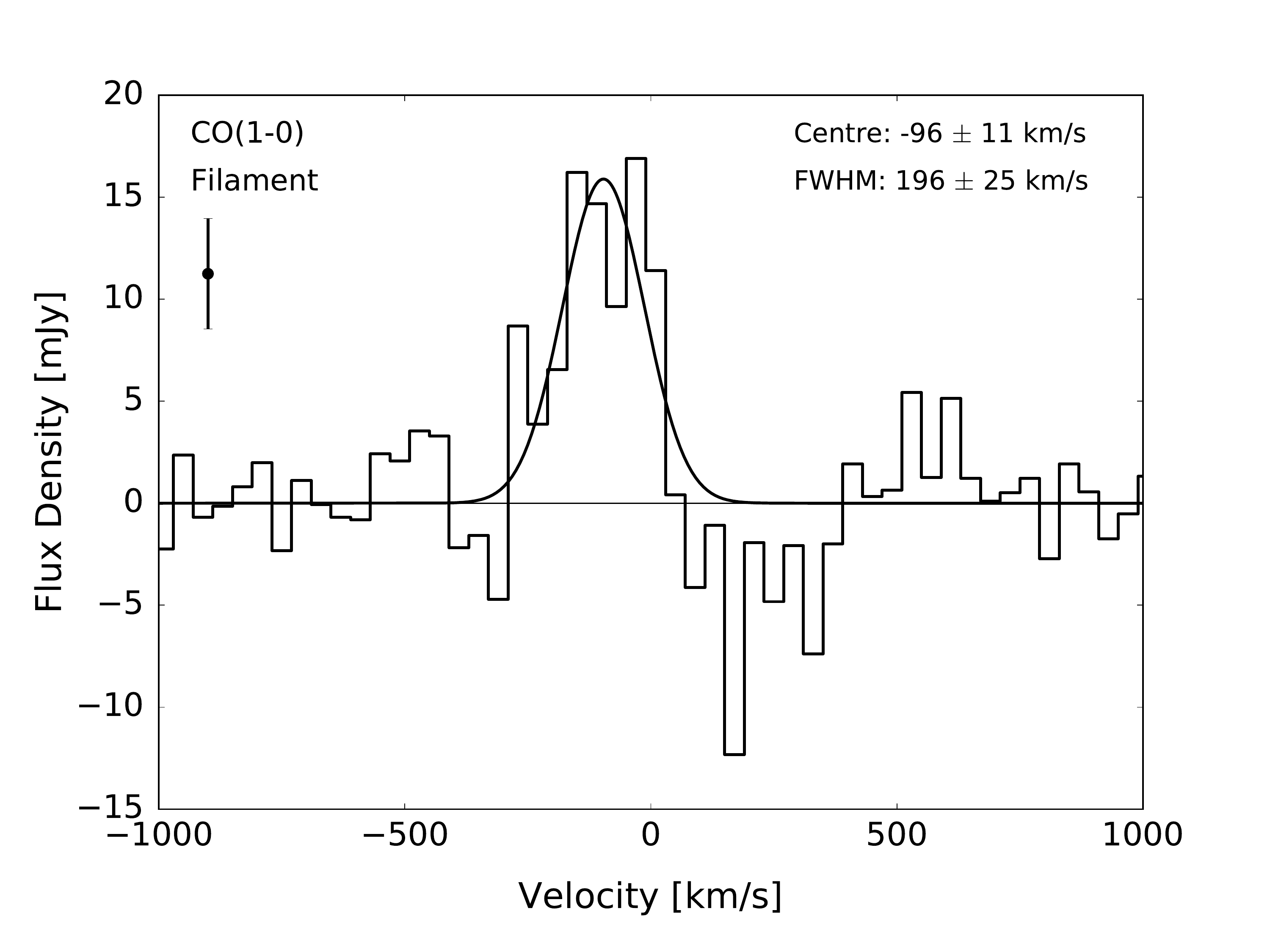}
  \caption{
    CO(1-0) spectra from a $4.5''\times 4.5''$ box centred on the BCG (top)
    and an $8''\times 8.5''$ box enclosing the filament (bottom). The spectra 
    are each fit by a single Gaussian component, and the best-fit parameters
    are given in Table \ref{tab:fitparameters}. The error bars indicate the rms 
    variation in the line-free channels.
  }
  \label{fig:BCGandFil}
\end{figure}

\begin{figure}
  \includegraphics[trim=0.0cm 0.0cm 0.0cm 2.0cm, clip=true, width=0.5\textwidth]{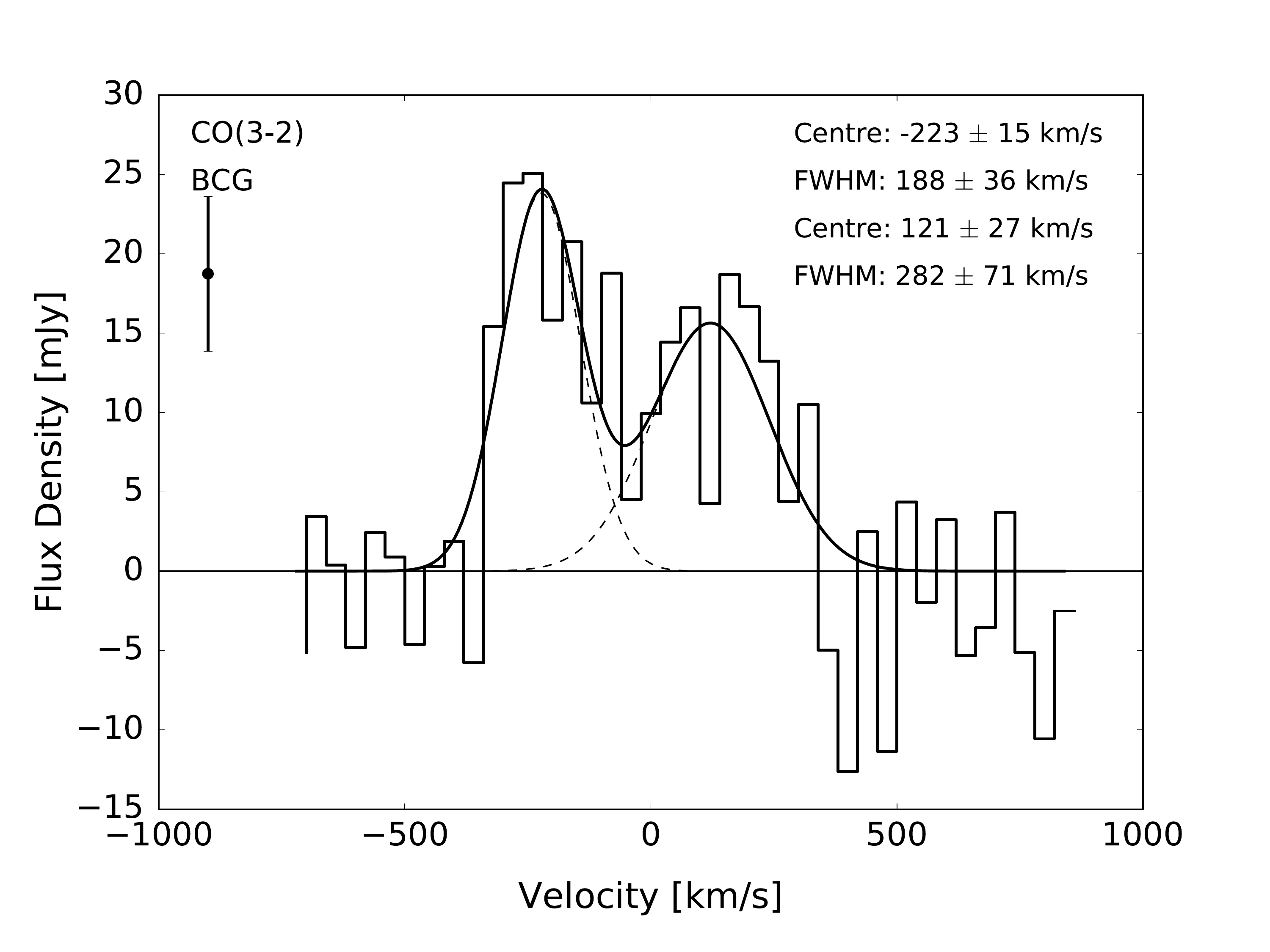}
  \caption{
    CO(3-2) spectrum of the BCG extracted from the same region as Fig. 
    \ref{fig:BCGandFil} (top panel). The best-fit parameters are given
    in Table \ref{tab:fitparameters}. The error bar indicates the rms variation
    in the line-free channels.
  }
  \label{fig:BCGCO32}
\end{figure}

\subsubsection{Gas in the BCG}

Near the nucleus of the BCG the molecular gas is distributed amongst a series of
clumps of varying size. At CO(1-0) the emission is mostly concentrated in an
unresolved clump located south of the nucleus. At CO(3-2) this feature is further
resolved into two clumps of comparable size. The brighter of these clumps is
coincident with significant dust extinction in the optical imaging, which is 
discussed further in Section \ref{sec:dust}.
No molecular gas is concentrated at the location of the AGN. 
Toward the northwest the CO emission extends in a small spur that breaks into
a series of small, faint clouds. These clouds are distributed in the direction of 
a nearby companion galaxy that itself hosts a small association of molecular gas.

A spectrum was extracted from a $4.5''\times 4.5''$ box centred on the nucleus
at both CO(1-0) (Fig. \ref{fig:BCGandFil} {\it top}) and CO(3-2) (Fig. \ref{fig:BCGCO32}).
This regions includes all of the southern emission and most of the clouds to
the northwest. Both spectra are best fit by two Gaussian components, with one
blueshifted to $\sim -200\kmps$ and one redshifted to $\sim100\kmps$ with respect 
to the systemic stellar velocity of the BCG. 
The total integrated flux in this region is $1.36\pm0.17\Jykmps$ at CO(1-0).

Additional spectra were extracted for the distinct structures within the BCG. At CO(1-0)
these include the southern unresolved clump as well as the spur extending to the
northwest, which are identified in Fig. \ref{fig:CO10subdiv}. The resulting spectra
are shown in the bottom panels of Fig. \ref{fig:CO10regions}. Both are well-fit
by a single velocity component that, when combined, accounts for the two peaks observed
in the total BCG spectrum in Fig. \ref{fig:BCGandFil}. The blueshifted emission
is observed exclusively in the southern clump, while the spur to the northwest is
entirely redshifted. Both are relatively broad, with FWHM of $297\pm38\kmps$ and 
$238\pm43\kmps$, respectively.

These regions can be subdivided further at CO(3-2), as shown in Fig. \ref{fig:CO32subdiv}.
The ``South BCG'' region is the same as in CO(1-0) (Fig. \ref{fig:CO10subdiv}) to
allow for a direct comparison, but has also been split into two regions that are not
shown in the figure, one for each clump. Toward the northwest the emission has also
been divided into two regions based on distance from the nucleus. Neither of these
regions correspond 1:1 with the ``North BCG'' region in Fig. \ref{fig:CO10subdiv}, 
which extends midway into the ``Diffuse Gas'' region.
All CO(3-2) spectra are shown in Fig. \ref{fig:CO32regions}, with all fit parameters
listed in Table \ref{tab:fitparameters}.
Multiple velocity components are required for several of the spectra. In the 
northwestern regions most of the linewidths are narrow ($<100\kmps$), so the multiple
peaks likely arise from giant molecular clouds or associations of different velocities.
South of the BCG, multiple velocity components are only required to fit the larger
of the two clumps, lying to the southwest. This clump is coincident with dust 
extinction, although the northeastern clump may still be coincident with dust if 
it is located on the far side of the BCG.

\subsubsection{Filament}

Beyond the companion galaxy the molecular gas is located in a 7~kpc long filament
that is coincident with an extended region of significant dust extinction.
Most of the emission is localized in the inner portion of the filament, with
a second, smaller region at the filament tail reaching $5\sigma$ significance.
These two regions are connected by a faint channel that is not evident in Fig.
\ref{fig:COmaps}, but is significantly detected in the maps presented in 
Section \ref{sec:velmaps}. This is because the channel has a narrow linewidth,
so is drowned out by the noise when integrated between $-500$ and $500\kmps$.
Fig. \ref{fig:BCGandFil} ({\it bottom}) shows the CO(1-0) spectrum of the filament
extracted from an $8''\times 8.5''$ box. We are unable to obtain a CO(3-2) spectrum
of the filament because it lies outside of the field of view. The CO(1-0) spectrum
is well-fit by a single Gaussian component with an integrated flux of $3.3\pm0.4\Jykmps$,
which is more than double the integrated flux within the BCG.

\begin{figure}
  \includegraphics[width=\linewidth]{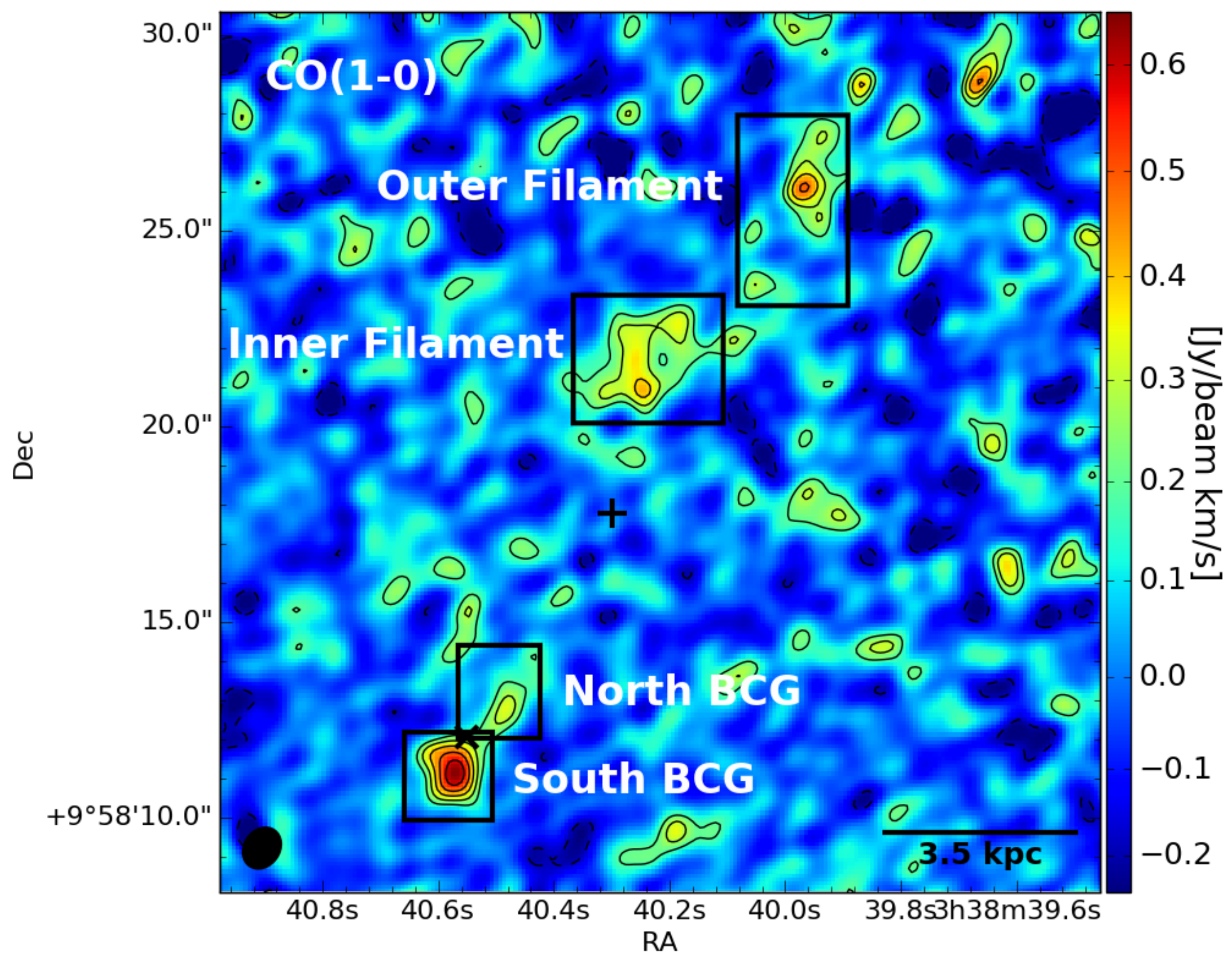}
  \caption{
    CO(1-0) integrated intensity map identifying regions with distinct molecular gas
    features. The spectra associated with these regions are shown in Fig. 
    \ref{fig:CO10regions}. The field of view is 22'' ($15.4\kpc$) on a side, showing
    the same area as in Fig. \ref{fig:COmaps}. The $\times$ and $+$ indicate the 
    centroids of the BCG and companion galaxy, respectively.
  }
  \label{fig:CO10subdiv}
\end{figure}

\begin{figure*}
\begin{minipage}{\textwidth}
  \centering
  \includegraphics[trim=0.0cm 0.0cm 0.0cm 2.0cm, clip=true, width=0.45\textwidth]{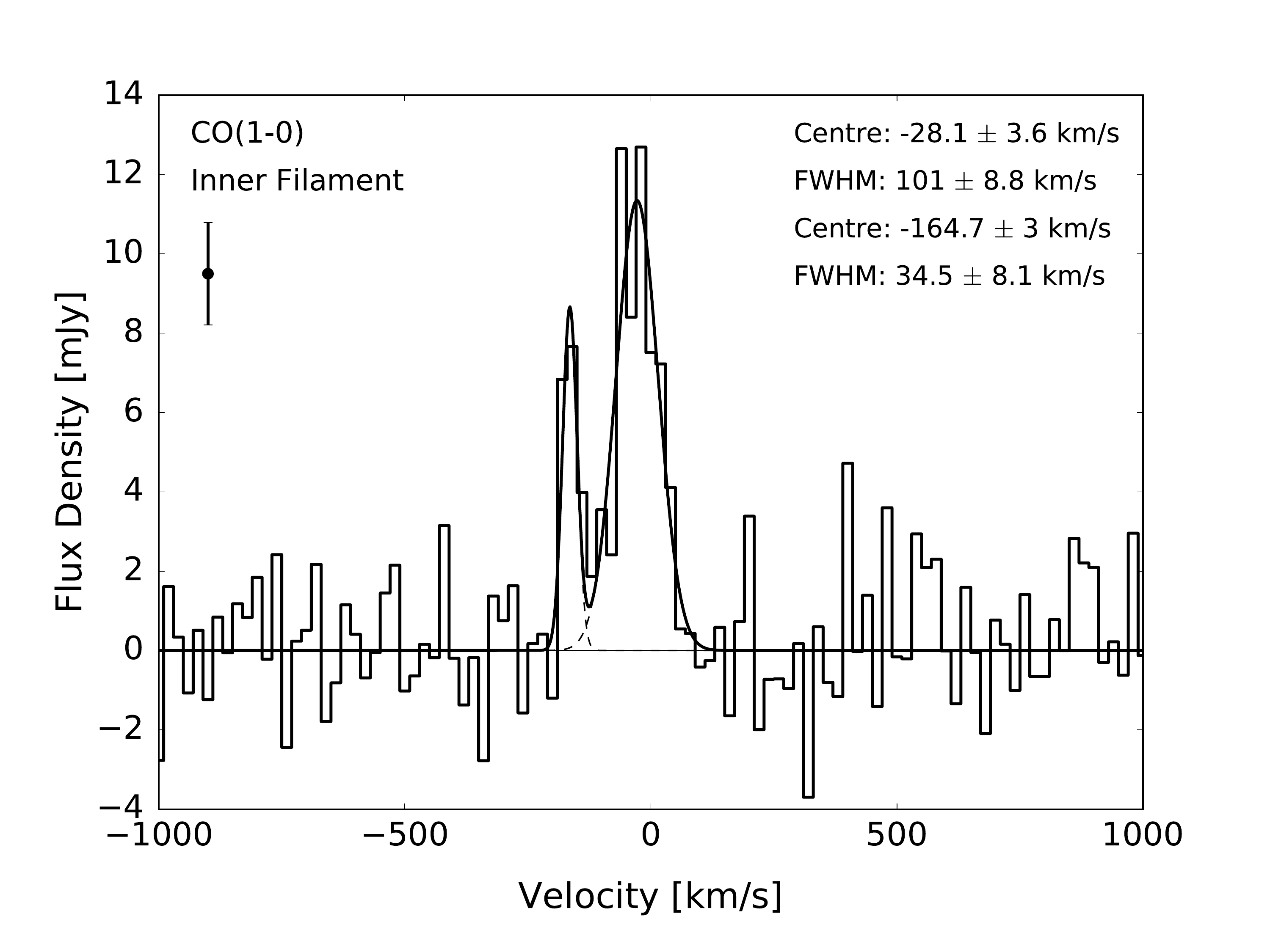}
  \includegraphics[trim=0.0cm 0.0cm 0.0cm 2.0cm, clip=true, width=0.45\textwidth]{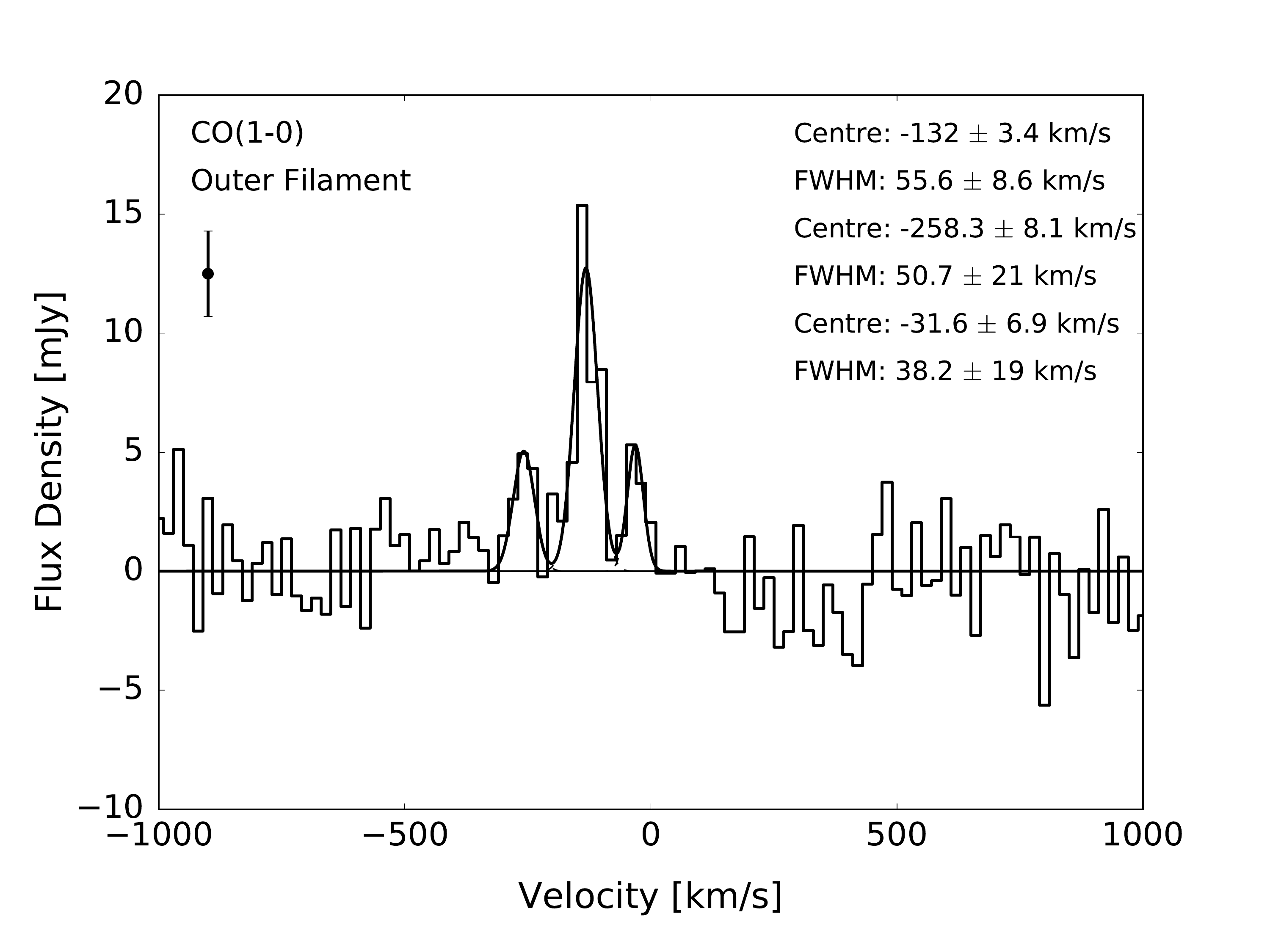}
\end{minipage}
\begin{minipage}{\textwidth}
  \centering
  \includegraphics[trim=0.0cm 0.0cm 0.0cm 2.0cm, clip=true, width=0.45\textwidth]{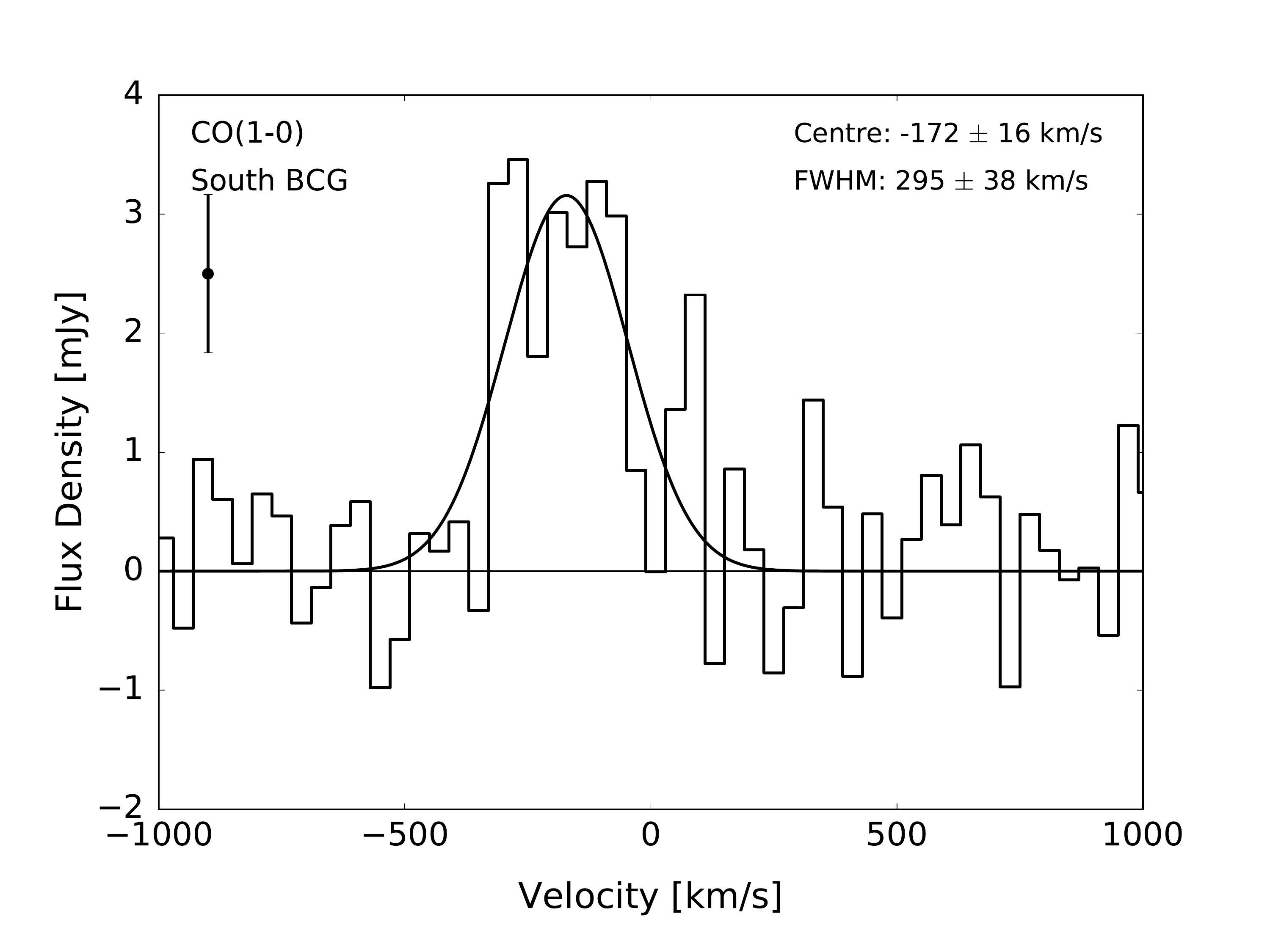}
  \includegraphics[trim=0.0cm 0.0cm 0.0cm 2.0cm, clip=true, width=0.45\textwidth]{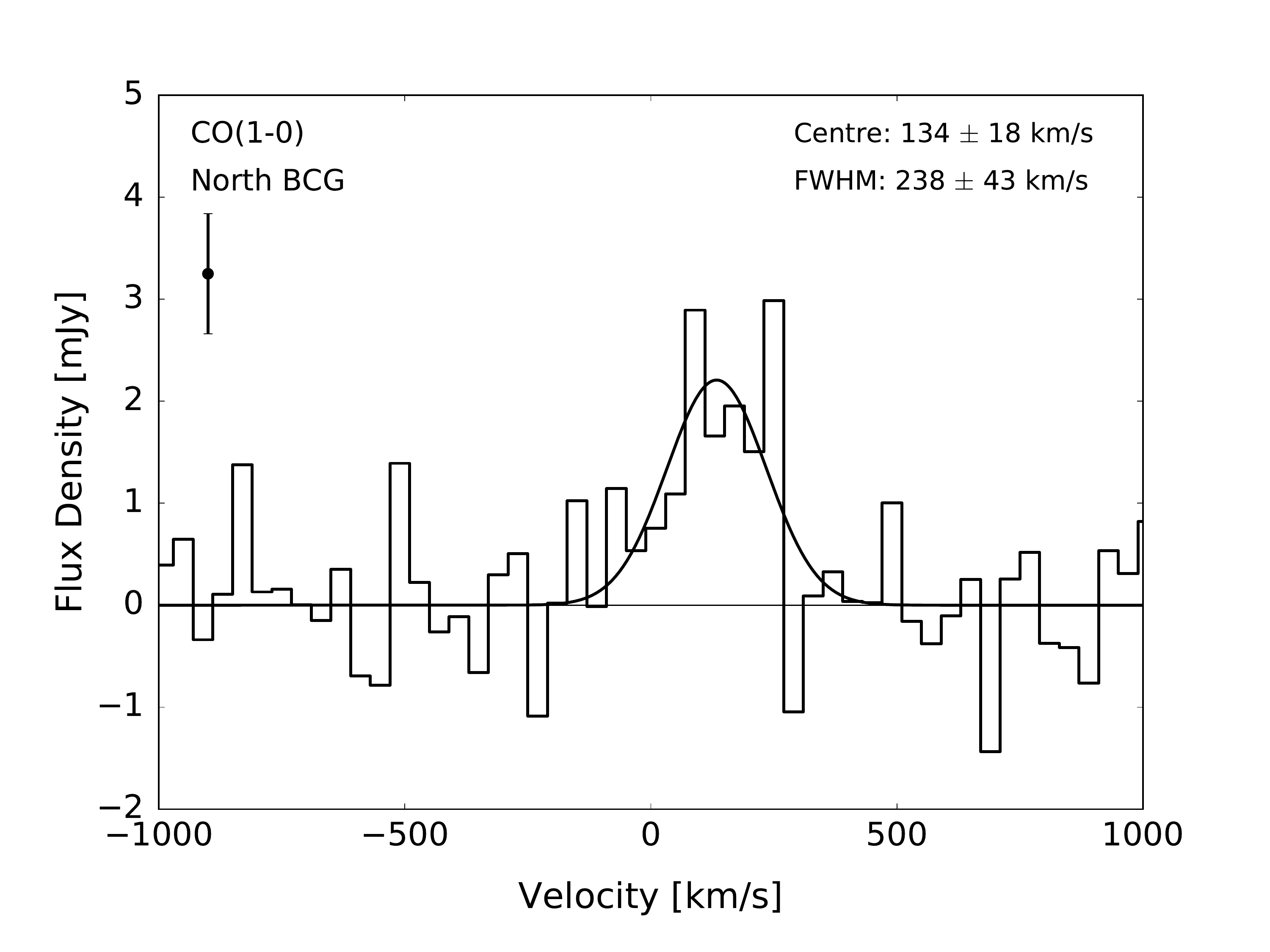}
\end{minipage}
  \caption{
    CO(1-0) spectra extracted from the regions shown in Fig. \ref{fig:CO10subdiv}.
    The best-fit parameters are given in Table \ref{tab:fitparameters}. The error 
    bar indicates the rms variation in the line-free channels. The ``Inner Filament''
    and ``Outer Filament'' spectra are presented with $20\kmps$ velocity channels,
    while the ``South BCG'' and ``North BCG'' have been binned up to $40\kmps$ to
    improve the signal-to-noise ratio.
  }
  \label{fig:CO10regions}
\end{figure*}

Spectra of the inner and outer clumps in the filament are shown in Fig. 
\ref{fig:CO10regions}, with the regions identified in Fig. \ref{fig:CO10subdiv}. 
Two velocity components are significantly detected in the inner filament. They 
are slightly blueshifted with respect to the BCG, with the primary component moving 
at $-28.1\pm3.6\kmps$ and the secondary at $-164.7\pm3.0\kmps$.
The outer filament consists of three velocity components, moving at
$-31.6\pm6.9\kmps$, $-132.0\pm3.4\kmps$, and $-258.3\pm8.1\kmps$.
The primary component is the central line, which is
blueshifted by $100\kmps$ relative to the main peak in the inner filament.
Along the entire filament the broadest component is the primary peak in the inner filament,
which still has a linewidth of only $101\pm9\kmps$ FWHM. All of the other peaks 
have widths in the range of $30-60\kmps$. These linewidths are much narrower than for
the gas in the BCG, reflecting the depth of the underlying gravitational potential.

\subsubsection{Companion Galaxy}

A small clump of gas detected in CO(3-2) is coincident with the nucleus of the companion
galaxy. This emission lies within the noise of the CO(1-0) data. The clump is also 
unresolved at CO(3-2), so spreading the total flux over the larger CO(1-0) beam 
will decrease the observed brightness.
A single-Gaussian fit to its CO(3-2) spectrum, shown in Fig. \ref{fig:CO32regions}, 
shows that the gas is redshifted to $224\pm19\kmps$ with respect to the systemic stellar 
component of the BCG. This is consistent with its stellar velocity of $212\pm58\kmps$, 
implying that the molecular gas is bound to the companion galaxy. 
This molecular gas linewidth of $256\pm46\kmps$ FWHM is fairly typical of small elliptical
galaxies, so the molecular gas may be virialized within the galaxy.
Following the mass conversion discussed in Section \ref{sec:mass} and assuming a line flux 
ratio of ${\rm CO(3-2)/CO(1-0)} \approx 7$, the total molecular gas mass within the 
companion galaxy is $5.6\pm0.9\e{7}\Msun$.

\begin{figure}
  \centering
  \includegraphics[width=\linewidth]{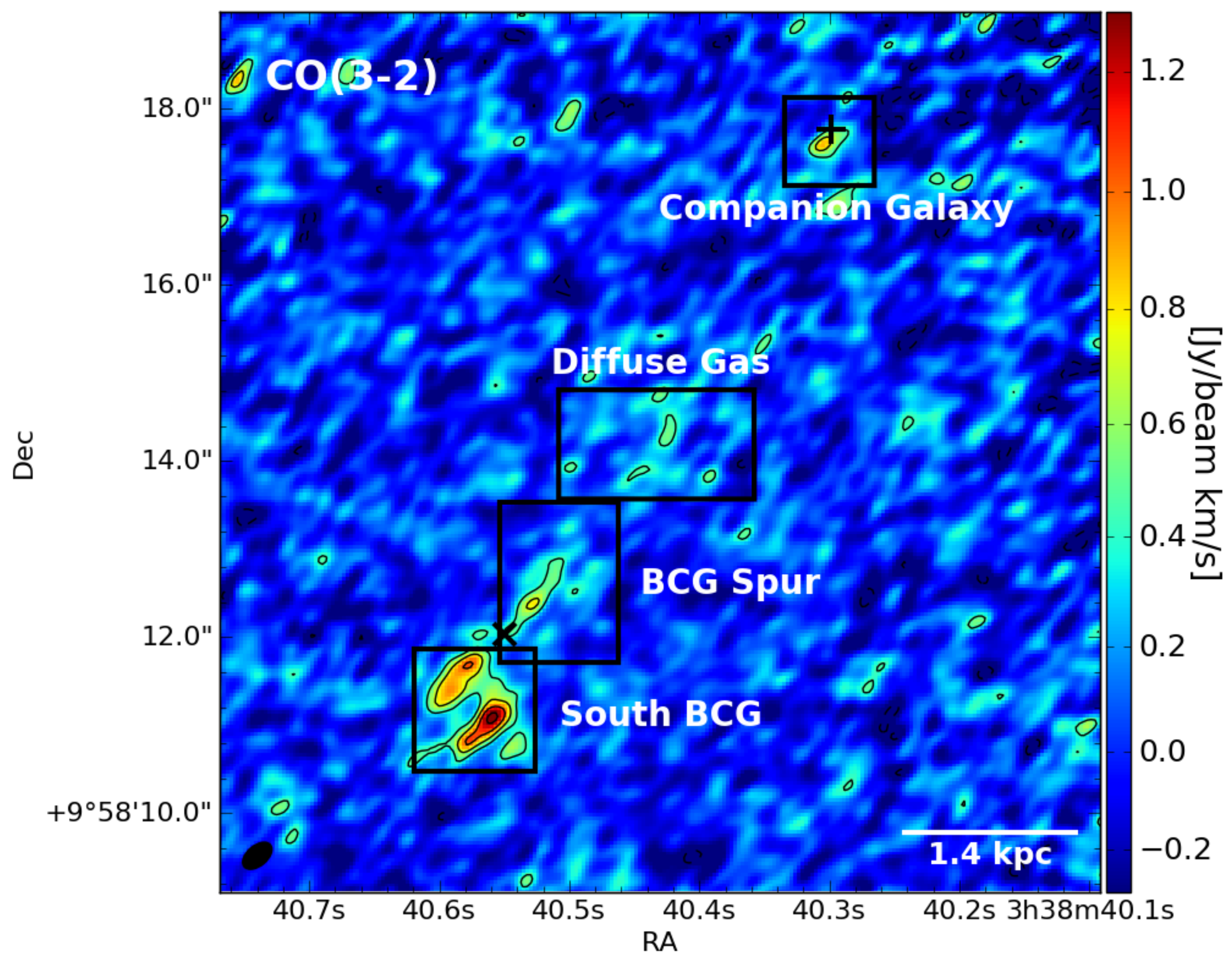}
  \caption{
    CO(3-2) integrated intensity map with regions highlighting the different structures
    seen in the BCG and the companion galaxy. The associated spectra are shown in
    Fig. \ref{fig:CO32regions}. The field of view is $10''$ ($7\kpc$) on a side, showing
    the same area as in Fig. \ref{fig:COmaps}.
  }
  \label{fig:CO32subdiv}
\end{figure}

\begin{figure*}
\begin{minipage}{\textwidth}
  \centering
  \includegraphics[trim=0.0cm 0.0cm 0.0cm 2.0cm, clip=true, width=0.45\textwidth]{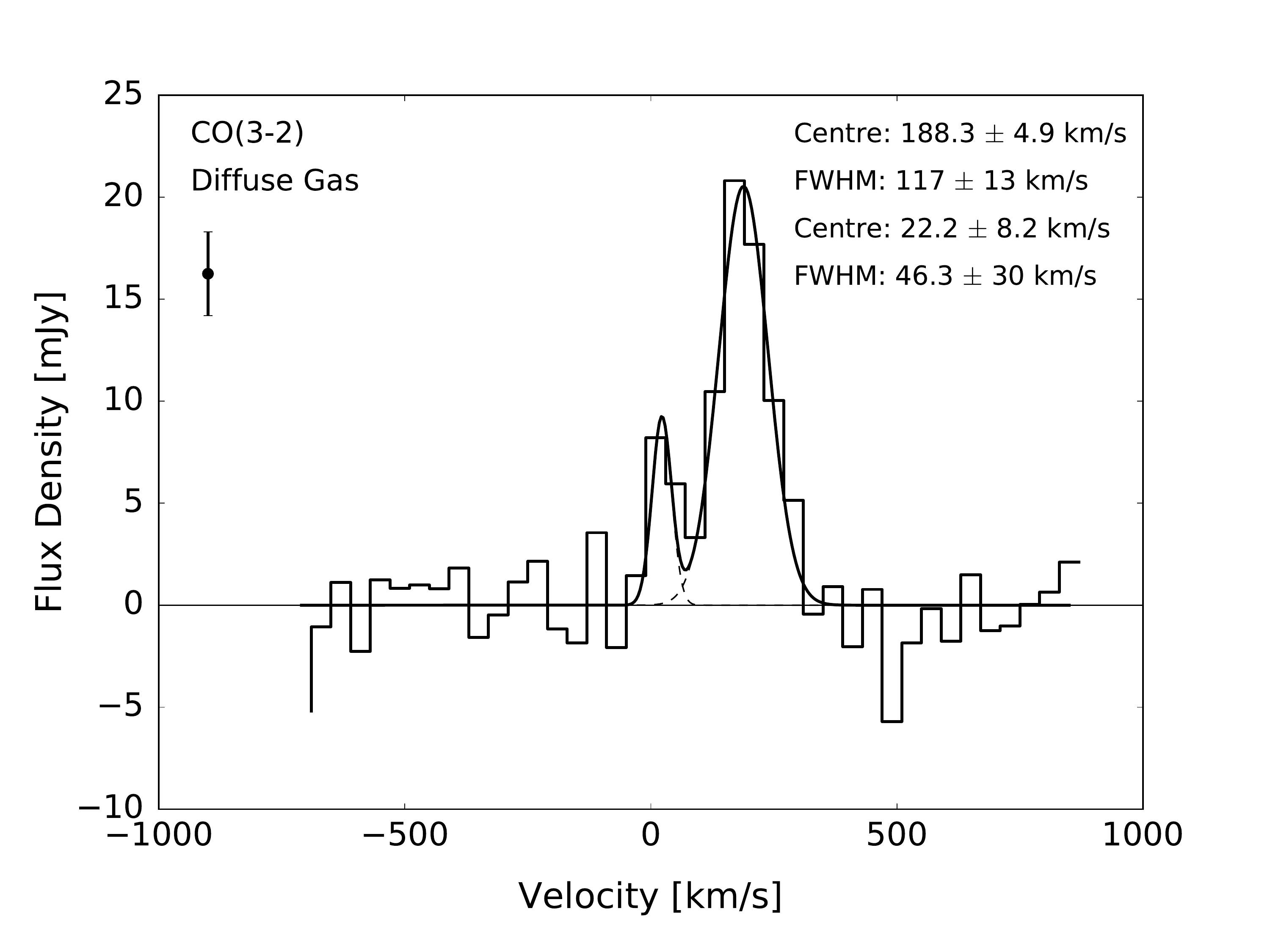}
  \includegraphics[trim=0.0cm 0.0cm 0.0cm 2.0cm, clip=true, width=0.45\textwidth]{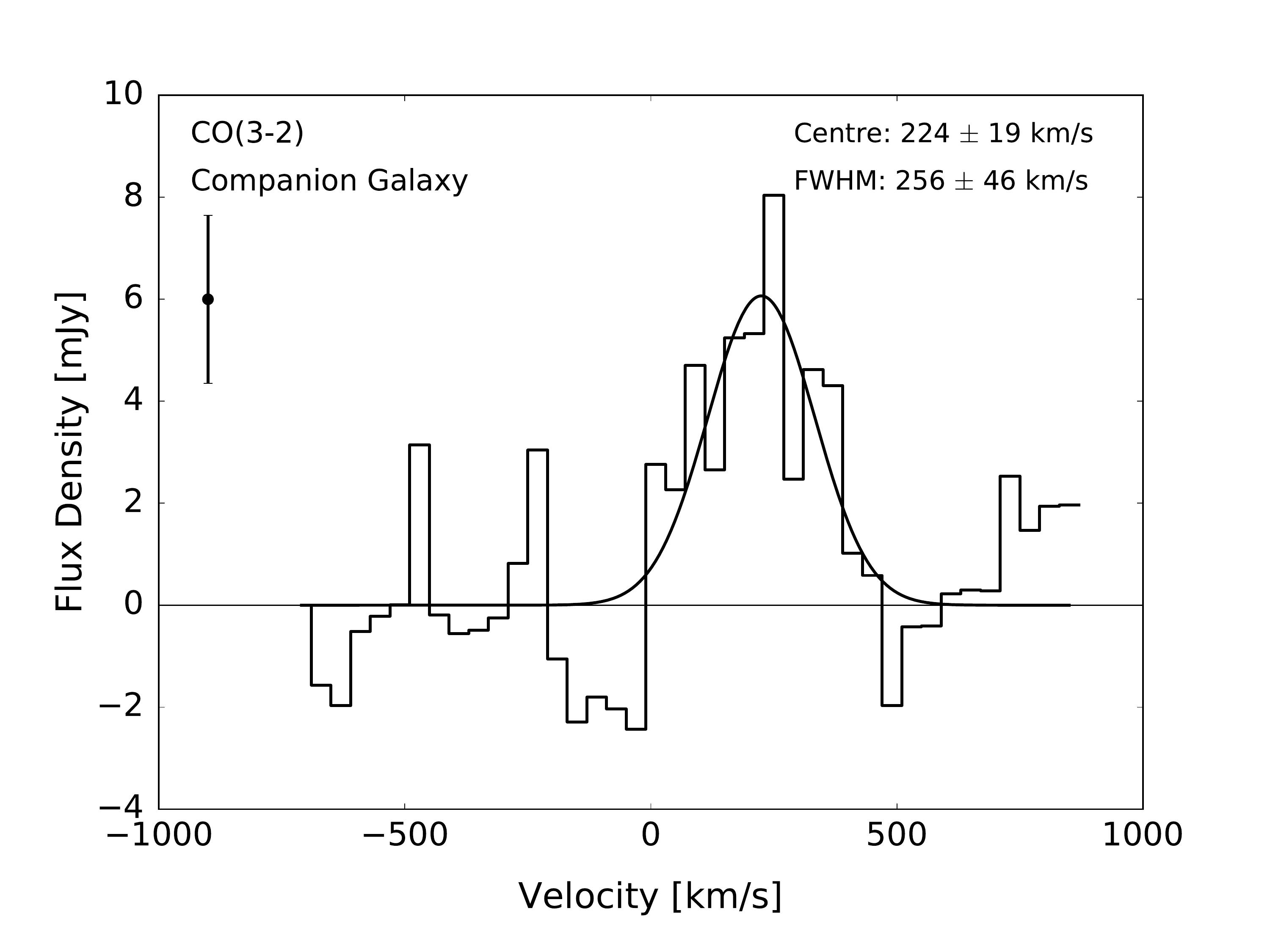}
\end{minipage}
\begin{minipage}{\textwidth}
  \centering
  \includegraphics[trim=0.0cm 0.0cm 0.0cm 2.0cm, clip=true, width=0.45\textwidth]{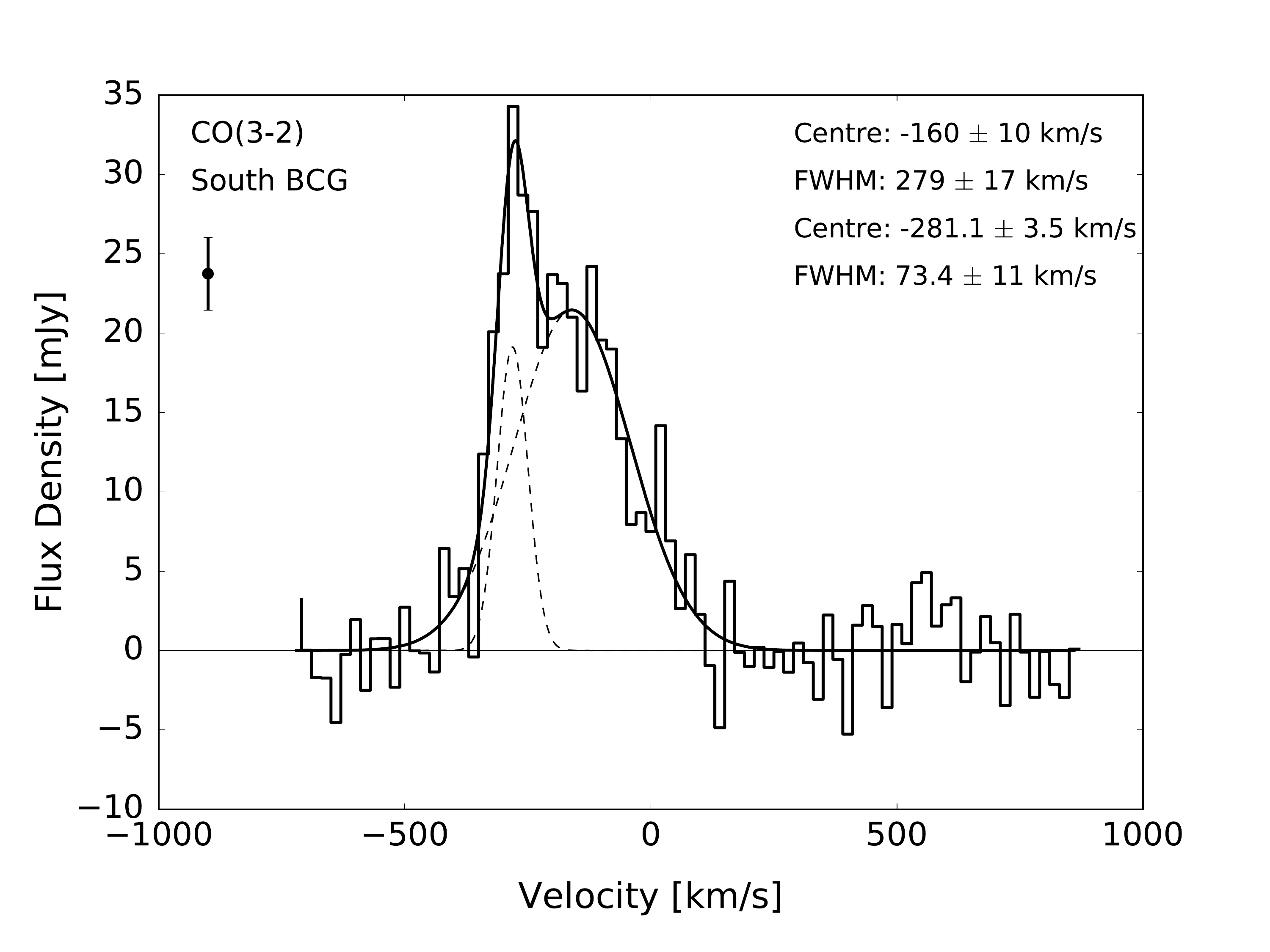}
  \includegraphics[trim=0.0cm 0.0cm 0.0cm 2.0cm, clip=true, width=0.45\textwidth]{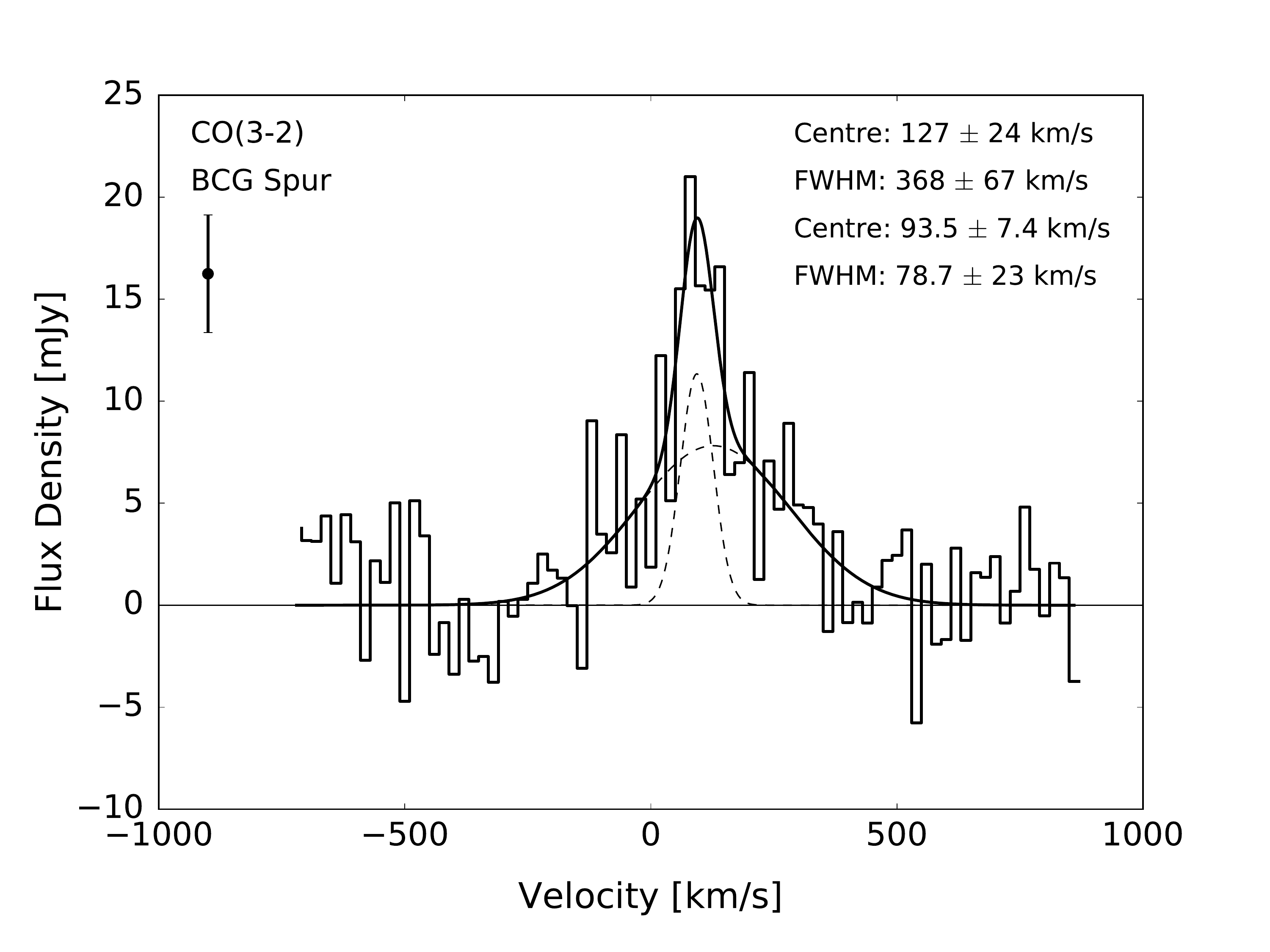}
\end{minipage}
\begin{minipage}{\textwidth}
  \centering
  \includegraphics[trim=0.0cm 0.0cm 0.0cm 2.0cm, clip=true, width=0.45\textwidth]{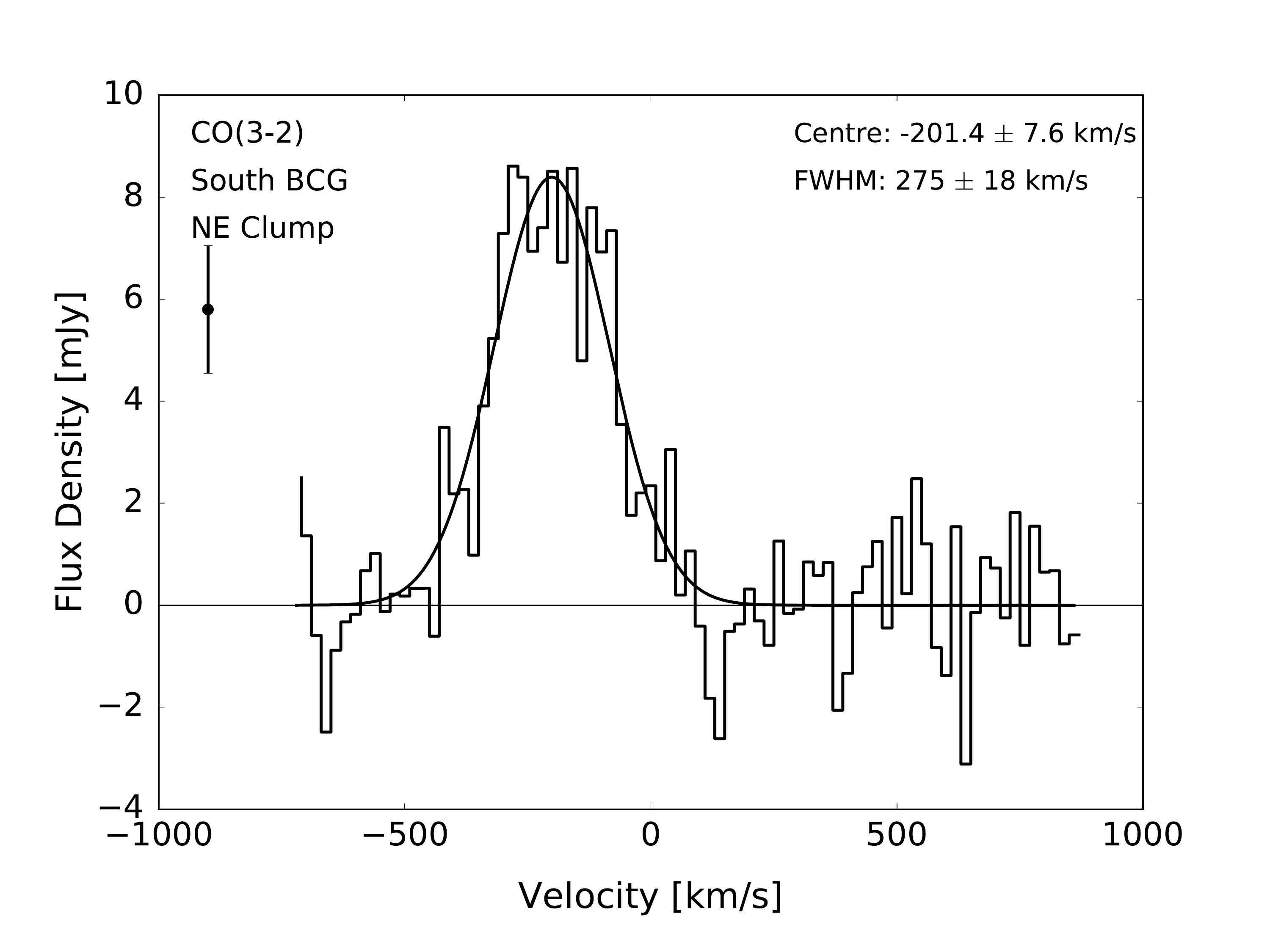}
  \includegraphics[trim=0.0cm 0.0cm 0.0cm 2.0cm, clip=true, width=0.45\textwidth]{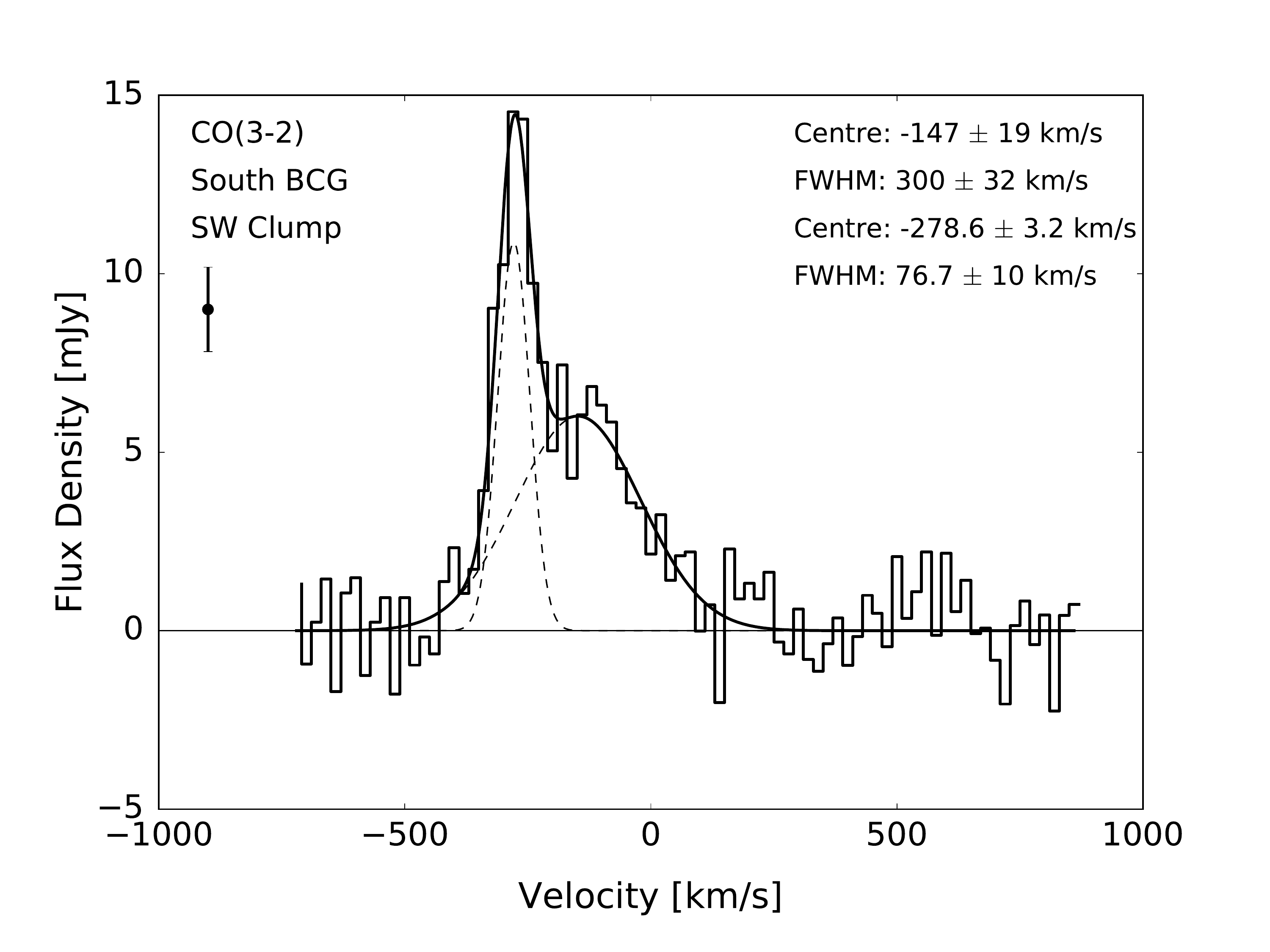}
\end{minipage}
  \caption{
    CO(3-2) spectra extracted from the regions shown in Fig. \ref{fig:CO32subdiv}.
    Additional spectra are shown for each of the two clumps within the South BCG
    region, which were extracted from $0.6''\times 0.6''$ boxes. The best-fit 
    parameters are given in Table \ref{tab:fitparameters}. Each spectrum was initially
    extracted with $20\kmps$ velocity bins. The ``Diffuse Gas'' spectrum was binned 
    up to $40\kmps$ channels to improve the signal-to-noise ratio, and ``Companion 
    Galaxy'' was binned to $60\kmps$. The error bar indicates the rms variation in 
    the line-free channels.
  }
  \label{fig:CO32regions}
\end{figure*}

\subsection{Molecular Gas Mass}
\label{sec:mass}

The integrated flux of the CO(1-0) line ($S_{\rm CO}\Delta v$) can be 
converted to a molecular gas mass, assuming a constant CO-to-H$_2$ conversion 
factor ($X_{\rm CO}$), according to the equation \citep{solomon87,solomon05,bolatto13}
\begin{equation}
M_{\rm mol} = 1.05\e{4} X_{\rm CO, Gal} \left( \frac{S_{\rm CO}\Delta v \,D_L^2}{1+z} \right)  \Msun.
\end{equation}
Here $S_{\rm CO}\Delta v$ is expressed in $\Jykmps$, $D_L$ is the luminosity
distance in $\Mpc$, and $z$ is the redshift of the BCG.
We assume a Galactic CO-to-H$_2$ conversion factor, 
$X_{\rm CO, gal} = 2\e{-20} \pcmsq (\K\kmps)^{-1}$,
which is typical of molecular clouds in the disk of the Milky Way. 
However, the conversion factor is not universal. In particular, low metallicity 
regions tend to have high values of $X_{\rm CO}$. 
The Galactic conversion factor can be approximately applied down to metallicities 
of $\sim0.5\Zsun$ \citep{bolatto13}.
In 2A~0335+096, the metallicity inferred from the intracluster medium peaks at 
$0.95\pm0.06\Zsun$ at the cluster centre \citep{sanders09}, so the Galactic
$X_{\rm CO}$ should be appropriate.
On the other hand, LIRG and starburst galaxies are known to have very low values
of $X_{\rm CO}$. However, the Spitzer-derived total IR luminosity of the BCG,
$6.7\e{9}\Lsun$ \citep{donahue11}, is well below the $10^{11}\Lsun$ threshold
for a LIRG.
Narrow absorption features observed in NGC5044 \citep{david14} and A2597 
\citep{tremblay16} suggest that molecular clouds in BCGs have similar linewidths 
to those of the Milky Way.
We therefore expect that the Galactic value is appropriate here. Our broad conclusions
are not affected by a factor of few difference in the adopted conversion factor.

The total CO(1-0) flux detected in our observations, $4.8\pm0.6\Jykmps$, corresponds
to a molecular gas mass of $1.13\pm0.15\e{9}\Msun$. Of this, $3.2\pm0.4\e{8}\Msun$
is localized to the BCG and $7.8\pm0.9\e{8}\Msun$ is contained in the elongated
filament north of the companion galaxy. Table \ref{tab:fitparameters} lists
the molecular gas mass associated with each CO(1-0) spectrum.

Masses are derived from CO(3-2) assuming a constant CO(3-2)/CO(1-0) flux ratio. 
This ratio was determined by smoothing the CO(3-2) data cube to the CO(1-0) resolution 
and extracting the spectrum from a 5$\times$5 arcsec box centred on the BCG for 
each line. The spectra were each fitted with two Gaussian components. Summing the 
fluxes of each component, the resulting flux ratio is ${\rm CO(3-2)/CO(1-0)}=7.2\pm1.4$. 
We therefore adopt a factor of 7 difference between the two integrated fluxes, with 
the caveat that the conversion is only accurate to $\sim20\%$. The ratio of integrated 
brightness temperature (in units of $\K\kmps$) is $0.80\pm0.16$. For optically thick CO 
emission this ratio indicates that the gas is approximately thermalized, implying gas 
densities above $10^{4}\pcmcu$.

\begin{figure*}
\begin{minipage}{\textwidth}
  \centering
  \includegraphics[trim=0cm 1.5cm 0cm 0cm, clip=true, width=6.5cm]{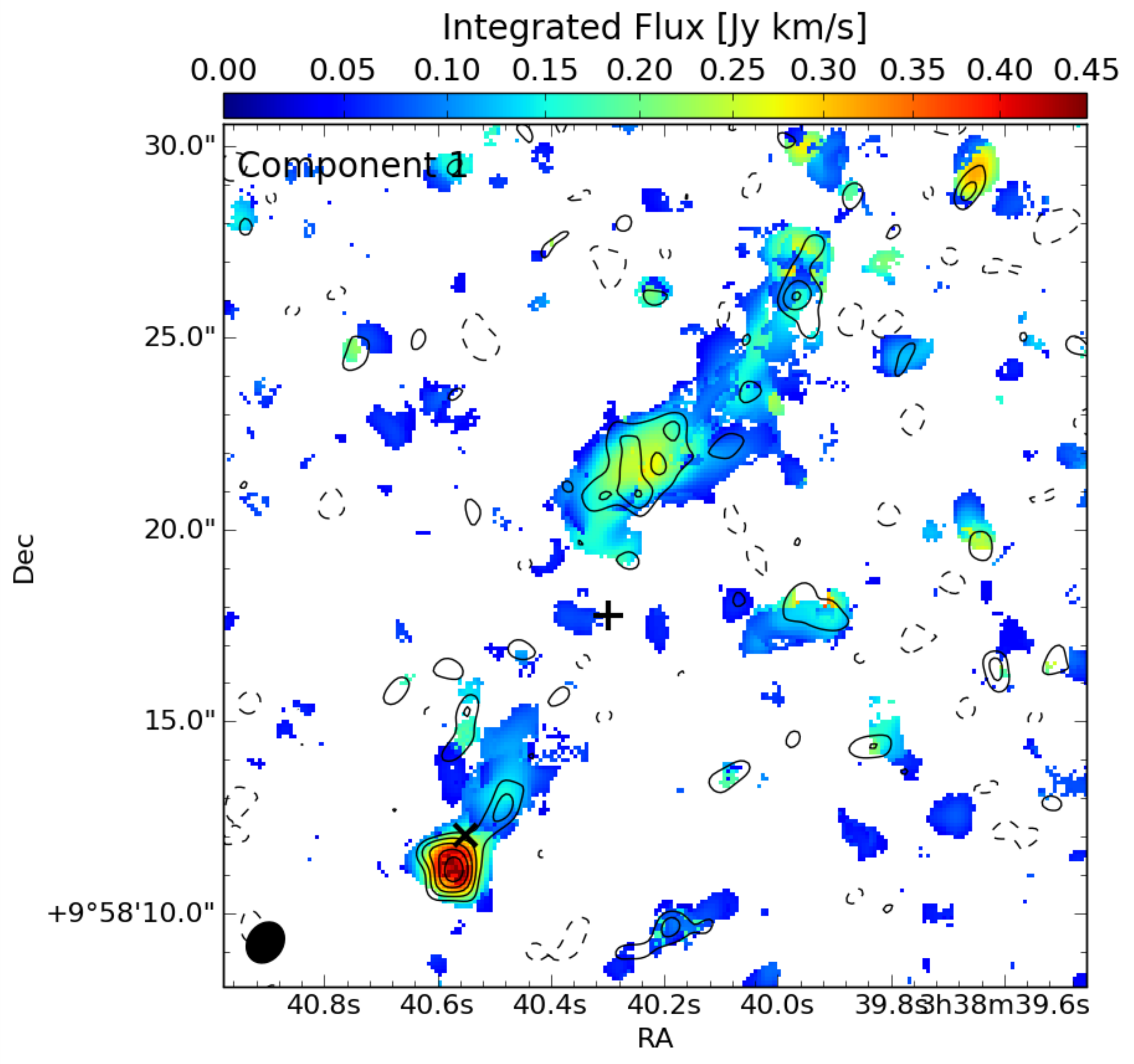}
  \includegraphics[trim=4.5cm 1.5cm 0cm 0cm, clip=true, width=5.125cm]{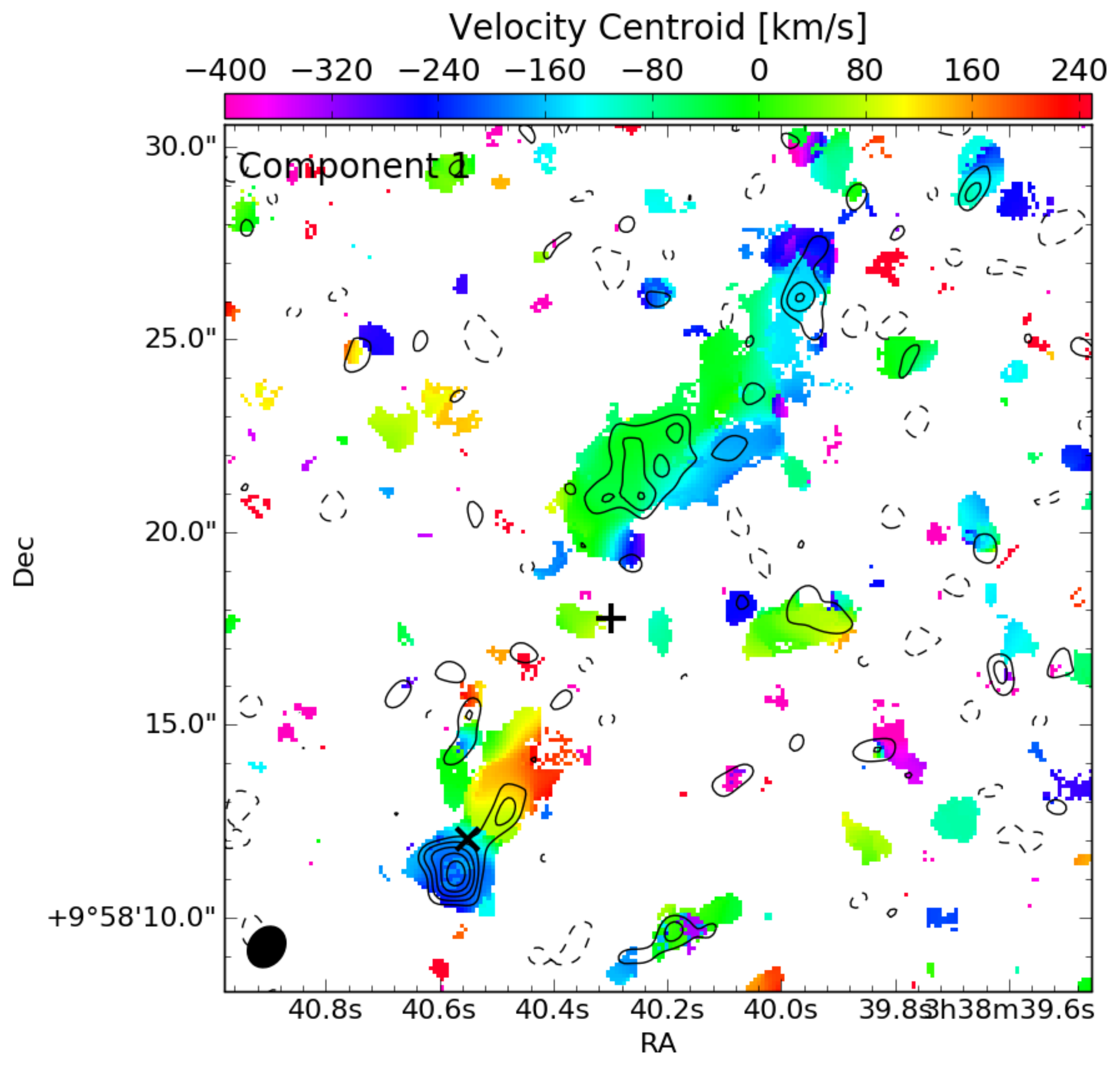}
  \includegraphics[trim=4.5cm 1.5cm 0cm 0cm, clip=true, width=5.05cm]{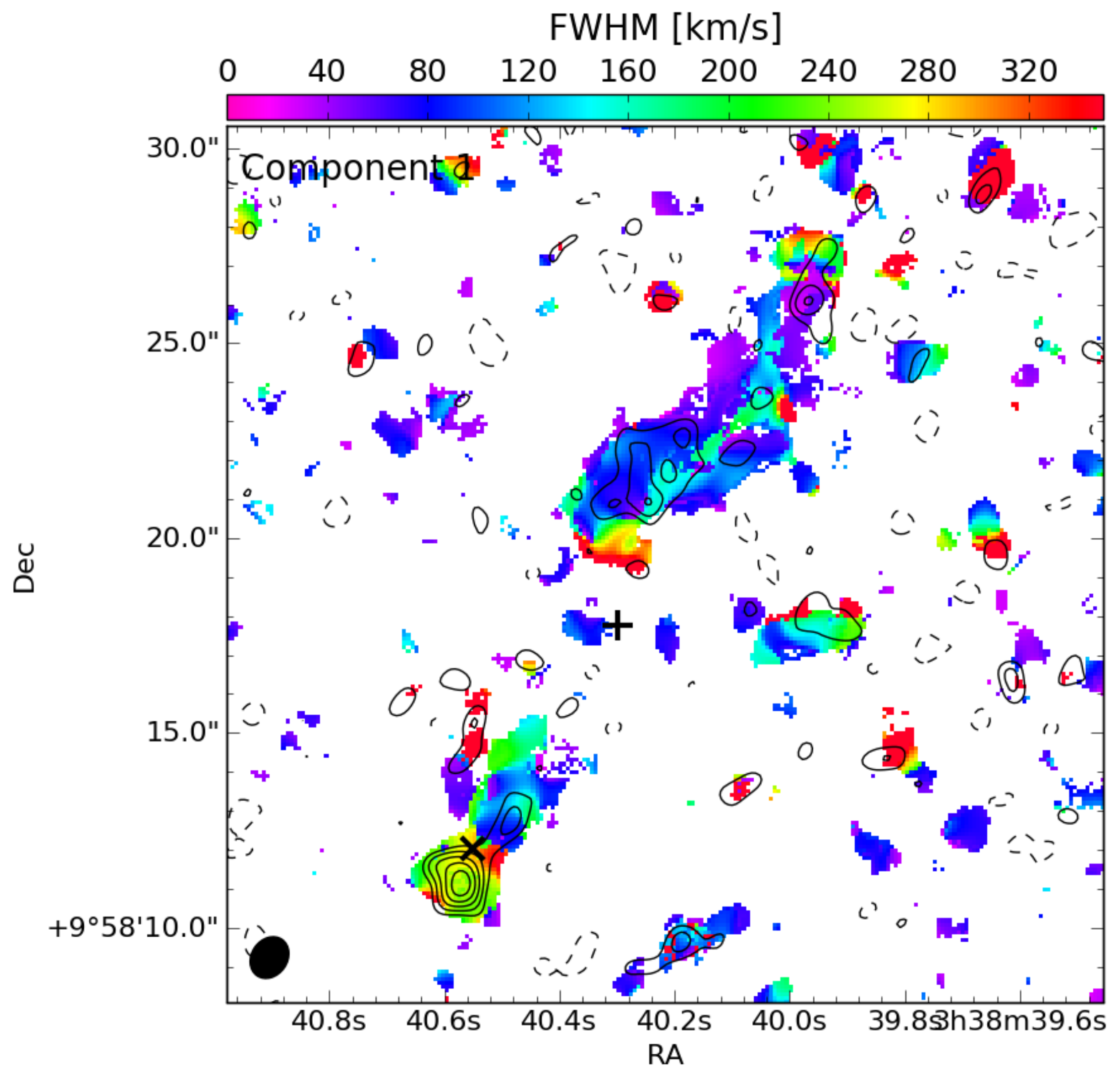}
\end{minipage}
\begin{minipage}{\textwidth}
  \centering
  \includegraphics[trim=0.0cm 0cm 0cm 2.5cm, clip=true, width=6.5cm]{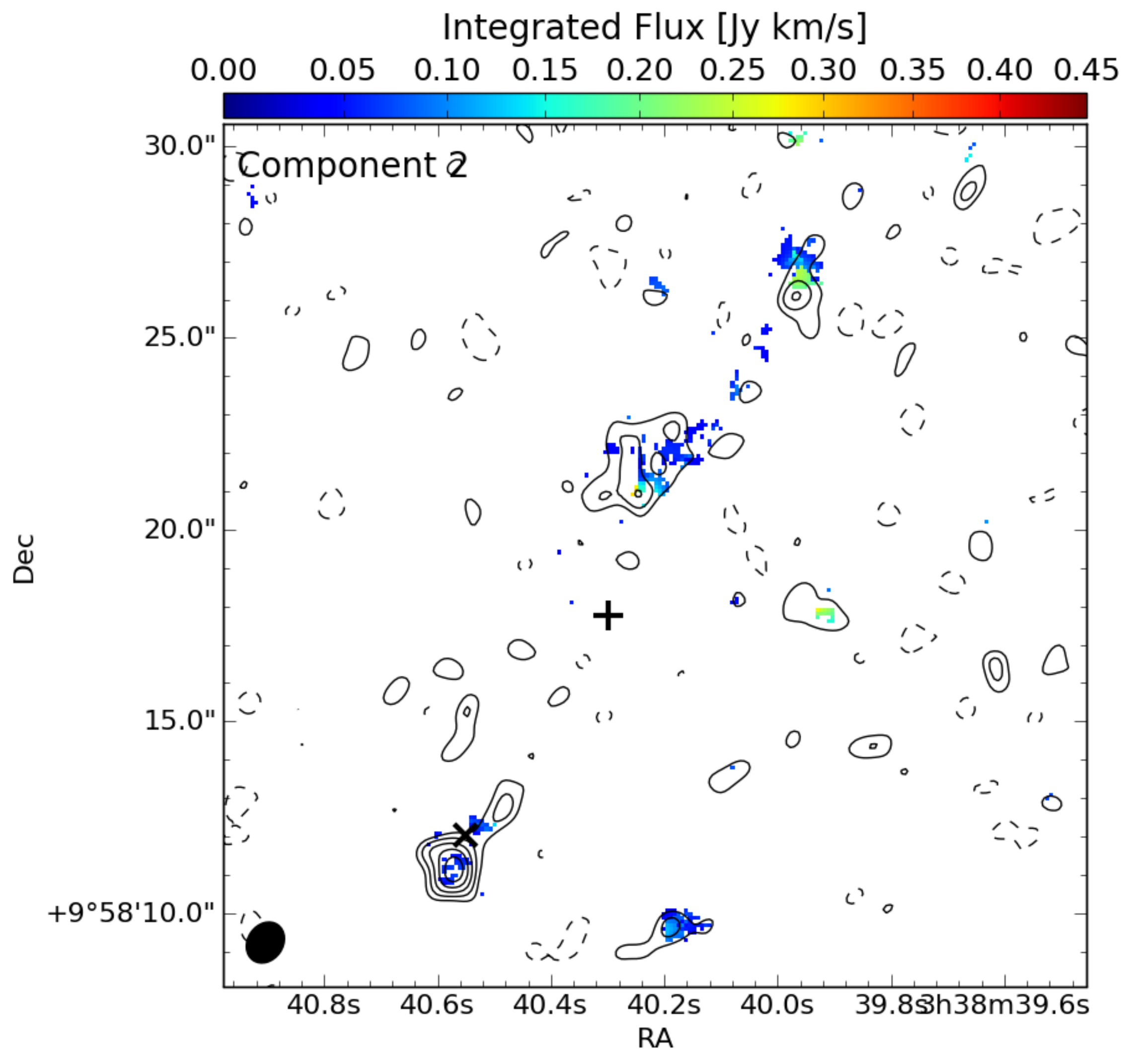}
  \includegraphics[trim=4.5cm 0cm 0cm 2.5cm, clip=true, width=5.125cm]{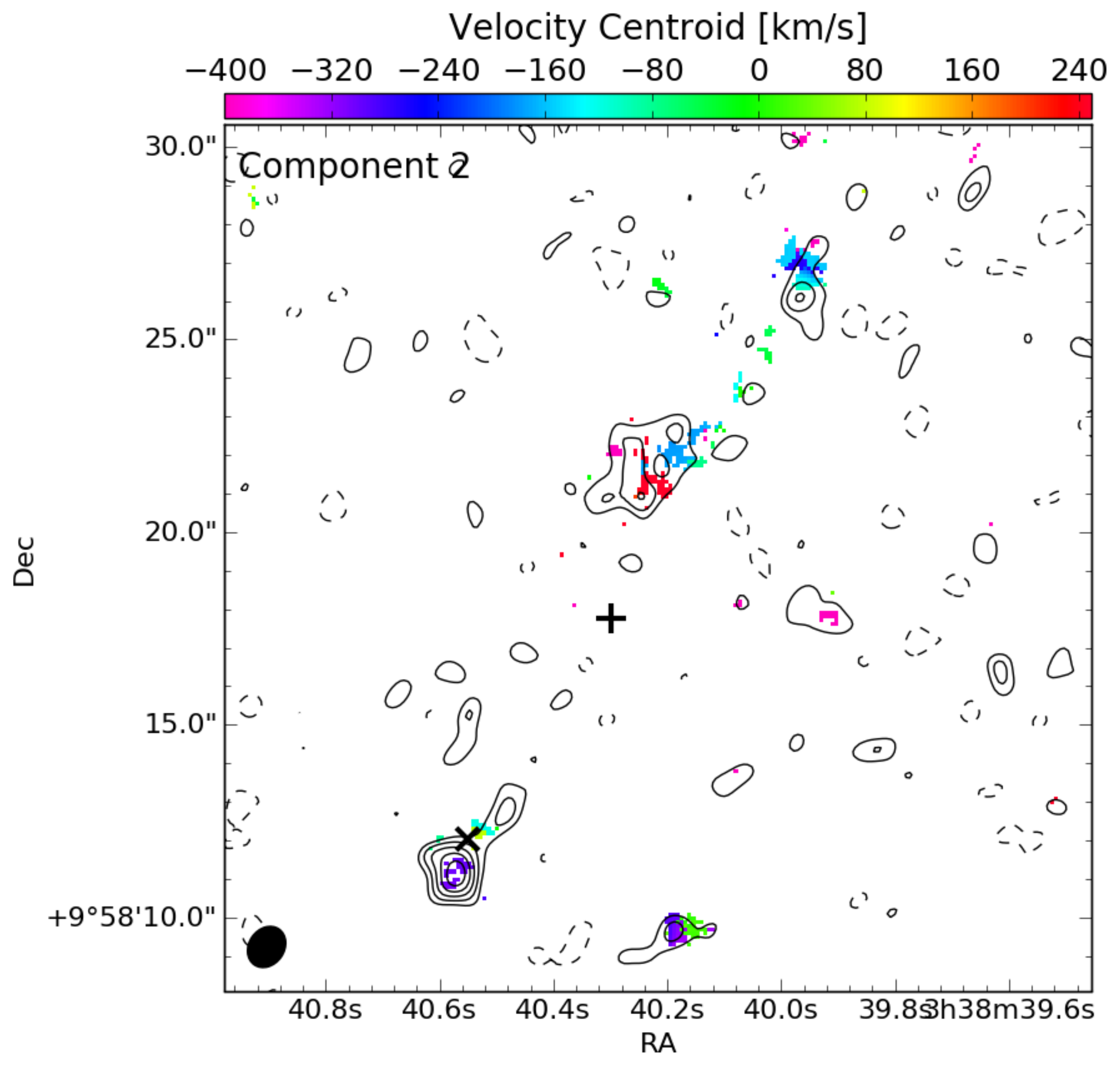}
  \includegraphics[trim=4.5cm 0cm 0cm 2.5cm, clip=true, width=5.05cm]{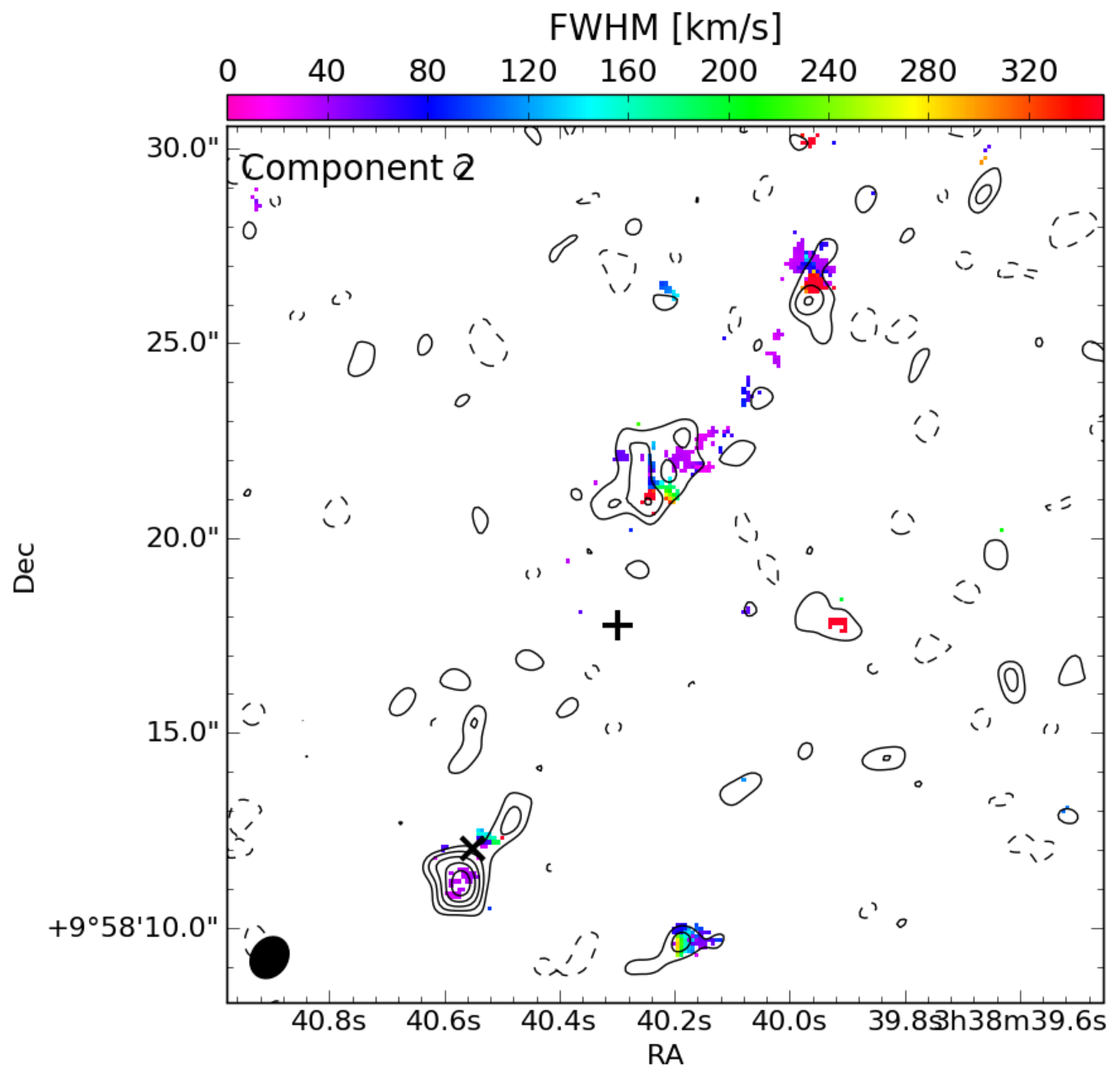}
\end{minipage}
  \caption{
    Maps of integrated flux ({\it left}), velocity centroid ({\it middle}), and 
    FWHM ({\it right}) obtained from pixel-by-pixel fitting of the CO(1-0) datacube, 
    as described in the text. Only pixels containing flux detected at $>2\sigma$ are 
    shown. The contours are the same as in Fig. \ref{fig:COmaps} (top).
    The $\times$ and $+$ indicate the centroids of the BCG and companion galaxy, respectively.
  }
  \label{fig:CO10vel}
\end{figure*}

\begin{figure*}
\begin{minipage}{\textwidth}
  \centering
  \includegraphics[trim=0cm 1.5cm 0cm 0cm, clip=true, width=6.5cm]{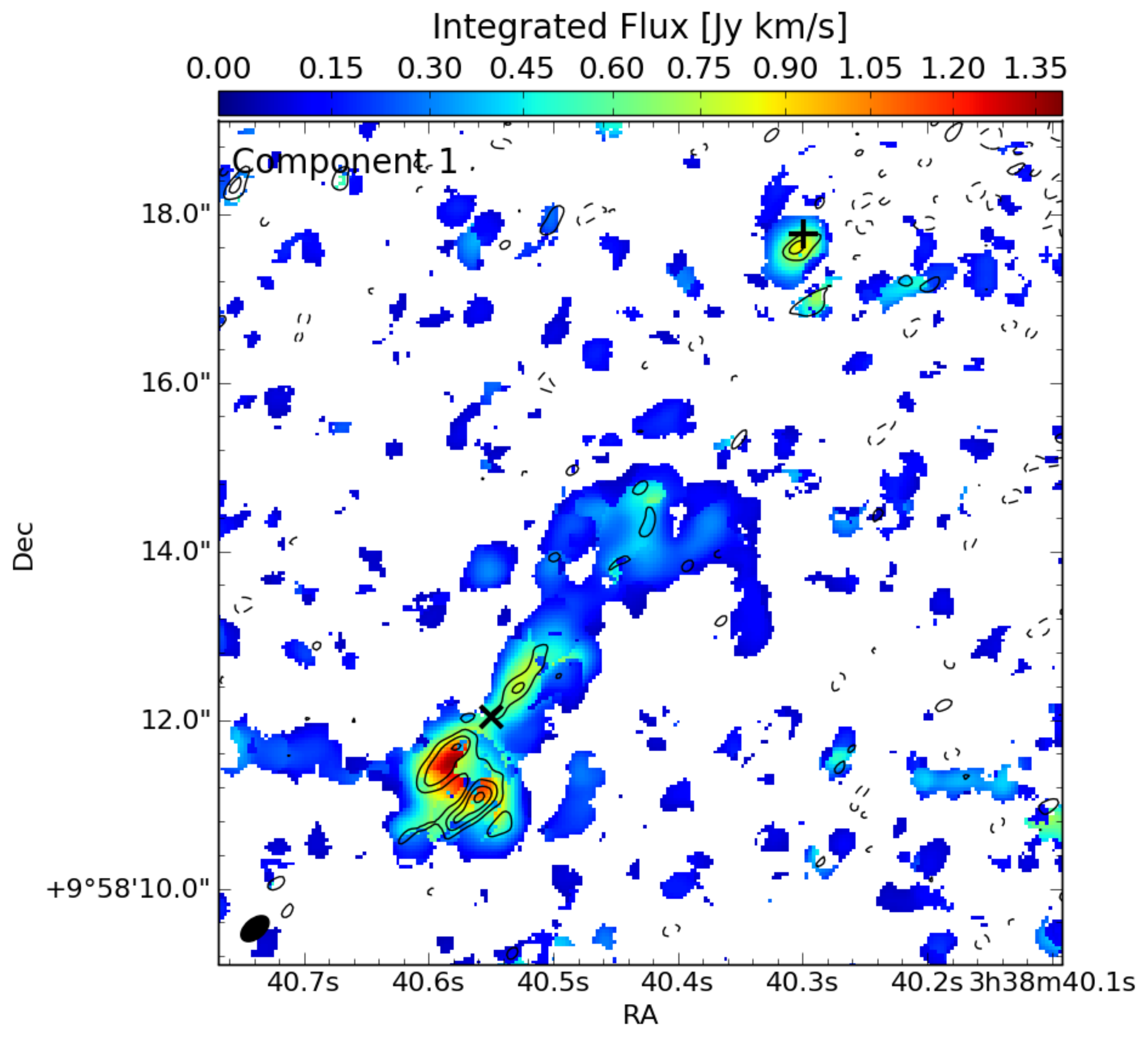}
  \includegraphics[trim=4.5cm 1.5cm 0cm 0cm, clip=true, width=5.3cm]{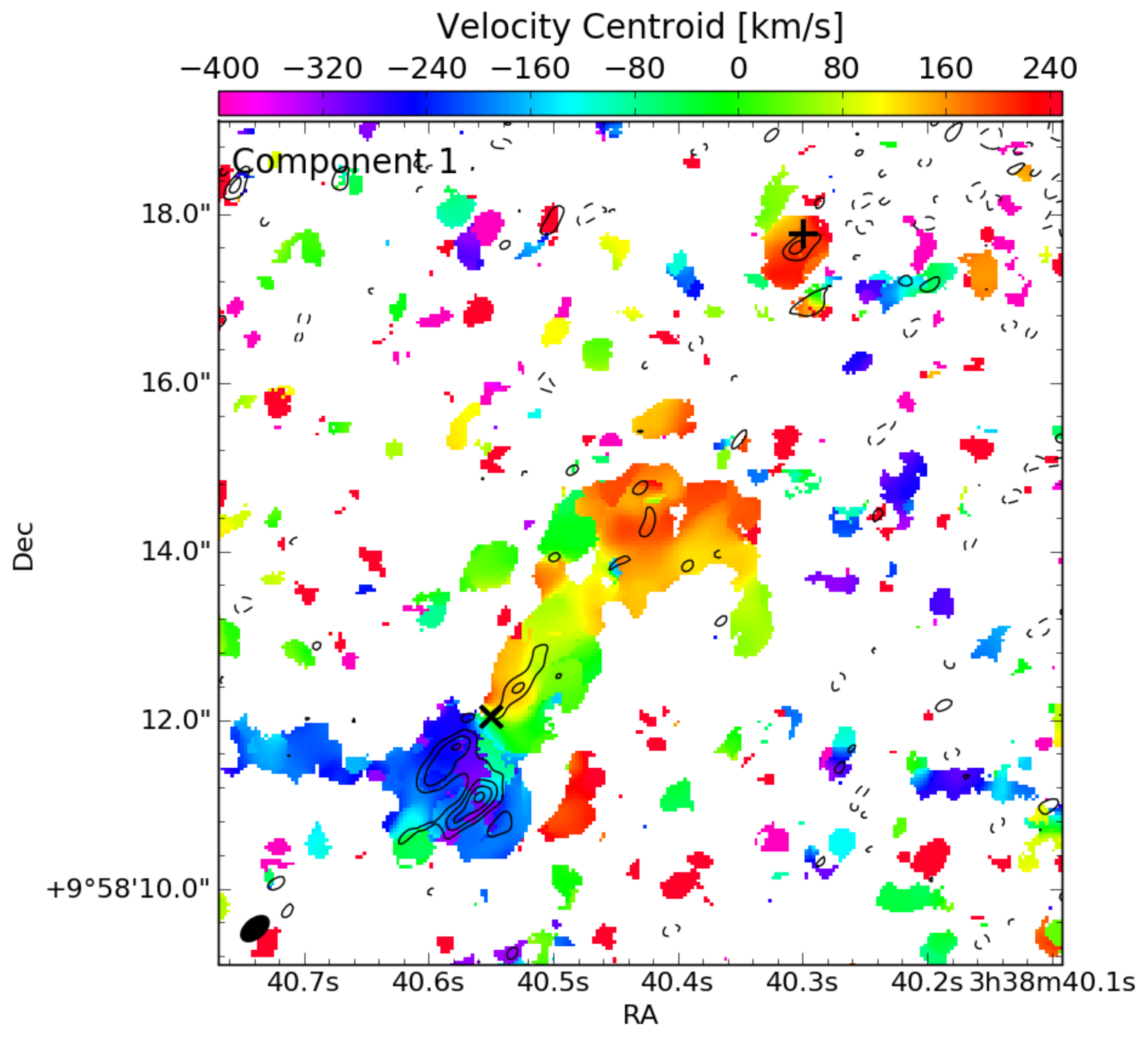}
  \includegraphics[trim=4.5cm 1.5cm 0cm 0cm, clip=true, width=5.3cm]{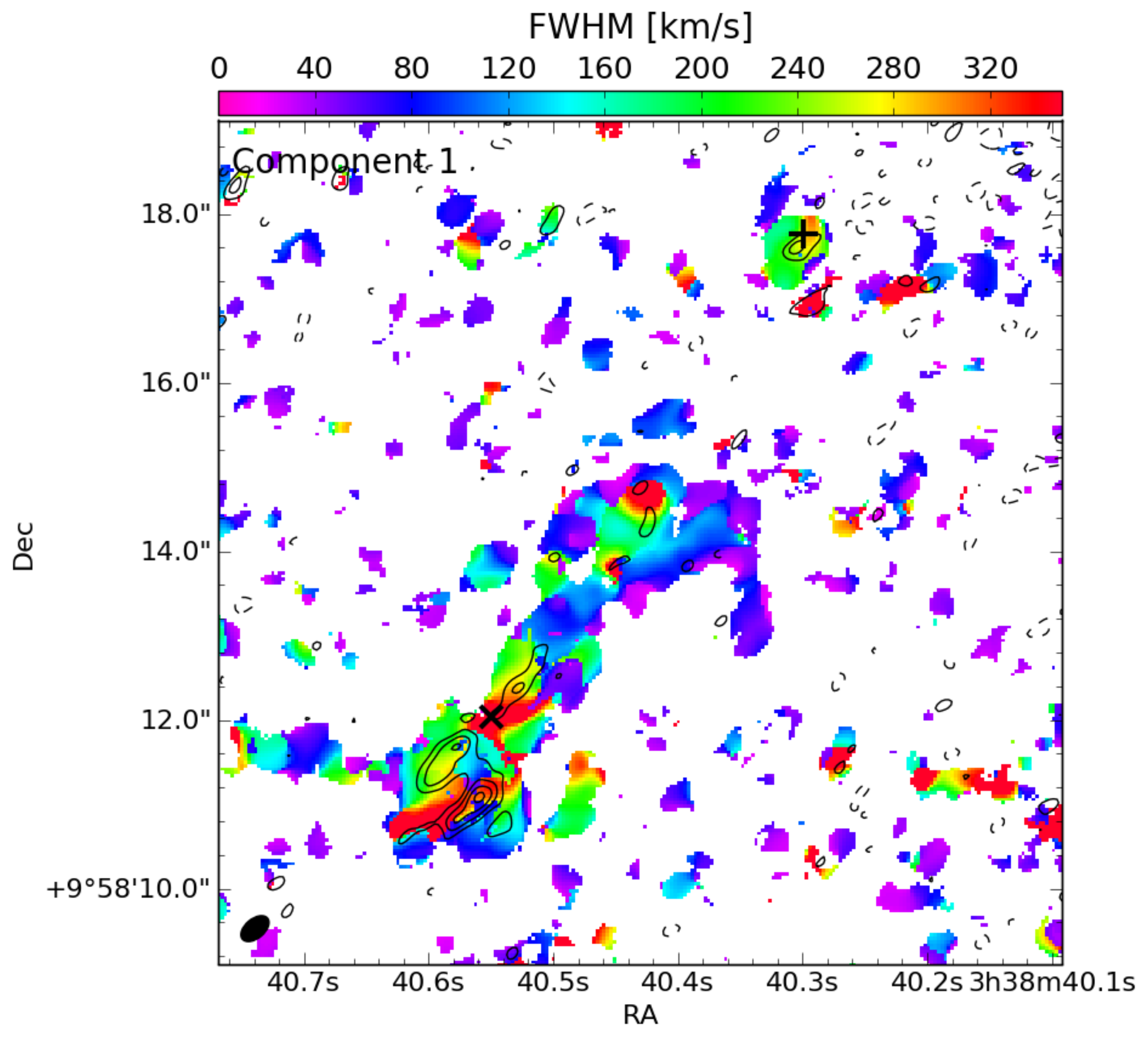}
\end{minipage}
\begin{minipage}{\textwidth}
  \centering
  \includegraphics[trim=0.0cm 0cm 0cm 2.5cm, clip=true, width=6.5cm]{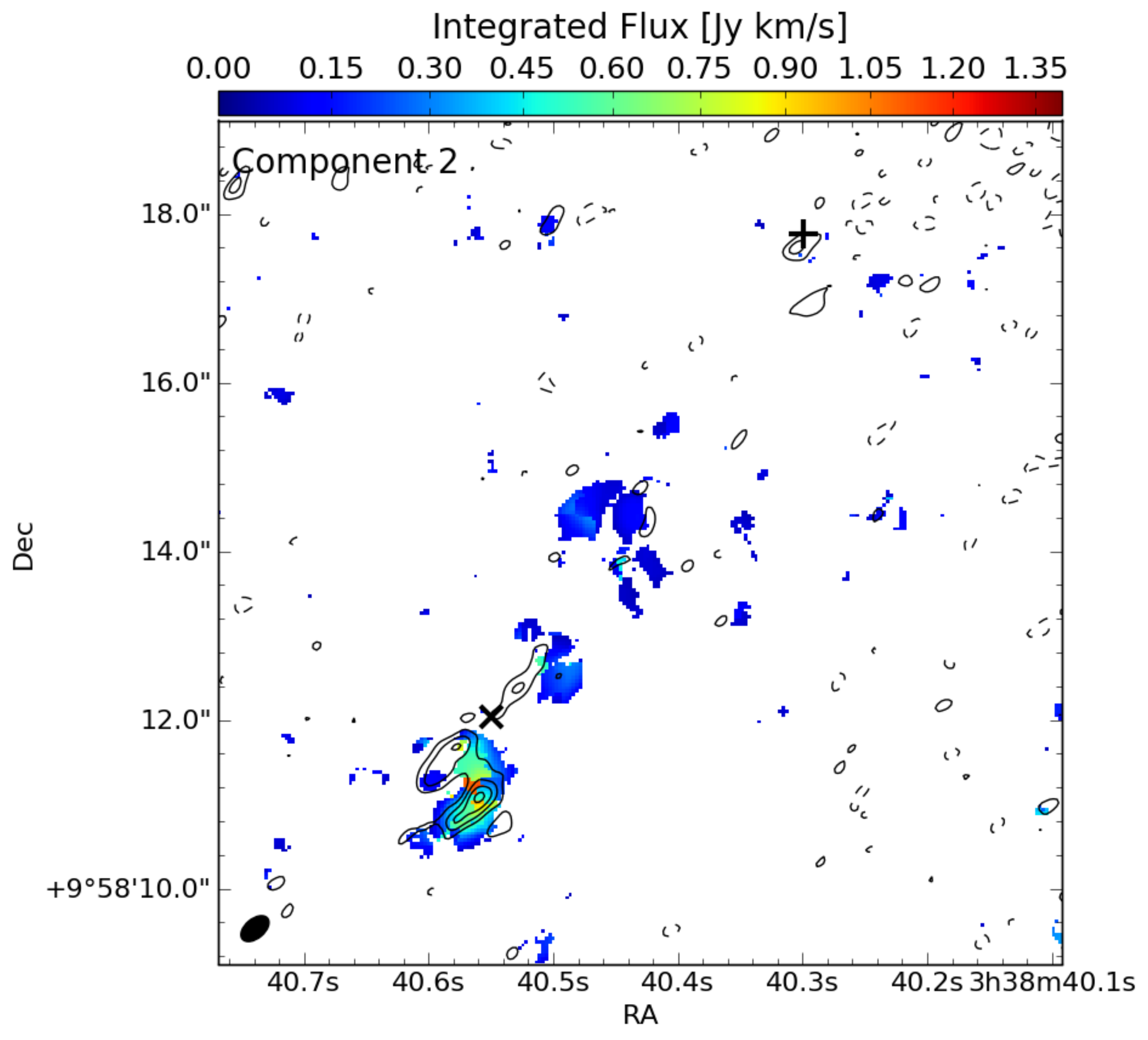}
  \includegraphics[trim=4.5cm 0cm 0cm 2.5cm, clip=true, width=5.3cm]{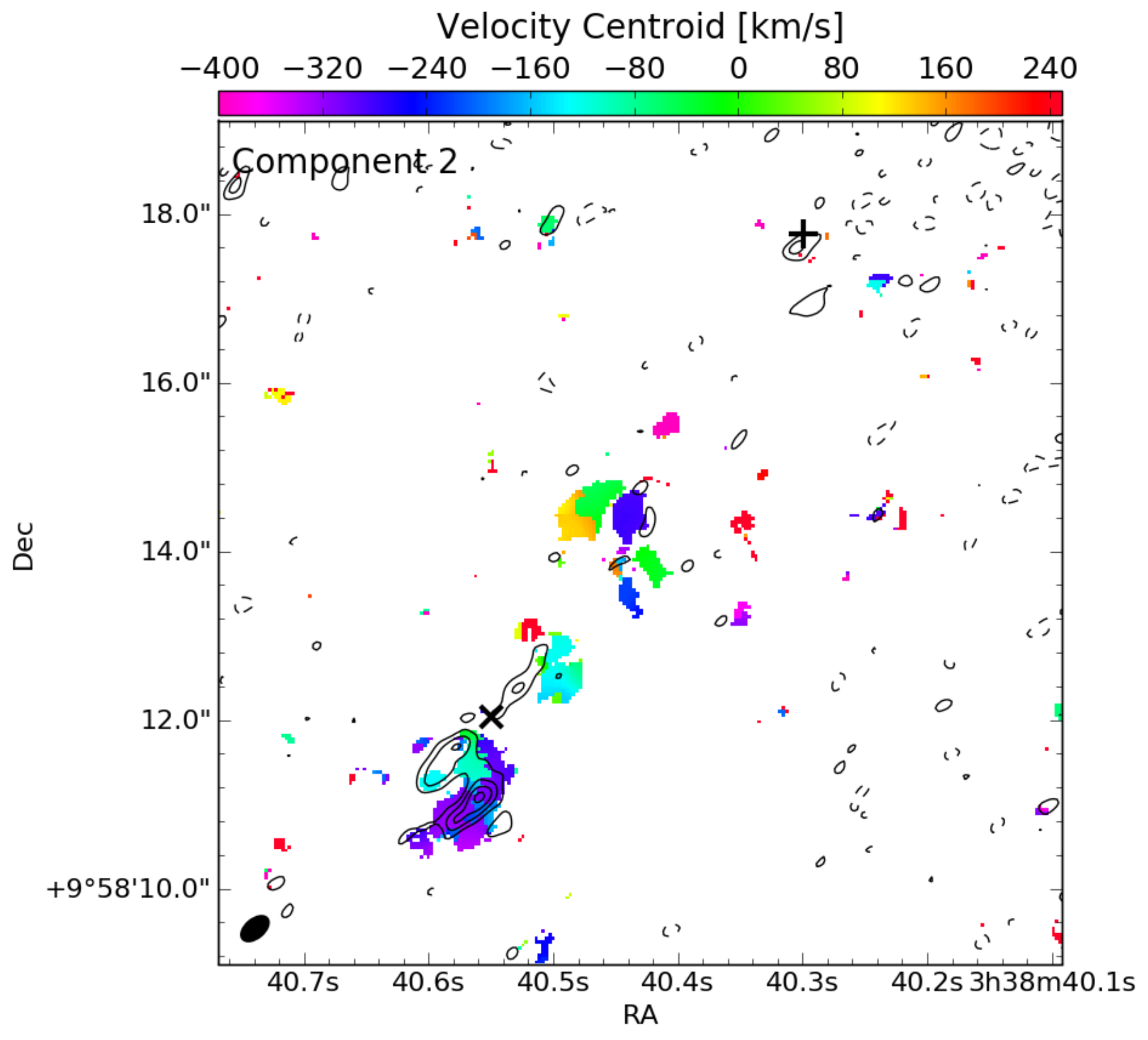}
  \includegraphics[trim=4.5cm 0cm 0cm 2.5cm, clip=true, width=5.3cm]{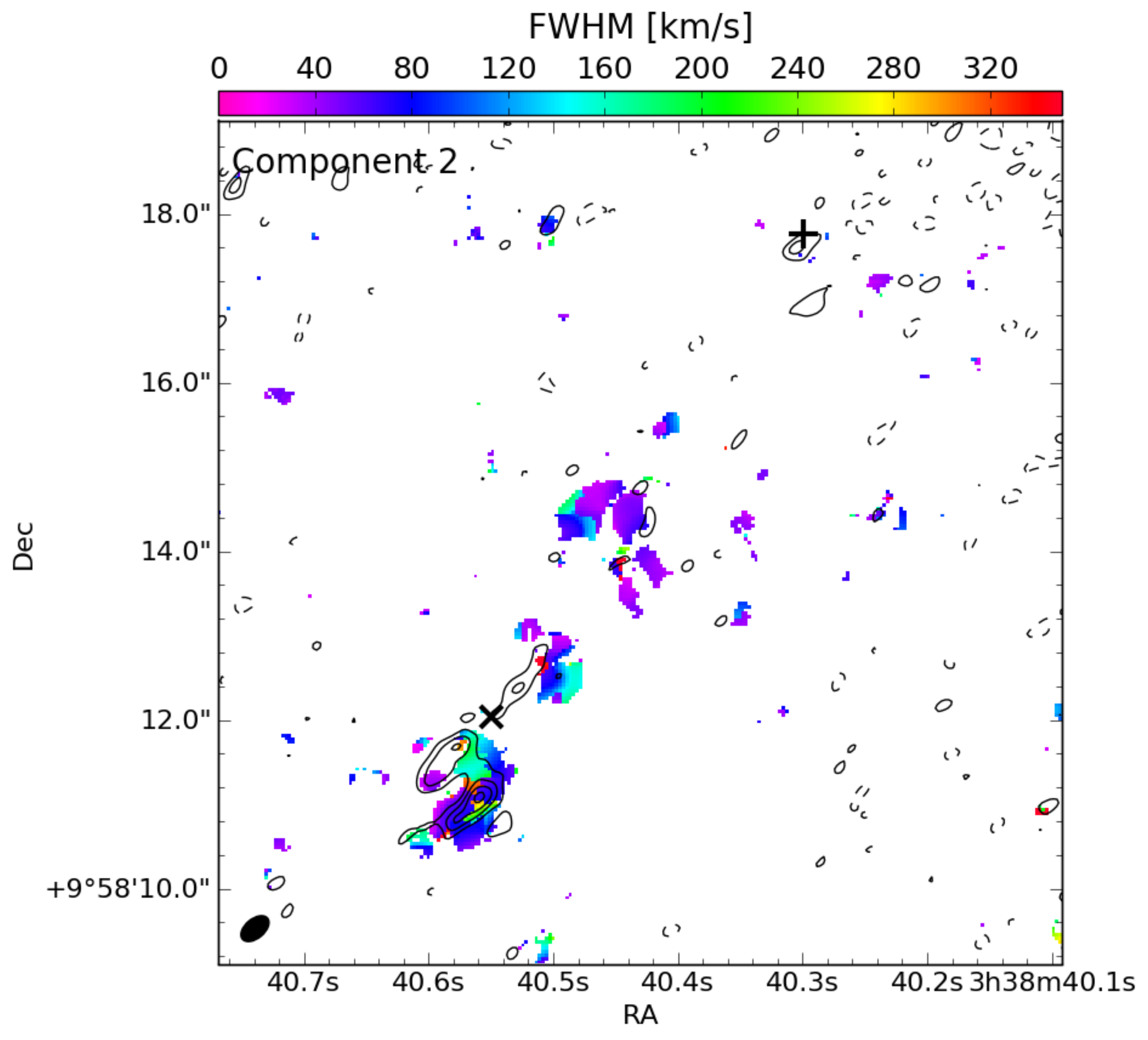}
\end{minipage}
  \caption{
    Maps of integrated flux ({\it left}), velocity centroid ({\it middle}), and 
    FWHM ({\it right}) of a two-component fit to the CO(3-2) emission line. 
    Only pixels containing flux detected at $>2\sigma$ are shown.
    The contours are the same as in Fig. \ref{fig:COmaps} (bottom).
    The $\times$ and $+$ indicate the centroids of the BCG and companion galaxy, respectively.
  }
  \label{fig:CO32vel}
\end{figure*}

\subsection{Velocity Distribution}
\label{sec:velmaps}

In order to study the velocity structure of the molecular gas, we extracted spectra 
from each pixel of the datacube averaged over a box the size of the synthesized beam. 
Each spectrum was fit with a single Gaussian component. The significance of the line
was tested using a Monte Carlo analysis following the prescription of 
\citet[][see Section 5.2]{protassov02}, with detections requiring a $2\sigma$ significance.
Spectra containing a line detection were then tested with a second component following
the same prescription. 
The resulting velocity (centroid and FWHM) maps are presented alongside the 
integrated flux for the corresponding component in Fig. \ref{fig:CO10vel} and 
\ref{fig:CO32vel} for CO(1-0) and CO(3-2), respectively.
The CO(1-0) maps were created with $40\kmps$ bins, with each pixel additionally tested 
with $20\kmps$ bins to ensure that the narrow features toward the tail of the filament 
could be recovered. $20\kmps$ velocity bins were used for the CO(3-2) map. 
The maps have been overlaid with the corresponding contours from Fig. \ref{fig:COmaps}.

The spatial distribution of molecular gas recovered by these maps is consistent 
with the integrated maps in Fig. \ref{fig:COmaps}, with the molecular gas divided
between the BCG and a long filament north of the companion galaxy.
Significant emission is detected along the entire length of the filament, confirming 
that the two clumps seen in Fig. \ref{fig:COmaps} are connected by a faint channel.
The apparent disconnect between the inner and outer filament seen in Fig. \ref{fig:COmaps}
arises because the maps were integrated over a velocity range much broader than the
linewidth of this channel, drowning out the signal.

A large velocity gradient is present within the core of the BCG. South of the
nucleus the gas is blueshifted to $-210\pm10\kmps$, while the velocity of the spur 
extending to the north increases from $80\pm10\kmps$ near the nucleus up to about
$200\kmps$. The blueshifted emission is relatively broad, with a FWHM of $260\pm20\kmps$,
while the redshifted emission is considerably narrower, with a FWHM of $90-130\kmps$.
The smooth gradient observed between these regions results from the beam smearing 
together regions of disparate velocities. This velocity gradient is more pronounced 
at CO(3-2), where the velocity changes abruptly across the nucleus.

South of the BCG nucleus the two clumps identified at CO(3-2) share a broad
velocity component (FWHM $200-300\kmps$) with a shallow velocity gradient ranging 
from of $-120\kmps$ in the southwest to $-200\kmps$ in the northeast. A second 
velocity component is present in the southwestern clump, with a velocity of 
$-270\kmps$ and FWHM of $80\kmps$.

North of the BCG nucleus the CO(3-2) emission becomes more diffuse farther from 
the galactic centre, with significant detections obtained further from the nucleus
than at CO(1-0). This emission follows the same gradient as the CO(1-0) emission,
increasing from $80\kmps$ near the nucleus to $180-250\kmps$ in the diffuse clouds.
A small region with higher velocity ($170\kmps$), broad ($230\kmps$) emission is 
observed immediately north of the nucleus, beside the $80\kmps$ emission. 
This map also reveals a significant detection coincident with the companion galaxy,
which has a velocity of $240\kmps$ and a FWHM of $200-300\kmps$.

Several distinct velocity structures are observed along the filament. Most
of the emission originates from the ``Inner Filament'' (Fig. \ref{fig:CO10subdiv}), 
which occupies a very narrow range in velocity ($-20$ to $-40\kmps$) with a FWHM 
of only $70-100\kmps$. Toward the tail of the filament the velocity becomes
increasingly blueshifted to $-145\kmps$, where it then bends northward and the
velocity further increases in magnitude to $-230\kmps$. 
Finally, a narrow tendril protrudes westward from the southern edge of the inner 
filament, with a velocity of $-120\kmps$ to $-170\kmps$ and a linewidth comparable 
to the inner filament. Multiple velocity components are detected at the interface 
between this tendril and the inner filament, extending all the way along the filament.

\subsection{Velocity Profiles}

\begin{figure*}
  \begin{minipage}{\textwidth}
    \centering
    \includegraphics[width=0.4\textwidth]{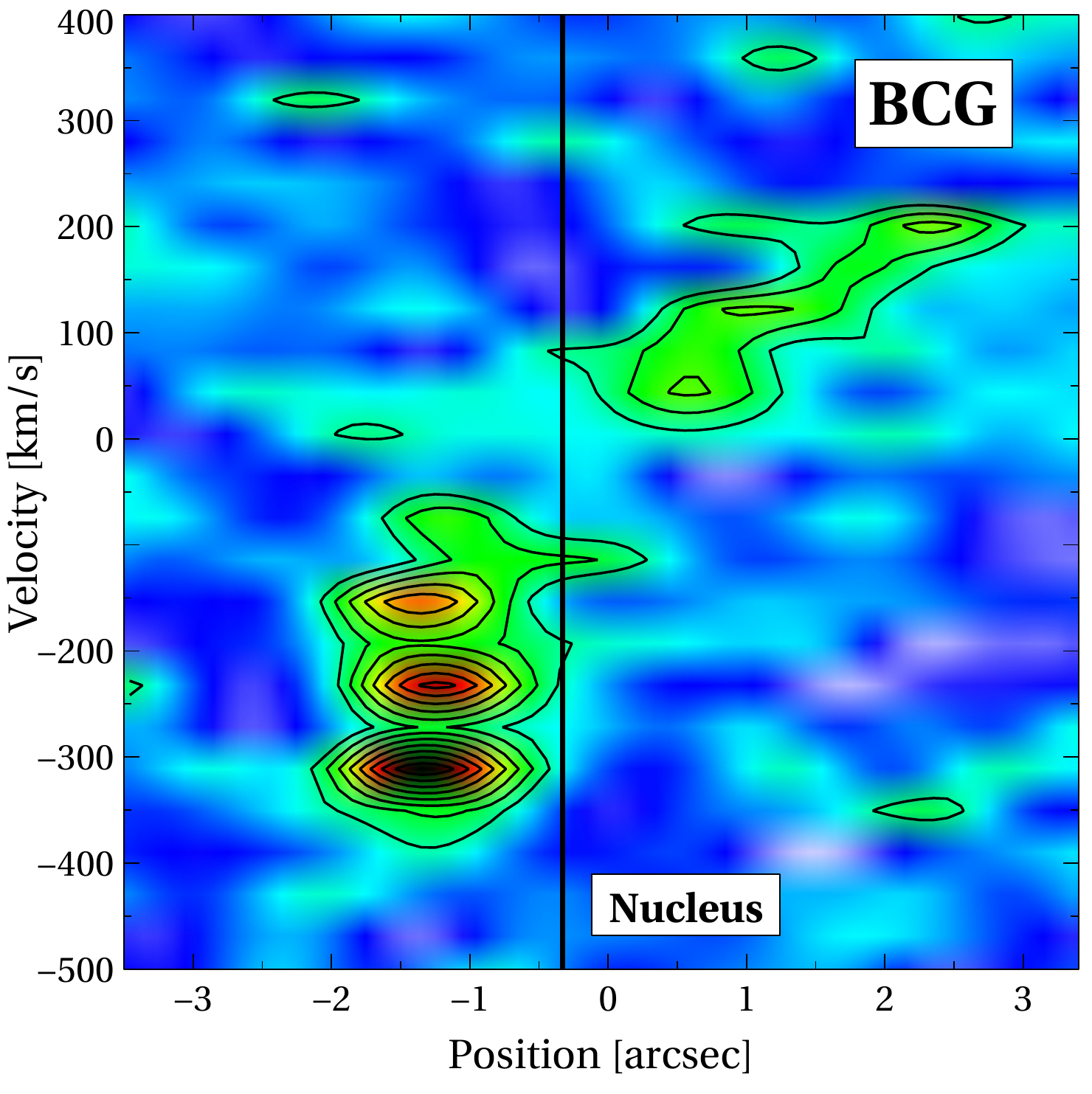}
    \includegraphics[width=0.4\textwidth]{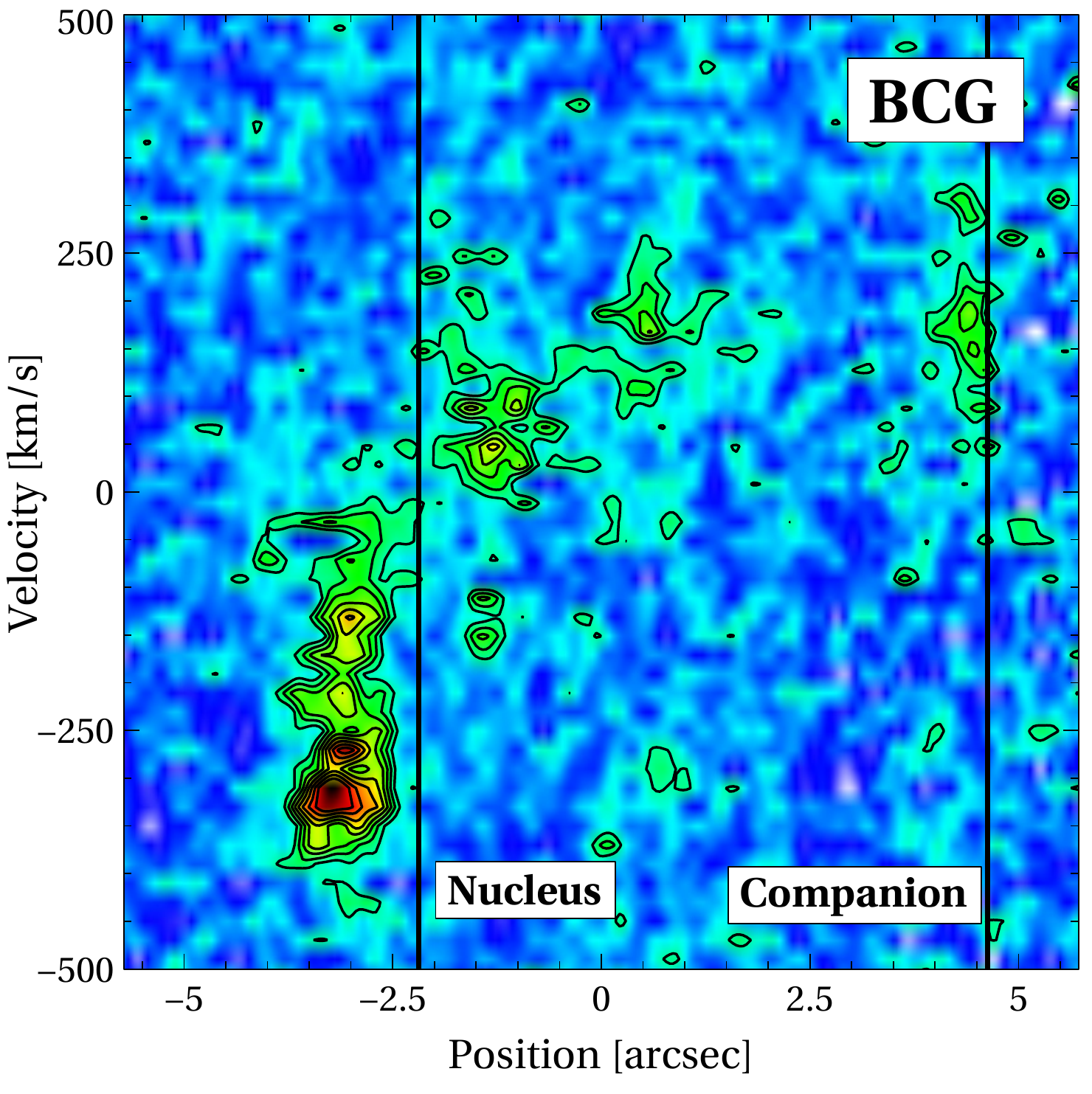}
  \end{minipage}
  \begin{minipage}{\textwidth}
    \centering
    \hspace{-0.05\textwidth}
    \includegraphics[trim=0.0cm 0cm 0cm 0cm, clip=true, width=0.45\textwidth]{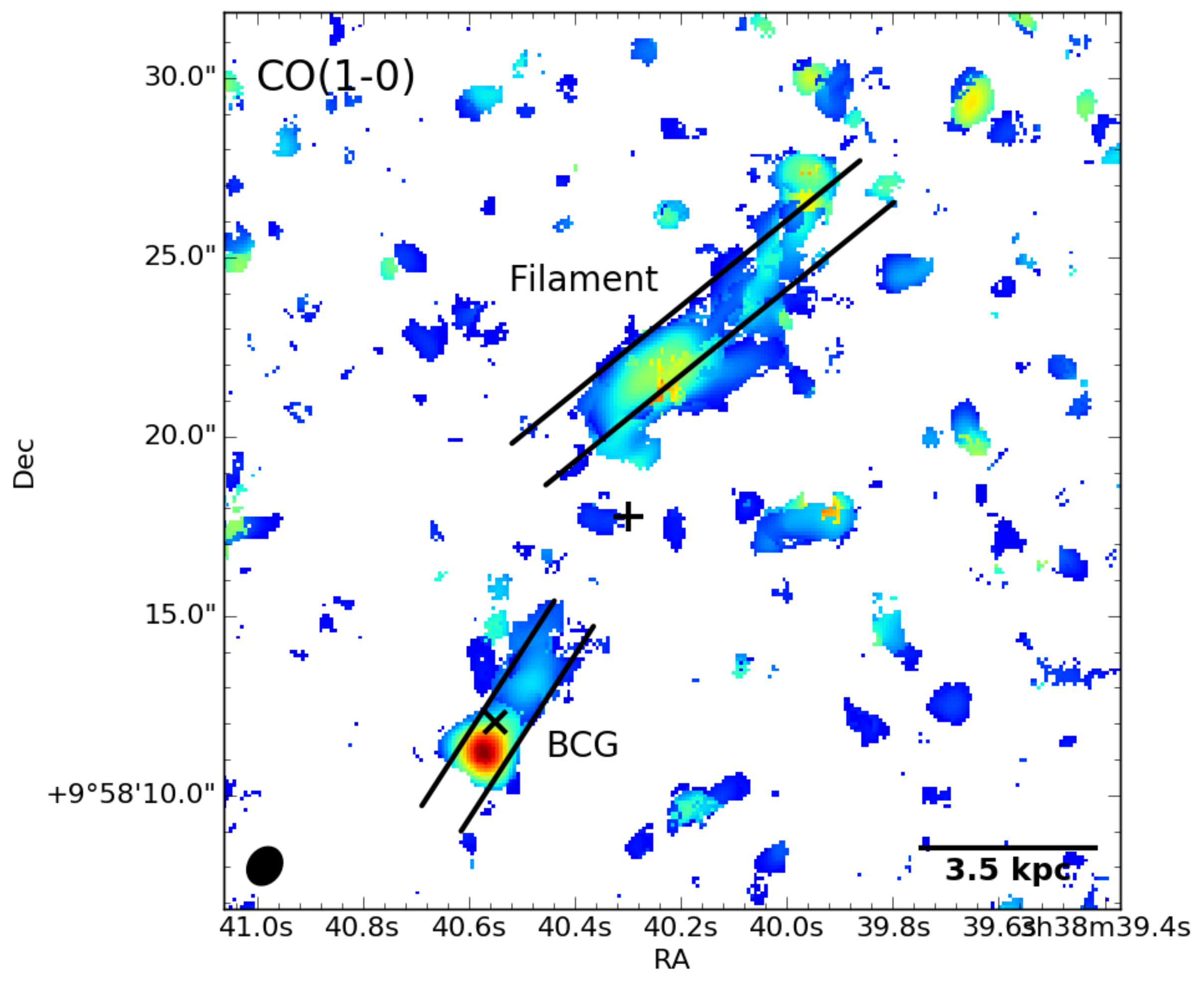}
    \includegraphics[trim=3.0cm 0cm 0cm 0cm, clip=true, width=0.37\textwidth]{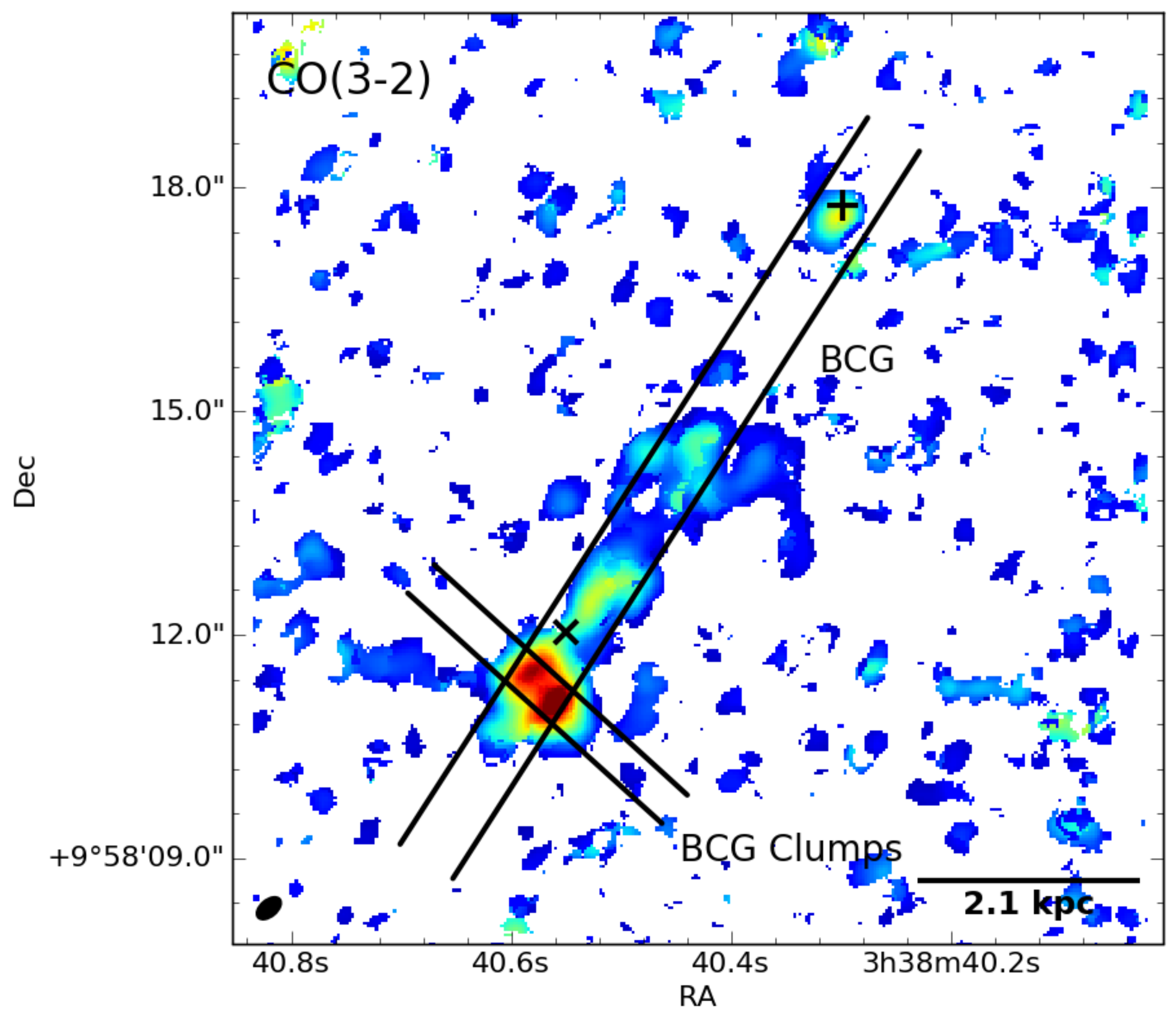}
  \end{minipage}
  \vspace{0.5cm}
  \begin{minipage}{\textwidth}
    \centering
    \includegraphics[width=0.4\textwidth]{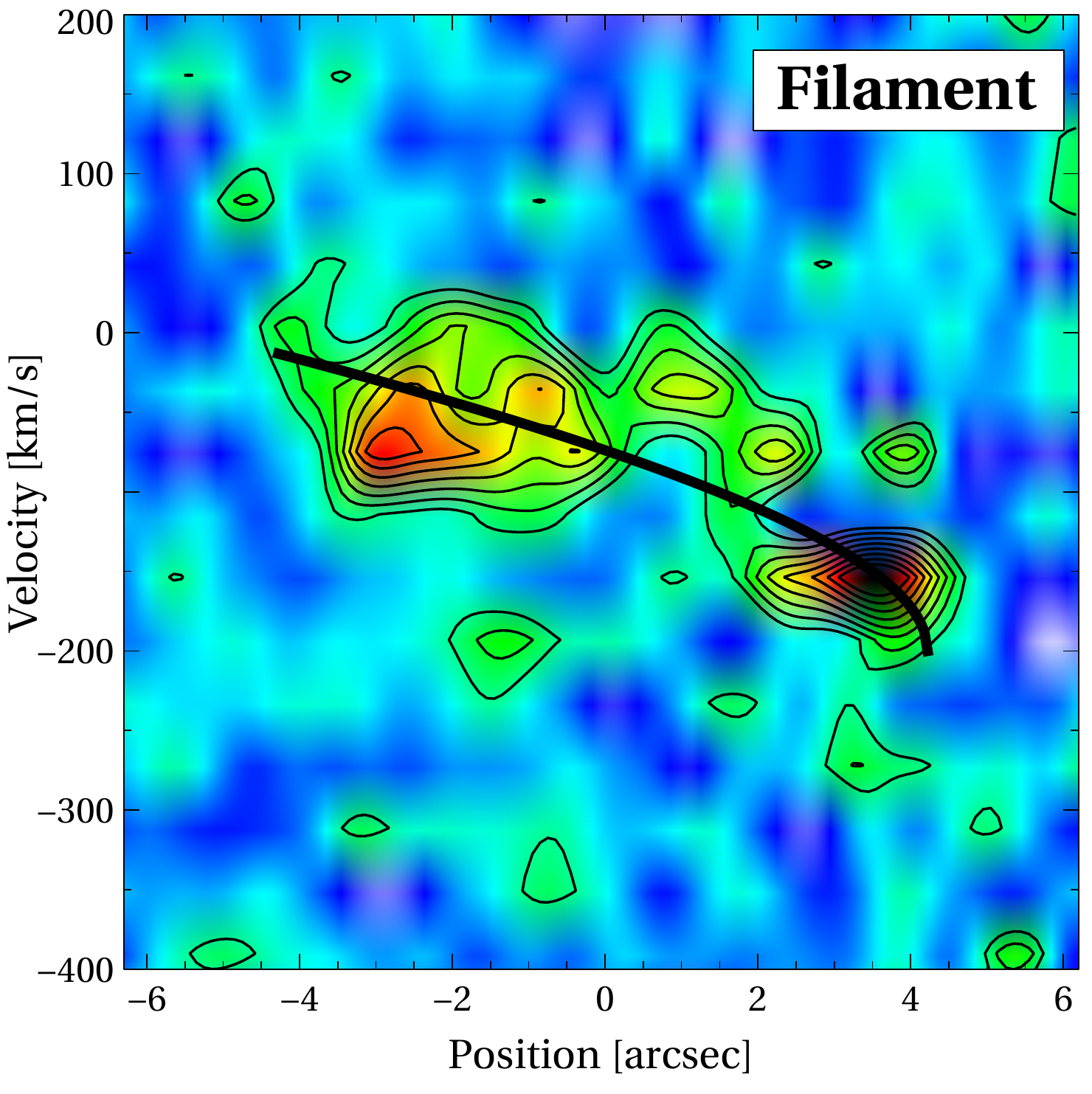}
    \includegraphics[width=0.4\textwidth]{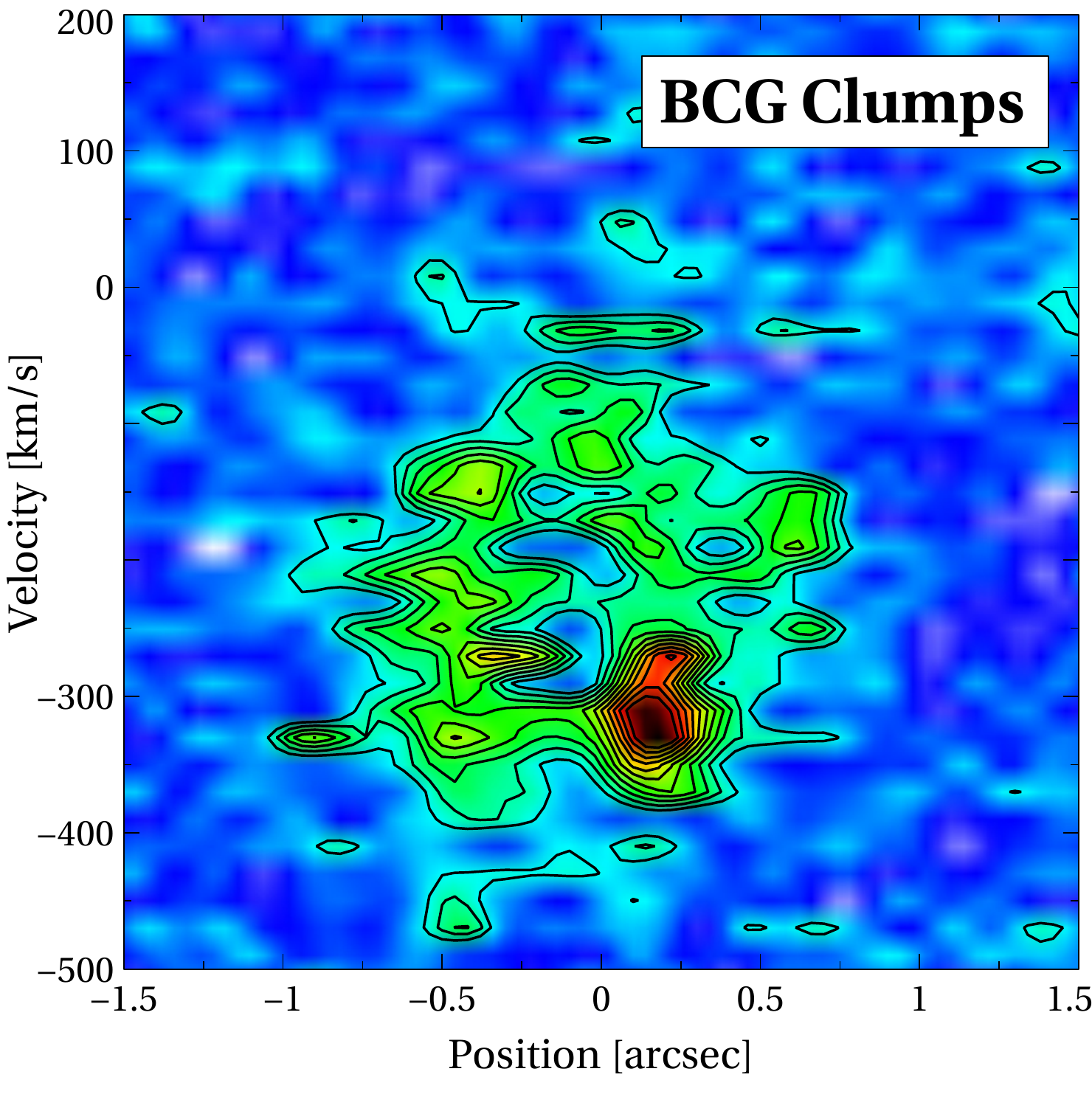}
  \end{minipage}
\caption{
  CO(1-0) ({\it left}) and CO(3-2) ({\it right}) position-velocity (PV) diagrams along
  the BCG ({\it top}), elongated filament ({\it bottom-left}), and a perpendicular
  cut across the BCG ({\it bottom-right}). 
  In all cases the eastern (left) end of the slit corresponds to a negative position.
  The middle row shows the location of the slits overlaid on the integrated
  fluxes determined from the velocity maps of Figs. \ref{fig:CO10vel} and \ref{fig:CO32vel}.
  The vertical lines in the PV diagrams indicate the locations of the BCG nucleus
  and companion galaxy.
  The PV diagram of the extended filament in CO(1-0) (bottom-left) is overlaid with
  a gravitational free-fall model, as discussed in Section \ref{sec:freefall}.
}
\label{fig:PV}
\end{figure*}

Position-velocity (PV) diagrams of the molecular gas in both the BCG and the 
elongated filament are presented in Fig. \ref{fig:PV}. For each PV diagram the 
flux was averaged over the width of the slit as a function of both slit position 
and velocity channel.
One slit was placed along the extended axis of the molecular gas within the BCG, 
encompassing both the clump to the SE and spur to the NW.
This slit was $1.3''$ wide at CO(1-0) and $0.84''$ wide at CO(3-2), both with a
position angle (PA) of $147^{\circ}$. 
A second, $0.52''$ wide slit was placed roughly orthogonal to this (PA $47^{\circ}$) 
in order to separate the two clumps identified at CO(3-2) but unresolved at CO(1-0).
Finally, a $1.5''$ wide slit (PA $130^{\circ}$) was placed lengthwise along the 
filament at CO(1-0).
The position indicated in the PV diagrams runs from the bottom edge of the slits
shown on the integrated flux maps to the top edge, with zero offset corresponding
to the midpoint of the slit.

The position-velocity diagrams along the long axis of the BCG (Fig. \ref{fig:PV}
top) are qualitatively consistent between CO(1-0) and CO(3-2). The gas south of the
nucleus is blueshifted and very broad. Extending toward the other side of the
nucleus the velocity increases roughly linearly until the edge of the diffuse
emission noted in CO(3-2) (see Fig. \ref{fig:CO32subdiv}). Gas coincident with
the companion galaxy is also detected at CO(3-2), where the velocity is comparable
to that of the diffuse emission.

The perpendicular cut across the nucleus in CO(3-2) shows the phase space information
for the two clumps south of the BCG nucleus. The velocities of both clumps are
$\sim -275\kmps$. The brighter clump is best fit by two velocity components in 
the velocity maps shown in Fig. 6. The narrow component is also visible in the 
PV diagram, and appears to extend between the two clumps. The similar velocities
in these structures indicates that the clumps are likely related dynamically.

All of the emission along the extended filament is confined to the range of $-200$ 
to $0\kmps$. However, the gas appears to be separated into two clumps.
Higher blueshifted velocities are observed toward the tail of the filament, with
velocities closer to systemic appearing at its base. The velocity of the inner
portion of the filament is relatively constant along its entire $4''$ ($3\kpc$)
length. An infall model has been overlaid on this PV diagram (see Section \ref{sec:freefall}).

\subsection{Spatial Correlation with Dust Extinction}
\label{sec:dust}

Two regions of significant extinction were noted by \citet{donahue07} in the
archival {\it HST} WFPC2 F606W image of 2A~0335+096.
A wedge-shaped region is located southwest of the BCG nucleus, and an elongated 
filament extends north of the companion galaxy as seen in Fig. \ref{fig:COmaps}. 
Both of these regions are coincident with significant amounts of molecular gas. 

The calibrated HST WFPC2 F606W image was obtained from the Hubble Legacy Archive.
All magnitudes quoted here have been transformed to the Johnson V-band using
$m_V({\rm Vega})-m_{{\rm F606W}}({\rm ST})=0.04$ derived for the \citet{kc96} elliptical 
galaxy template in the IRAF tool Synphot. The $V-I$ colour of the template galaxy, 
1.3, is similar to that of a BCG, 1.2, so this conversion should be appropriate 
\citep{whiley08}. The quoted magnitudes have also been corrected for foreground Galactic 
extinction (A$_{\rm F606W}$), K-correction, evolution of the stellar population, and 
surface brightness dimming. The applied corrections are shown in Table \ref{photometry}.

In order to quantify the dust extinction we model the 2D distribution of stellar light
using the galaxy-fitting code Galfit-M \citep{galfitm}, an extended version of Galfit
\citep{peng02,peng10}. A S{\'e}rsic profile significantly overestimates the flux in the
core of the BCG, an effect that is common in BCGs \citep{lauer07, mcn09} and is thought
to be caused by black hole scouring \citep{kormendy09}.
The BCG is better modelled by the core-S{\'e}rsic profile \citep{graham03}, which is a 
S{\'e}rsic profile that transitions to a power law below the break radius $r_b$ and is given by
\begin{equation}
I = I' \left[ 1+\left( \frac{r_b}{r} \right) ^\alpha \right]^{\gamma/\alpha} 
    \exp{ \left[ -\kappa \left( \frac{r^\alpha + r_b^{\alpha}}{r_e^\alpha} \right)^{1/(\alpha n)} \right]}.
\label{corser}
\end{equation}
Here $r_e$ and $n$ are the effective radius and S{\'e}rsic index defined in the typical 
S{\'e}rsic profile, and the coefficient $\kappa$ is a function of $n$. The transition and
inner power law are described by the indices $\alpha$ and $\gamma$, respectively.s
The companion galaxy is modelled using a normal S{\'e}rsic profile and fit simultaneously
with the BCG.

\begin{table}
\caption{Photometric Corrections}
\begin{center}
\begin{tabular}{l c c c}
\hline
Galactic extinction$^a$ & K$^b$  & Evolution$^b$ & $(1+z)^4$ \\
(mag)                   & (mag)  & (mag)         & (mag)     \\
\hline
\hline
0.989                   & $0.073$ & $-0.047$     & $0.148$     \\
\hline
\end{tabular}
\end{center}
Notes: $^a$ \citet{schlafly11}
$^b$ \citet{poggianti97}
\label{photometry}
\end{table}

Structures unrelated to the BCG and companion galaxies' stellar light were masked out prior
to fitting. These included background galaxies, a bright foreground star in the SE, and the 
obvious dust features in the filament and the BCG nucleus. A wide detector artefact at the 
edge of the CCD was also masked out, and limits the fittable area considerably.
The resulting fit parameters are shown in Table \ref{opticalfit}. 
The companion galaxy is well-modelled by a S{\'e}rsic profile with index $1.8$ and total 
V-band magnitude $16.8$. Using the same photometric corrections as the BCG, its absolute
magnitude of $-19.1$ is slightly brighter than a dwarf galaxy.

Several biases are present in this fitting procedure. First, the fittable area is small due
to the foreground star and a large detector artefact near the edge of the CCD. Since 2A~0335+096
is a nearby cluster, these fits do not extend far into the envelope of the galaxy. The full 
shape of the BCG's light profile cannot be traced, so its effective radius is unconstrained. 
This also results in a large S{\'e}rsic index, since a power law provides a sufficient fit
beyond the core.
Second, the light profiles of galaxies are best fit when their position angle and ellipticity
are allowed to vary as a function of radius. Isophotal variations are not currently supported
by Galfit-M. Using the {\sc ellipse} task in IRAF to extract elliptical isophotes of the stellar 
light favours a position angle of $-5^{\circ}$ to $-10^{\circ}$ in the centre, twisting to 
$-30^{\circ}$ beyond $10\arcsec$. Similarly, the axis ratio $b/a$ is consistent at large radii
with the $0.7$ measured here, but decreases to $0.65$ in the core.

\begin{figure}
  \centering
  \includegraphics[width=0.9\columnwidth]{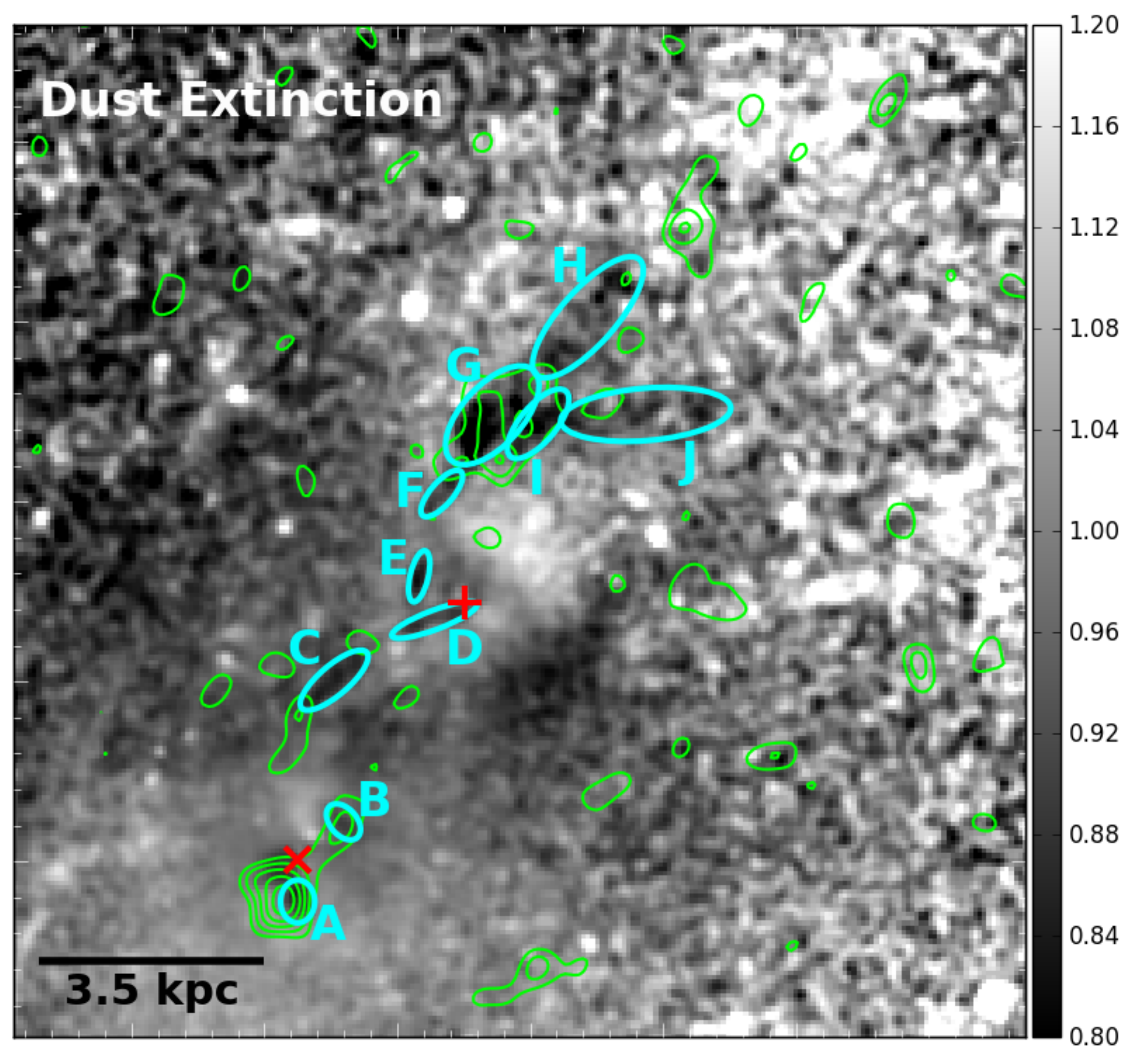}
  \caption{
    Dust extinction map of the HST F606W image. Regions with significant dust
    extinction are identified by the green ellipses, with statistics shown in
    Table \ref{tab:dust}. The BCG and companion galaxy centroids are identified
    by the red $\times$ and $+$, respectively. The image is shown in the same field 
    of view as the ALMA CO(1-0) image (Figs. \ref{fig:COmaps} and \ref{fig:CO10vel}).
    The CO(1-0) contours from Fig. \ref{fig:COmaps} have been overlaid for reference.
  }
\label{fig:dustmap}
\end{figure}

In Fig. \ref{fig:dustmap} we present a map of the dust extinction, showing the ratio of 
image surface brightness to the model brightness ($I/I_0$). Regions with 
significant extinction are highlighted by the labelled ellipses, with the corresponding 
extinction (both the peak extinction in a pixel and mean within the ellipse) tabulated 
in Table \ref{tab:dust}. These regions were confirmed by eye in the original image. 
Systematic errors are clearly visible in the residuals of the extinction map, primarily 
perpendicular to the BCG--companion axis. Positive and negative residuals near the BCG 
centroid (red $\times$) correspond to a double-peaked nucleus, so the extinction northwest 
of region $A$ does not correspond to dust. The extinction surrounding the labelled dust 
features is generally $\lesssim 5\%$, though increases to $\sim10\%$ around the filament. 
Model residuals surrounding the identified regions also typically have $I/I_0 \geq 1$, 
so the tabulated optical depths and dust masses may be underestimated.

\begin{table}
\caption{2D Optical Fitting Results}
\begin{center}
\begin{tabular}{l l c c}
\hline
                & Units             & BCG                   & Companion \\
\hline
\hline
$I(r_b)^{1}$    & mag arcsec$^{-2}$ & $19.310\pm0.005$      & -- \\
$r_b$           & arcsec            & $0.924\pm0.007$       & -- \\
$\alpha$        &                   & $2.56\pm0.05$         & -- \\
$\gamma$        &                   & $0.081\pm0.004$       & -- \\
$m_{\rm tot}$   & mag               & --                    & $16.736\pm0.003$ \\
$r_e$           & arcsec            & Unconstrained$^{2}$   & $0.997\pm0.002$ \\
$\frac{{\rm d}\ln I}{{\rm d}\ln r}$ & & $-1.35^{3}$         & -- \\
$n$             &                   & $19.6\pm2.8$          & $1.838\pm0.005$ \\
$b/a$           &                   & $0.7010\pm0.0004$     & $0.884\pm0.001$ \\
PA              & degrees           & $-21.85\pm0.06$       & $7.8\pm0.5$ \\
\hline
\end{tabular}
\end{center}
Notes: $^{1}I(r_b)$ is related to $I'$ in Eqn \ref{corser} via
$I' = I(r_b)~2^{-\gamma / \alpha}~\exp \left[ \kappa~(2^{1/\alpha}~r_b/r_e)^{1/n} \right]$.
$^{2}$The BCG radii probed by the HST imaging only show a bend at the break radius. The
S{\'e}rsic component is close to a power law, with the corresponding effective radius lying
outside of the HST field of view. 
${^3}$Since the effective radius is unconstrained, we quote the logarithmic derivative
at a radius of $20''$ instead.
\label{opticalfit}
\end{table}

The mean optical depth $\tau$ is computed from the mean extinction via 
$I/I_0 = {\rm e}^{-\tau}$. This is converted into column density ($N_H$), assuming a 
standard V-band Galactic extinction curve \citep{cardelli89}, through 
$N_H = 2.05\e{21} \tau_V \cmsq$ and finally to dust mass. Region $A$ is coincident with 
a clump of molecular gas identified at CO(3-2). Taking ${\rm CO(3-2)/CO(1-0)}=7$, 
the corresponding molecular gas mass is $9.5\e{7}\Msun$ and gas-to-dust ratio is 
$130$, which is similar to the Milky Way value.
Regions $F$ through $I$ are coincident with the extended filament. The total dust 
mass in these regions is $1.07\pm0.07\e{7}\Msun$. With a molecular gas mass of 
$7.8\pm0.9\e{8}\Msun$, the local gas-to-dust ratio is $66$. This is a factor of two 
lower than the gas-to-dust ratio in the BCG.
Regions $B$ and $D$ are coincident with molecular gas in the northern spur and 
companion galaxy, respectively, although neither traces the gas distribution well. 
Attributing all of the molecular gas in the spur to region $B$ results in a gas-to-dust 
ratio of $290$, although the differing spatial distributions add considerable uncertainty
to this value.

\begin{table*}
\begin{minipage}{\textwidth}
\caption{Dust Extinction}
\begin{center}
\begin{tabular}{l c c c c c c}
\hline
Region & Area & $(I/I_0)_{\rm min}$ & $I/I_0$  & $<\tau>$        & $N_H$         & $M_{\rm dust}$    \\
       & ($10^{42}\cmsq$) &          &          &                 & ($10^{20}\psqcm$) & ($10^5\Msun$) \\
\hline\hline
 A     & 2.58 & $0.82\pm0.03$ & $0.923\pm0.005$ & $0.080\pm0.005$ & $3.42\pm0.23$ & $7.4\pm0.5$ \\
 B     & 2.0  & $0.88\pm0.04$ & $0.938\pm0.006$ & $0.064\pm0.006$ & $2.75\pm0.27$ & $4.6\pm0.5$ \\
 C     & 4.52 & $0.80\pm0.06$ & $0.896\pm0.007$ & $0.110\pm0.007$ & $4.69\pm0.32$ & $17.7\pm1.2$ \\
 D     & 3.12 & $0.81\pm0.02$ & $0.944\pm0.005$ & $0.057\pm0.005$ & $2.45\pm0.23$ & $6.4\pm0.6$ \\
 E     & 1.58 & $0.72\pm0.05$ & $0.884\pm0.010$ & $0.123\pm0.012$ & $5.25\pm0.50$ & $6.9\pm0.7$ \\
 F     & 2.3  & $0.76\pm0.09$ & $0.922\pm0.014$ & $0.081\pm0.015$ & $3.46\pm0.63$ & $6.7\pm1.2$ \\
 G     & 13   & $0.48\pm0.08$ & $0.884\pm0.007$ & $0.123\pm0.008$ & $5.26\pm0.35$ & $57.0\pm3.8$ \\
 H     & 14.9 & $0.56\pm0.10$ & $0.962\pm0.009$ & $0.039\pm0.009$ & $1.66\pm0.39$ & $20.7\pm4.8$ \\
 I     & 5.24 & $0.56\pm0.09$ & $0.886\pm0.012$ & $0.121\pm0.013$ & $5.16\pm0.23$ & $22.6\pm2.5$ \\
 J     & 15.6 & $0.51\pm0.10$ & $0.959\pm0.008$ & $0.042\pm0.009$ & $1.79\pm0.36$ & $23.3\pm4.7$ \\
\hline
Total  &      &             &                   &                 &               & $173\pm8$ \\
\hline
\end{tabular}
\end{center}
\label{tab:dust}
\end{minipage}
\end{table*}

Treating the dust as a thin screen, for regions G--J, where the minima of $I/I_0$ 
are about 0.5, the dust must lie on or in front of the midplane of the BCG.
The velocity of the molecular gas in these regions is within a few tens of $\kmps$ 
from the systemic value, and the FWHM is $100\kmps$ at its broadest. Taken together, 
these suggest that the filament is oriented roughly perpendicular to the line of 
sight, possibly flowing on a nearly radial trajectory toward the BCG. 
The lesser peak extinction within the BCG, where the minimum of $I/I_0 \simeq 0.82$,
means that the location of the dust along our line of sight is poorly constrained.

Overall the most significant associations of molecular gas are coincident with significant 
dust extinction. Dust shielding in these regions may be promoting the formation of 
molecular gas. It is also possible that the more extended associations (such as along 
the spur) do not have detected dust extinction because the dust has been spread out 
over a larger area, resulting in too low a column density for visible dust
extinction.

\subsection{Spatial Correlation with X-ray and H{$\alpha$} Filaments}
\label{sec:spatialcorr}

2A~0335+096 hosts a bright H$\alpha$ nebula with total luminosity 
$L_{\rm H\alpha}=0.8\e{42}\ergps$ \citep{donahue07}. While the H$\alpha$ nebula is 
not associated with any structures in the $0.5-7\keV$ {\it Chandra} X-ray image, 
multi-temperature fits reveal a 0.5 keV component in the ICM that is spatially 
coincident with the H$\alpha$ emission \citep{sanders09}. 
Similar spatial correlations have been observed in several other cool core clusters, 
including Perseus \citep{fabian03, fabian08}, M87 \citep{sparks04, werner13}, and 
several nearby giant ellipticals \citep{werner14}.
Filaments of molecular gas have been detected along the H$\alpha$ filaments in
Perseus \citep{salome06, lim08, salome08, lim12}, implying that the co-spatial gas 
in the filaments occupies 5--6 decades in temperature.

\begin{figure}
  \includegraphics[width=\columnwidth]{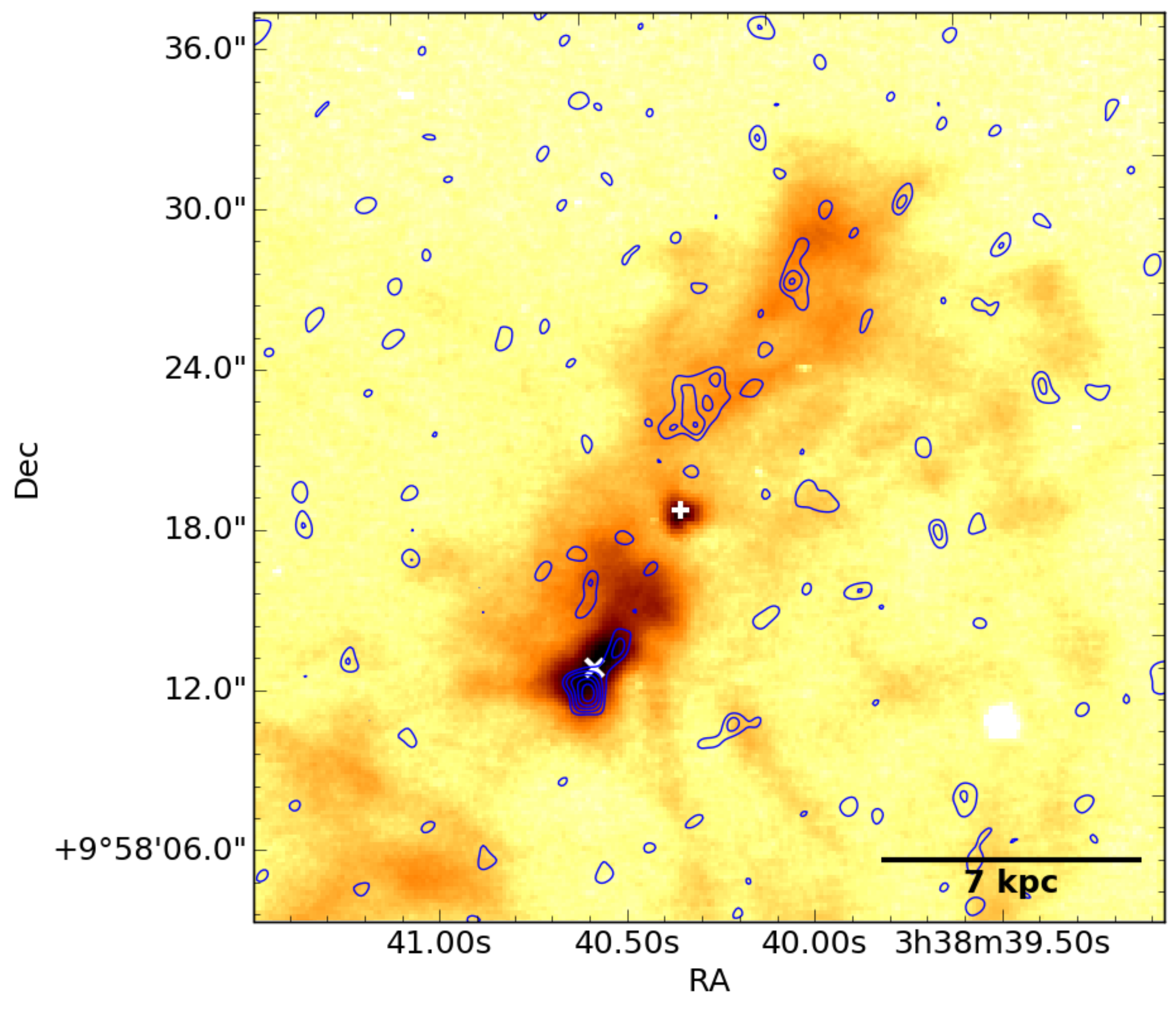}
  \caption{
    H$\alpha$ emission from Donahue (2007) overlaid with the CO(1-0) contours from Fig. 
    \ref{fig:COmaps}. The $\times$ and $+$ indicate the nuclei of the BCG and 
    companion galaxy, respectively. The coordinate reconstruction of the H$\alpha$
    image is accurate to $\sim 1\arcsec$.
  }
  \label{fig:halpha}
\end{figure}

Multiphase gas spanning many decades in temperature is also observed here. Fig.
\ref{fig:halpha} shows the H$\alpha$ nebula from \citet{donahue07} overlaid with 
contours of the CO(1-0) emission. The two distributions are qualitatively 
very similar. Bright H$\alpha$ emission near the BCG extends in a spur toward the 
companion galaxy. The emission extends beyond the companion galaxy into a filament 
coincident with the extended filament of molecular gas seen in CO(1-0) emission. 
However, the 
most luminous nebular emission is in the BCG, while the molecular gas is
observed preferentially in the filaments. This may be due to line emission that 
has been resolved out by the interferometer. Only 40\% of the single dish flux has 
been recovered here, so a diffuse component of cold gas may be present in the BCG.

Integral field spectroscopy of the H$\alpha$ nebula reveals that the [N~{\sc ii}] 
$\lambda$6583 line and the molecular gas are co-located in both position and velocity
space \citep{farage12}. Broad, blueshifted emission is slightly offset to the south of
the BCG with a velocity of about $-120\kmps$ ($-240\kmps$ in our adopted frame).
Near the companion galaxy the H$\alpha$ nebula becomes redshifted with respect to the
systemic velocity, with blueshifted emission extending along the direction of the 
filament. \citet{farage12} also detected nebular emission $5-10\kpc$ southeast of
the BCG nucleus. Its redshifted velocity led them to conclude that the nebula is in
rotation about the centre of the BCG with a velocity amplitude of $130\kmps$.
Molecular gas either does not extend along the southeastern arm of the filament or is 
too faint to detect, so we are unable to corroborate this finding.
The lack of molecular gas to the southeast may arise if the surrounding atmosphere 
lacks the dense regions required to form H$_2$, or if H$_2$ production is enhanced
by the presence of dust grains to the northwest.
Since our observations trace only the high density molecular gas (see 
Section \ref{sec:mass}), molecular gas may lie to the southeast but falls below our 
detection limit.

\section{Discussion}
\label{sec:discussion}

\subsection{Origin of the Molecular Gas}

With a total molecular gas mass exceeding $10^{9}\Msun$, 2A~0335+096 harbours
significantly more molecular gas than is typically observed in early-type 
galaxies. Identifying the origin of this gas is critical in understanding
the evolution of the BCG. Two primary mechanisms could be contributing this
gas: stripping of merging galaxies or cooling from the hot atmosphere.

Large supplies of molecular gas are observed preferentially in BCGs at the
centres of galaxy clusters with cooling times falling below a Gyr. These
systems are associated with bright emission-line nebulae and enhanced star
formation. Thus the presence of molecular gas should in general be linked to 
residual cooling of the hot atmosphere.

The presence of a companion galaxy located 5~kpc from the BCG in projection 
raises the possibility of a merger origin for the molecular gas in this system. 
Its low stellar velocity \citep[$\sim 200\kmps$;][]{gelderman96} relative to 
the BCG suggests that the galaxies also have a small radial separation.
The trail of molecular gas orientated toward the companion galaxy may 
then be indicative of an interaction between the two galaxies.

Before delving further into the origin of the molecular gas, we emphasize 
the difference in dynamics between the two components. The inner edge of the 
filament is slightly blueshifted with respect to the stellar component of the 
BCG, while all nuclear gas north of the BCG is redshifted. Since 
the components are not smoothly connected in phase space, we treat them separately 
in our discussion.

Mergers with donor galaxies can supply gas to the BCG through several distinct
avenues. Repeated mergers between the BCG and gas-rich donor galaxies are unlikely given 
the low number of member galaxies in this low richness cluster \citep{schwartz80}, 
and because the high velocity dispersion of a cluster decreases the merger rate. 
We instead focus only on a merger with the nearby companion galaxy. 
A tidal interaction between the galaxies
may result in either the stripping of cold gas from the companion or the disruption
of a pre-existing gas supply within the BCG. Alternatively, ram pressure stripping 
would form a tail that may correspond to the filament.

\subsubsection{Tidal Interaction}
\label{sec:tidal}

In the nucleus of the BCG, the opposed redshifted and blueshifted velocities are
suggestive of rotation. This velocity pattern is consistent with the H$_2$ kinematics
derived from $K$-band integral field spectroscopy \citep{wilman11} as well as 
lower resolution H$\alpha$ integral field spectroscopy of the nuclear barred structure
\citep{hatch07, farage12}. The peak-to-peak velocity difference observed in CO(1-0)
is $\sim400\kmps$ over a spatial scale of about $2\kpc$. This is expected of molecular
gas in merger remnants, which generally exhibits large-scale rotation with high 
velocities \citep{ueda14}.

Although ordered motion seems plausible at CO(1-0), the higher resolution CO(3-2) maps
show a highly asymmetric distribution of cold gas that is inconsistent with being a
rotationally-supported disk. Two times more molecular gas is located south of the 
nucleus versus the north side, and the gas velocity jumps abruptly across the nucleus.
This asymmetry indicates that any ordered structure must be in the process of forming, 
as the gas is not in an equilibrium structure.
If the gas is orbiting the BCG, then it will form one complete spiral when the 
gas on the innermost orbit has circled the galaxy one more time than the gas on 
the outermost orbit. With resolved scales of $1-2\kpc$ and velocities of 
$\sim200\kmps$, the corresponding disk formation timescale is $\sim3\e{7}\yr$. 
Since no disk-like structure is observed, the gas is either moving mainly in
the plane of the sky or we are observing the system very recently after the
stripping began.

Gas clumps oriented toward the companion galaxy have velocities that increase roughly
linearly with projected radius to values matching the stellar velocity of the companion
galaxy. An interaction between the companion
galaxy and the BCG can account for the molecular gas via two scenarios: either the gas
is tidally stripped from the companion, or the passage of the companion through a
pre-existing gas supply has dredged up the gas. Differentiating between the 
original source of the molecular gas is difficult with the current data.
Given the greater recession velocity of the companion, if it has already 
interacted with the BCG, it is now receding from the BCG, moving outward in projected 
radius and away behind the midplane of the BCG.

In the first scenario,
direct stripping from the companion would initially maintain the sign of the velocity,
with the magnitude decreasing as the gas is slowed by the gravity of the BCG. At later
times the gas will reverse direction and fall back onto the BCG. The persistent redshifted
velocities observed here then imply that the interaction was recent, as the clouds have
not yet reversed in direction. If even half of the gas distributed between the galaxies
originated in the companion, then its molecular gas mass would have initially been 
$\sim1.2\e{8}\Msun$, double its current mass. This is relatively gas-rich compared 
to the ellipticals in Virgo \citep[e.g.][]{young11}. A merger with a gas-rich donor
galaxy is certainly possible, but is unlikely to expect a priori.

Following the second scenario,
gas originating in the BCG that is pulled outward by the companion would have a velocity
that increases with radius as it is accelerated away from the BCG, as observed.
When the companion is far enough away, the BCG will again dominate the potential and any
gas that is not bound to the companion will rain back onto the BCG. The 1.5~kpc gap 
between the companion and the farthest extent in the spur suggests that this has occurred
recently and the clumps between galaxies may now be dominated once again by the BCG.
Although no stellar dispersion is available for the companion, the linewidth of its 
molecular gas ($260\kmps$) is typical of a normal galaxy, suggesting that the molecular
gas has had time to settle into its gravitational potential well.
From the ``BCG'' spectrum presented in Fig. \ref{fig:BCGandFil}, the masses of the
redshifted and blueshifted peaks are consistent, containing $1.5\pm0.3\e{8}\Msun$ and
$1.7\pm0.3\e{8}\Msun$, respectively. If these originated from the same reservoir, then
the tidal disruption would have removed $\sim50\%$ of the pre-existing gas supply.

With a projected separation of 4.5~kpc, an interaction between these galaxies would 
have occurred roughly 20~Myr ago, assuming the line of sight speed of the companion 
is representative of its 3D velocity. However, if the velocity is this low near the 
cluster core, then the companion cannot be on its first passage through the cluster, as
several orbits would be required to decelerate the galaxy. If the companion is on its 
first passage then it must be travelling near the plane of the sky. Its transverse 
velocity would be several times greater than its line of sight velocity, decreasing 
the interaction time by a factor of a few.

Importantly, tidal interactions affect all matter within the galaxy, without regard
to its phase. Stars should therefore be affected just as strongly as the molecular
gas. This is not observed in the {\it HST} imaging, as the stellar light does not
show strong deviations from a smooth profile. Removing half of a pre-existing gas 
reservoir via tidal forces is therefore unlikely. It is possible that an interaction
between these galaxies has not yet begun, potentially from a radial offset between 
them, but the similarity in velocity between the companion and the series of clumps
appears too striking to be a coincidence.

We again emphasize that the tidal stripping discussed here only attempts to
account for the molecular gas within the BCG; the filament must be formed separately.
The molecular filament cannot have
originated from an older merger, as the merger rate is low and the gas would have 
already fallen onto the BCG. Instead we must invoke a separate mechanism entirely, 
such as cooling from the hot atmosphere. It is far more likely that the molecular
gas is produced via one primary mechanism, and that the companion galaxy has 
merely disrupted a pre-existing gas supply.

\subsubsection{Ram Pressure Stripping}

In the dense cores of galaxy clusters, the ram pressure exerted on atomic gas in 
an infalling galaxy is enough to overcome its gravitational binding energy, allowing
the gas to be efficiently stripped from its host galaxy \citep{haynes84, diSerego07, 
grossi09}. 
Molecular gas is more difficult to strip than atomic gas because molecular clouds
have a much smaller surface area than atomic nebulae and they reside deeper within the 
galaxy's gravitational potential. 
Galaxies in the Virgo cluster with strong H{\sc i} deficiencies have 
minimal molecular gas deficiencies \citep{kenney89}, and the CO detection rate in 
the cluster is the same as in the field \citep{young11}. However, ram pressure stripping 
of molecular gas has still been observed in the Virgo \citep[e.g.][]{vollmer08}, Norma
\citep{jachym14}, and Coma clusters (Jachym et al. in prep).
If the companion galaxy is indeed falling into the cluster, then ram pressure stripping
might be a viable source of the $7.8\e{8}\Msun$ of molecular gas in the filament.
This, however, is not a viable mechanism for supplying the large reservoirs of molecular
gas in BCGs in general, since molecular gas is observed preferentially in BCGs
residing at the centres of clusters with short central cooling times.

The Norma cluster galaxy ESO~137-001 is the first system found to contain a large
amount of molecular gas in a ram pressure stripped tail, with a total molecular gas
mass exceeding $10^9\Msun$ \citep{jachym14}. This molecular gas is associated with
a 40~kpc long tail of H$\alpha$ emission and an 80~kpc long X-ray tail, with a total
H{\sc i} upper limit of $\sim2\e{9}\Msun$. The presence of soft X-ray emission in the
tail is expected to arise from the mixing of the cold, stripped interstellar medium 
with the hot intracluster gas \citep{sun06, tonnesen11}. \citet{jachym14} suggest that
the molecular gas in the outer extent of the tail has formed in situ out of stripped
atomic gas, while the inner reaches of the tail may also be comprised of molecular 
clouds that were stripped directly from the infalling galaxy.

The spatial correlation of molecular gas with H$\alpha$ and X-ray emission in 
ESO~137-001 resembles the filament trailing the companion galaxy in 2A~0335+096. 
Evidently, ram pressure stripping of gas from a high velocity infalling galaxy is 
able to form a filament similar to that observed in 2A~0335+096. This offers a natural
explanation for the significant dust extinction observed along the filament, which would
be stripped from the companion galaxy along with the atomic and molecular gas, and would
offer a seeding site for in situ molecular gas formation. Since the line of sight 
speed of the molecular gas along the filament is $\lesssim 200\kmps$, the high infall 
velocity required to form the filament via ram pressure stripping would require that 
most of the motion is along the plane of the sky.

Ram pressure stripping requires a very high relative velocity in order to
form the observed filament. Orbit modelling in ESO~137-001 favours infall velocities
of $\gtrsim 3000\kmps$. The necessary relative velocity in 2A~0335+096 is not this
severe, since the companion galaxy is situated closer to its cluster centre so is in
a higher density environment, lessening the demands on velocity. Extrapolating the 
\citet{sanders09} gas density profile assuming a power law of $n_e\propto r^{-1}$,
the ICM density at the base of the filament ($\sim13\kpc$) is $\approx 0.07\pcmcu$.
This is 35 times greater than the ICM density at the location of ESO~137-001, using
the $\beta$-model parameters in Table 7 of \citet{jachym14}. Assuming that the force 
exerted by ram pressure, $\rho v^2$, is the same here as it is in ESO~137-001, the 
infall velocity required to produce the tail is $\sim500\kmps$. Since the line of sight
velocity of the companion galaxy is $230\kmps$, the inclination of the filament would
need to be about $30^{\rm o}$.

As discussed in Section \ref{sec:tidal}, the series of redshifted clumps located 
between the BCG and the companion suggests that the two galaxies have already 
interacted. If this is the case, then the companion galaxy must have passed 
through the BCG on a northwestward trajectory. This scenario is mutually exclusive
with the ram pressure stripping origin of the filament, which requires that the
companion is on its first passage through the cluster and is travelling to the
southeast.

Additionally, the morphology of the filamentary emission differs from what is
expected from ram pressure stripping. First, the orientation of the filament does 
not coincide with the companion galaxy. Extending the inner edge of the filament 
to the southeast, the shortest distance between the filament and the companion is 
2~kpc. Ram pressure stripping should form a tail in a direction directly opposing 
the direction of motion, which is not the case here.
Second, H$\alpha$ emission and soft X-rays are not confined to the region trailing 
the companion. Significant H$\alpha$ emission is observed on both sides of the 
BCG, with the brightest emission surrounding the molecular gas within the BCG. This
emission is visible in the lower left corner of Fig. \ref{fig:halpha} as well as
the contours of Fig. \ref{fig:xray}, but its full extent is obscured due to the presence
of a bright foreground star. If the filament is formed by ram pressure stripping then
the H$\alpha$ emission should be localized to the tail of the companion galaxy, in
addition to whatever emission is associated with the molecular gas in the BCG.

\subsubsection{Cooling of the Hot Atmosphere}
\label{sec:ICM}

Cooling from the hot atmosphere naturally explains the spatial correlation of gas
over 5--6 decades in temperature. The upper limit on the mass deposition rate of 
the 0.5~keV X-ray emitting gas is $30\Msunpyr$ \citep{sanders09}, which can form 
the $7.8\e{8}\Msun$ of molecular gas in the filament in $24\Myr$ or the entire gas 
reservoir in $40\Myr$. Depletion of the molecular gas via star formation proceeding 
at a rate of $2\Msunpyr$ \citep{odea08} increases the accumulation time by $<10\%$. 
We also note that the cooling gas is distributed on larger scales than the molecular
gas. Roughly half of the 0.5~keV gas is located on the side of the BCG opposite
the molecular gas. This increases the accumulation time by roughly a factor of two,
resulting in a total time of $50-90\Myr$. This is approaching the mean outburst 
interval between generations of AGN outbursts of $10^{8}\yr$ \citep[e.g.][]{vantyghem14}.

Most of the baryonic content in clusters is contained in the hot atmosphere,
offering a vast reservoir which can supply the molecular gas. Approximating the 
central hot gas density profile in \citet{sanders09} with $\rho \propto r^{-1}$,
the extrapolated gas mass within 13~kpc, which encloses all of the molecular gas, 
is $7.9\e{9}\Msun$. This is about 7 times more than is needed to form the entire
$1.13\e{9}\Msun$ of molecular gas. However, cooling does not occur over the full 
azimuth, and the local supply of hot gas is much more restrictive. On the other hand,
gas cooling out of the hot atmosphere can be replenished by gas from higher altitudes,
providing more than the $7.9\e{9}\Msun$ of gas within the central 13~kpc.

The presence of multiple cavities in the hot atmosphere \citep{mazz03, sanders09} 
indicates that 2A~0335+096 has undergone several cycles of AGN feedback, which
would include multiple cycles of cooling. However, if cooling has persisted over 
multiple cycles then we should expect to see several filaments, similar to those in
Perseus \citep{hatch06, salome11}, while only one is observed. Filaments from previous
cooling cycles may have fallen back onto the BCG, forming the observed reservoir 
within the BCG. Conversely, the presence of a single filament is reminiscent of the 
cooling wake observed in A1795 \citep{fabian01, crawford05}, where cooling is stimulated by 
the motion of the cD galaxy through the hot atmosphere. Sloshing motions in 2A~0335+096
indicate that the BCG is in motion with respect to the cluster, which may establish
the preferred direction of cooling.

Gas cooling from the hot atmosphere is expected to be relatively dust-free 
\citep{donahue93}. This is because dust grains are rapidly sputtered in the ICM, 
and can only form when the gas is shielded by UV and X-ray irradiation \citep{draine79}.
As noted in Section \ref{sec:dust}, significant dust extinction is present over much
of the length of the filament. Since dust production from cooling gas is difficult,
the dust likely originated from the within the BCG. \citet{hatch07} suggested that 
the dusty nebulae observed in several BCGs have been drawn out of the central 
molecular gas reservoir, where the high densities can provide shielding long enough 
for the gas to become polluted with dust.

Recent simulations have suggested that thermal instabilities in the hot atmosphere 
are induced along the direction of the radio jet \citep{gaspari13,li14a}, which is 
not the case in 2A~0335+096. A bipolar radio jet observed at 1.5~GHz extends 12~arcsec 
to the northeast and southwest of the BCG \citep{sarazin95, donahue07}, which is 
orthogonal to the filament. Instead, we argue in Section \ref{sec:stimulated} that the
preferred direction of gas cooling has been imposed by uplift from an X-ray cavity.
This enables dust to be lifted out of the BCG, providing seeding sites for the
production of molecular gas.

\subsubsection{Summary: Gas Origin}

While galaxy mergers are unable to account for the high molecular gas masses in
cool core galaxy clusters in general, the presence of a close companion in this
system has the potential to supply the cold gas.
Tidal stripping from the companion galaxy can supply the BCG with cold gas, but does 
not account for the filament. Instead, the merger may have disrupted a pre-existing 
gas supply within the BCG, dredging up cold gas as it passed through. However,
we do not see evidence for a tidal disruption in the stellar light of either galaxy.
Alternatively, the filament observed in 2A~0335+096 resembles the ram pressure 
stripped tails in other systems (e.g. ESO~137-001). Ram pressure stripping is feasible
if the companion is infalling from the northwest with a high relative velocity, and
can account for the spatial coincidence of molecular gas, dust extinction, H$\alpha$
emission, and soft X-rays. However, it does not explain the inclined orientation
of the filament or the H$\alpha$ emission and soft X-ray located southeast of the
BCG nucleus. Furthermore, if gas has in fact been dredged up by the companion
galaxy, then its implied orbit rules out the possibility of a ram pressure stripping
origin of the filament.
Cooling of the hot atmosphere provides a feasible mechanism for supplying the molecular 
gas that is tenable in a much broader sample of cool core clusters. The hot atmosphere 
harbours more than enough gas to produce the filament, and the 0.5~keV phase is cooling 
rapidly enough to form the entire molecular gas supply in $\sim10^{8}\yr$.
The cold gas within the BCG may then correspond to an older cycle of cooling from 
the hot atmosphere that has fallen onto the BCG.

\subsection{Origin of the Cooling}

\begin{figure}
  \centering
  \includegraphics[width=\columnwidth]{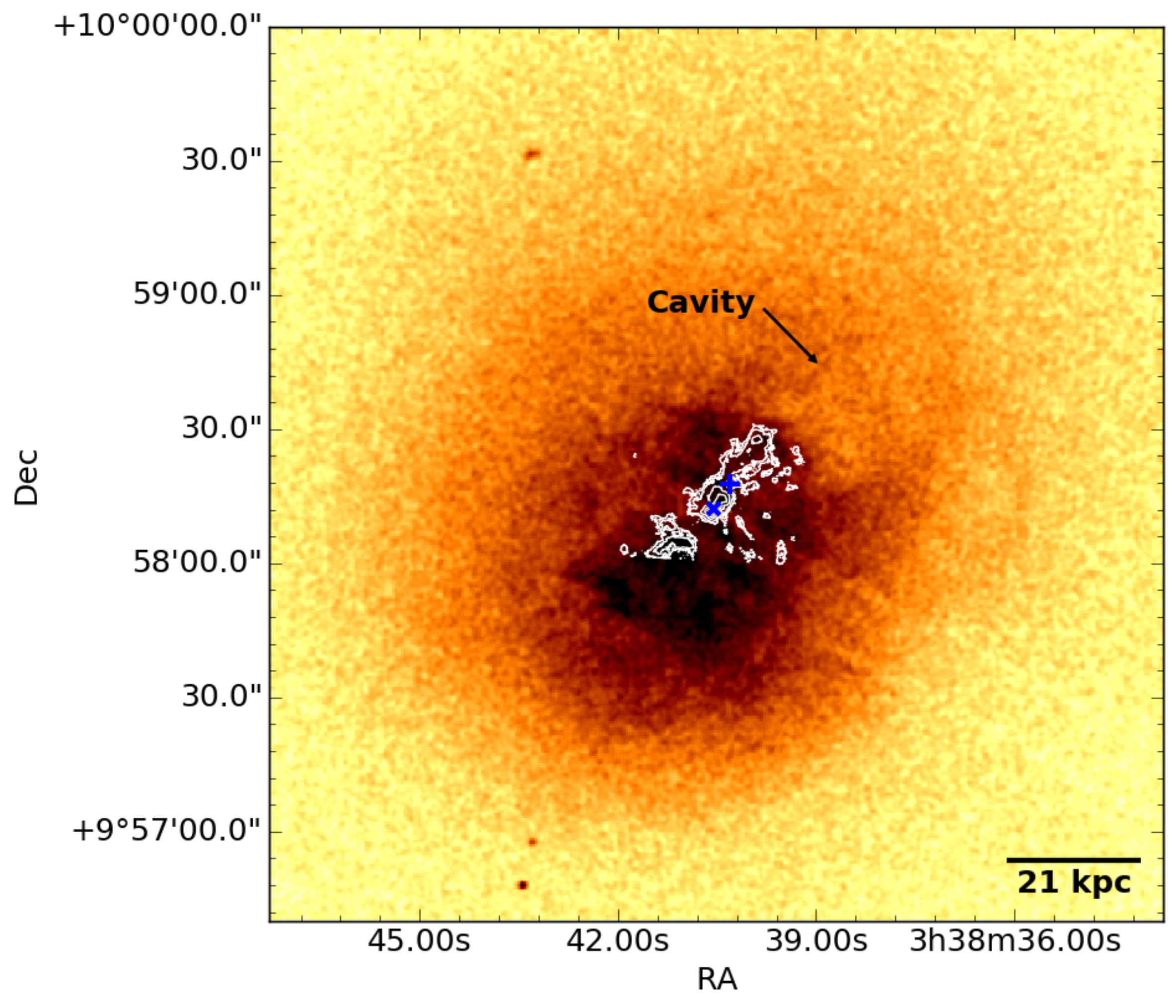}
  \caption{
    Chandra X-ray image of the hot atmosphere of 2A~0335+096. Several X-ray cavities and clumps 
    of cool gas are visible in the image. The H$\alpha$ nebula, shown in white contours,
    extends toward the most energetic cavity. Any emission to the south of the contours
    has been masked out due to contamination by a foreground star.
  }
  \label{fig:xray}
\end{figure}

\subsubsection{Cooling Stimulated by the AGN}
\label{sec:stimulated}

Following recent ALMA observations of PKS0745-191 \citep{almaPKS0745} and A1835 
\citep{almaA1835}, \citet{mcn16} proposed the ``stimulated feedback'' model. In
this model, molecular gas condenses from lower entropy gas that is lifted outward 
from the cluster core by X-ray bubbles, away from the location where the heating rate 
matches its cooling rate. Unless the overdensity falls back to its original position 
within its cooling time, it will condense out of the intracluster medium \citep{nulsen86, 
revaz08, gaspari13, li14b}, forming molecular gas and emission-line nebulae.

The molecular gas in the PKS0745-191 BCG is distributed along three narrow filaments
that are $3-5\kpc$ in length \citep{almaPKS0745}. Two of the filaments are oriented 
behind X-ray cavities, while the third is coincident with UV emission from young stars.
The velocities of the cold gas, which lie within $\pm100\kmps$ of the systemic velocity,
are too low to arise from steady inflow of clouds condensing out of the hot atmosphere.
Instead, the velocity distribution of the two filaments oriented toward X-ray cavities
are consistent with the majority of the gas flowing outward from the galaxy centre.

A similar outflow was observed in the A1835 BCG, where $10^{10}\Msun$ of cold gas
is located in a bipolar flow behind two X-ray cavities \citep{almaA1835}. This
system hosts comparably low velocities, which were initially interpreted as a 
face-on disk. However, low velocities have been observed in a growing sample of 
BCGs observed with ALMA \citep{david14, almaA1664, almaPKS0745, tremblay16}, 
indicating that velocities well below the stellar velocity dispersion are common
amongst BCGs. 
Similarly low velocities were also noted in the filamentary optical-line nebulae 
of BCGs \citep[e.g.][]{heckman89, jaffe05, hatch06, oonk10}.
Evidently, the gas must either be dynamically young, having not had time to settle 
into its gravitational potential, or the clouds are supported by pressure from either 
magnetic fields or the hot gas.

Redshifted CO absorption lines have been observed in NGC5044 \citep{david14} and A2597 
\citep{tremblay16}. The widths of these lines are small ($\sim5\kmps$), and are 
comparable to individual giant molecular clouds within the Milky Way. Since the lines
are seen in absorption, the clouds must lie in front of the AGN, blocking its continuum 
emission. The redshifted velocities indicate that the clouds are falling toward the
black hole, and are possibly in the process of being accreted.
The presence of both inflow and outflow in BCGs suggests that gas lifted by from the
cluster core eventually returns in a circulation flow, or molecular ``fountain''.

The results presented here are consistent with this stimulated feedback model.
{\it Chandra} imaging of 2A~0335+096, shown in Fig. \ref{fig:xray}, reveals a 
series of X-ray cavities corresponding to multiple generations of AGN feedback 
\citep{mazz03, sanders09}. The H$\alpha$ nebula, overlaid in white contours in 
Fig. \ref{fig:xray}, shows a striking connection to the northwestern cavity. 
The emission extends from the centre of the BCG directly toward the X-ray cavity 
identified in Fig. \ref{fig:xray}, with fainter emission along the cavity's inner 
edge.
As noted in Section \ref{sec:spatialcorr}, the H$\alpha$ filament is coincident
with soft X-ray emission from 0.5~keV gas as well as $20-30\K$ molecular gas. 
This indicates that the gas is cooling over 5 decades in temperature in the
wake of a rising X-ray cavity. 
Dust grains situated along the filament have presumably been uplifted from the
BCG along with the low entropy ICM. This resembles the scene in PKS0745-191,
where the two filaments trailing X-ray cavities are correlated with dust extinction.
The uplifted dust potentially enhances the formation of molecular gas by offering 
seeding sites.

A bipolar radio jet observed at 1.5~GHz extends 12~arcsec to the northeast and 
southwest of the BCG \citep{sarazin95, donahue07}. This radio jet traces the
most recent generation of AGN feedback, which is orthogonal to the emission-line
filament. No molecular gas is observed along this direction, indicating that
the radio jet itself has not disrupted the central supply of molecular gas or
created thermally-unstable overdensities in the hot atmosphere. 
Similarly, any cavities formed along this jet have yet to induce a significant
amount of thermally unstable cooling.

\subsubsection{Cooling Wake from Galaxy Motion}

Cooling of the ICM can also be induced by the gravitational wake of a galaxy in 
motion with respect to the cluster. For example, A1795 hosts a BCG with a 
velocity of $+150\kmps$ with respect to the mean of all galaxies in the cluster, 
or $+374\kmps$ compared to the galaxies within $270\kpc$ \citep{oegerle94}.
The cluster also harbours a luminous emission-line nebula extending in a $50\kpc$
long filament to the southeast of the BCG \citep{cowie83, crawford05, mcdonald09}.
This filament has also been identified in X-ray imaging \citep{fabian01, crawford05, 
ehlert15}, molecular gas \citep{mcdonald12}, U-band polarimetry \citep{mcn96}, and 
far-UV imaging \citep{mcdonald09}.
\citet{fabian01} argued that this filament was formed from a cooling wake, where the
gravitational attraction from the BCG moving through a region with a short cooling
time focuses cooling onto its wake. Since the gas cools from the ICM, its velocity
reflects that of the cluster instead of the BCG.

In principle either the BCG or the companion galaxy could induce this cooling wake.
However, it is more likely that the BCG causes the cooling wake due to its larger 
gravitational attraction, provided it is in motion with respect to the ICM. 
Furthermore, the presence of H$\alpha$ filaments in cool core clusters is not 
correlated with the presence of galaxies near the BCG \citep{mcdonald10}. 
If the companion galaxy in 2A~0335+096 is able to induce a cooling wake, then
we should expect similar filaments in all cool core clusters hosting galaxies
near their core, which is evidently not the case.

Without a direct measurement of ICM velocity it is difficult to determine if
the BCG is in motion with respect to the cluster. However, the X-ray atmosphere
of 2A~0335+096 hosts a series of cool clumps and a cold front, which are indicative
of an unrelaxed dynamical state \citep{mazz03, sanders09}. In particular, the X-ray
centroid of the cluster is difficult to pinpoint because of the series of cool clumps.
The brightest clump of X-ray emission in the $0.5-7\keV$ band is co-spatial with the
H$\alpha$ emission $6-10\kpc$ southeast of the BCG nucleus. The offset between the
X-ray peak and the BCG, as well as the presence of sloshing motions in the ICM, suggest
that the BCG is in motion relative to the cluster. We are unable to constrain the BCG
velocity in this system, although sloshing motions in Virgo have velocities of 
$\sim50-100\kmps$ \citep{roediger11}. The low velocity of the main component of the
inner filament, $-28\pm4\kmps$, may imply a small line of sight velocity offset
between the BCG and cluster, though multiple distinct velocities throughout the filament
point to a more complicated picture.
Similarly, we cannot confirm that the motion is along the direction of the filament.

As noted in Section \ref{sec:ICM}, gas cooling from the ICM is expected to be dust-free.
This appears to be the case in A1795, where dust extinction is only observed within
the central galaxy \citep{pinkney96}, and is likely not affecting star formation along 
the filament \citep{crawford05}. The presence of dust along the filament in 2A~0335+096
argues against the cooling wake scenario unless the filament has been enriched with
dust from the BCG or companion, perhaps through ram pressure stripping. This possibility
has been suggested by \citet{sparks04} for the filaments in M87. However, ram pressure
stripping requires velocities of several hundred $\kmps$, even in the dense cluster core.

\subsubsection{Cooling Along a Dark Matter Filament}

We also note an interesting similarity in the position angles of the BCG and filament 
and the apparent trajectory of the companion galaxy. BCGs are known to have a common
orientation with their host clusters, which are themselves aligned with the cosmological
filaments of dark matter they accreted from \citep{binggeli82, niederste10}. The companion
galaxy may have accreted along this axis, falling from the southeast and passing through
the gas supply in the BCG. Understanding how the filamentary emission is linked to the
dark matter distribution is more difficult. Any overdensity caused by the filament should
only be significant in the outer extent of the cluster, not in the inner tens of kpc where
the emission-line nebula is observed. Direct accretion from the cosmological filament is
similarly unlikely, since any accreted gas should be moving at speeds approaching the 
cluster velocity dispersion. The alignment noted here does not appear to persist in 
other cool core clusters. Perseus and Virgo, for example, harbour filamentary emission
with no preferred orientation \citep[e.g.][]{young02, salome06}. A more likely possibility
is that an infalling galaxy caused an overdensity in the ICM that led to a thermal 
instability. However, the only galaxy located along the filament is the companion, and
it cannot produce both the disrupted central gas and the filament. Thus, the alignment
between the major axis of the BCG and filament of molecular gas is most likely coincidental.

\subsubsection{Summary: Cooling Origin}

While stimulated cooling and the cooling wake are both feasible origins for gas 
condensation, we favour stimulated cooling throughout the remainder of our 
discussion. This is mainly because of the connection between the cooling gas
and X-ray cavity, which resembles a growing sample of objects that have been
observed with ALMA. As a result, stimulated cooling offers a more generally
applicable explanation for the presence of molecular gas in cool core clusters.

\subsection{Molecular Filament: Inflow or Outflow?}

We now consider the velocity profile of the molecular filament and investigate
whether it is consistent with inflow or outflow. Along the filament the gas 
velocity becomes increasingly blueshifted farther from the cluster centre. At 
its base the gas is blueshifted to $-30\kmps$, while the velocity at the tail 
is nearly $-200\kmps$. This can be consistent with either inflow or outflow,
depending on where the gas forms and where it is located along the line of sight.
With the present data we are unable to conclusively distinguish between the two.

Molecular gas condensing out of the hot atmosphere should trace the velocity of
the gas that it cooled from. In a cooling wake the molecular gas condenses from
a hydrostatic atmosphere, resulting in a mean velocity of zero with respect to 
the cluster. Our adopted frame is measured with respect to the BCG, so an offset
is expected if the BCG is indeed oscillating within the cluster.
Molecular gas condensing from low entropy gas lifted by an X-ray cavity should
initially be flowing away from the cluster centre. The observed velocity gradient
then depends strongly on both the inclination of the filament as well as its
shape. For example, a filament with a constant velocity may have an observed
gradient if it becomes progressively more inclined toward the line of sight.
We cannot differentiate between these morphologies, so preliminarily assume that
the filament is straight.

Eventually the dense molecular clouds are expected to decouple from the hot 
atmosphere, decelerating and falling back onto the central galaxy in a circulation 
flow. Recent results from {\it Hitomi} found that the velocity of the ICM in 
Perseus is very similar to the cold gas, suggesting that the two phases are held 
together by magnetic fields for some time \citep{hitomi16}.
\citet{lim08} argued in favour of infall for the filaments nearest the cluster core 
based on their velocity gradient, while inflowing molecular gas was observed directly
in NGC5044 \citep{david14} and A2597 \citep{tremblay16} based on the presence
of redshifted CO absorption lines. 
Distinguishing between inflow and outflow does not affect the consistency with
the stimulated cooling model, as both are expected to occur at some point in the 
feedback cycle.

\subsubsection{Outflow in the Wake of the Rising Cavity}
\label{sec:outflow}

If the clouds along the filament formed recently and are located in front of the 
midplane of the BCG, then their increasing velocity
with radius implies that the gas is being accelerated away from the BCG nucleus.
Localized condensation at the base of the filament followed by direct uplift of
the molecular gas is unlikely, since coupling the diffuse X-ray cavity to dense 
molecular clouds is difficult. This is exacerbated by the high mass of the filament,
so the coupling of diffuse to dense gas would need to be remarkably efficient.
Instead, the molecular clouds probably condensed in situ from the uplifted, low 
entropy gas.

The terminal velocity of X-ray cavities is generally $50-60\%$ of the speed of 
sound, $c_s=\sqrt{\gamma kT/\mu m_{\rm H}}$, where $kT$ is the temperature of the 
hot gas and we have taken $\gamma=5/3$ for an ideal gas and $\mu=0.62$ 
\citep[e.g.][]{birzan04}. 
In a 3.5~keV cluster $c_s \approx 900\kmps$, so a typical bubble velocity is 
$\sim500\kmps$. The molecular gas in the filament reaches a line of sight velocity 
(magnitude) of $200\kmps$, several times lower than the bubble velocity. In order 
for the two velocities to match, the filament must be inclined at $<25^{\circ}$ from 
the plane of the sky.

Accelerating the molecular gas to a speed comparable to the bubble terminal velocity 
of $\sim500\kmps$ requires that the AGN contribute $2\e{57}\erg$ to the kinetic 
energy of the cold gas. Assuming a gravitational acceleration of $g=2\sigma^2/r$ with 
$\sigma=255\kmps$ \citep{mcn90}, the potential energy required to lift the gas 
from 1~kpc to $R\approx13\kpc$ is $\sim5\e{57}\erg$. The total energy requirement is 
therefore $\sim7\e{57}\erg$, which is only 4\% of the enthalpy of the northwestern 
cavity, $1.6\e{59}\erg$ \citep{sanders09}. Lifting the molecular gas in the wake of 
the X-ray cavity is therefore energetically feasible.

\subsubsection{Clouds in Gravitational Free-fall}
\label{sec:freefall}

When molecular clouds decouple from the hot atmosphere, they should fall ballistically
under the influence of gravity. The clouds will initially have the same velocity
as the hot gas they cooled from. This is zero in a hydrostatic atmosphere, but
can be nonzero if the clouds have cooled from uplifted gas. 
Following \citet{lim08}, we assume that the gravitational potential can be modelled 
by a Hernquist profile \citep{hernquist90}. A gas cloud undergoing free fall should 
accelerate to a velocity of
\begin{equation}
v(r)^2 = v(r_0)^2 + 2GM \left( \frac{1}{r+a} - \frac{1}{r_0+a} \right)
\label{eqn:hernquist}
\end{equation}
with respect to the hot atmosphere. In the rest frame of the BCG, which has been
adopted for these observations, the velocities in the above equation are modified
to be $v(r) - v_{\rm ICM}$, where $v_{\rm ICM}$ is the velocity offset between
the BCG and the cooling gas. We assume that the initial velocity of the cloud is
the same as the ICM, so $v(r_0) = v_{\rm ICM}$.
In Eqn \ref{eqn:hernquist} $M$ is the total gravitating mass of the BCG, $a$ is 
the scale length, and $r_0$ is the radius where the cloud originally formed. 
The inclination angle of the cloud's trajectory and $v_{\rm ICM}$ are both free 
parameters in this model.

The scale length $a$ is related to the half-light radius $R_e$ according to
\begin{equation} R_e \approx 1.8153 a. \end{equation}
Taking an effective radius of $19.24\arcsec$ ($13.47\kpc$) from the 2MASS K-band 
catalog, the corresponding scale length is $7.4\kpc$. We note that 2MASS 
photometry is likely missing a significant amount of the total stellar light 
\citep{lauer07}, so this scale length is underestimated. Adopting a larger 
scale length would result in flatter velocity profiles.

We estimate the total BCG mass using the cluster mass profiles of \citet{main15},
who modelled the X-ray emission of the ICM with an NFW profile under the 
assumption of hydrostatic equilibrium. For 2A~0335+096 the total mass interior to 
$30\kpc$ is $2.1\e{12}\Msun$.
Alternately, 2MASS report a total K-band magnitude of $9.808\pm0.052$ that was 
extrapolated out to $45\kpc$. Assuming a K-band stellar mass-to-light ratio of
$0.8$ \citep{bell03,humphrey06}, the total stellar mass in the BCG of 2A~0335+096 
is $4.7\e{11}\Msun$. 
Combining these values gives a dark matter mass fraction within the BCG of
80\%, which is similar to that of Hydra A at a similar radius \citep{okabe16}.
The total gravitating mass controls the amplitude of the velocity profile, and
is degenerate with the inclination angle. Our results are therefore not sensitive
to the adopted value of total mass, as the inclination angle, which is not known,
can be adjusted to compensate.

From the PV diagrams presented in Fig. \ref{fig:PV}, infall along the extended
filament begins at a radius of $11.5\kpc$ and proceeds over a length of $6\kpc$ 
that is oriented $20^{\circ}$ from purely radial. The solid black line in Fig.
\ref{fig:PV} shows the velocities resulting from infall along this filament assuming
an inclination angle of $25^{\circ}$ and a velocity offset of $-200\kmps$. 
The position shown along the x-axis has been transformed into the appropriate radial
distance from the BCG nucleus. 
The velocity shift corresponds to a bulk offset between the ICM and the stellar
component of the BCG. The presence of a cold front seen in the X-ray emission
\citep{mazz03, werner06, sanders09} indicates non-zero motion between the two
components.

Over the $7\kpc$ length of the filament the gas is primarily located in two
clumps. Gravitational free-fall reproduces the observed increase in
velocity between these clumps. However, the clumps are not smoothly connected
in velocity, and the free-fall model fails to account for this feature. Furthermore,
the observed average velocity of the inner portion of the filament does not depend
on the radius, while the velocity of the free-fall model increases linearly through 
this region. Although the free-fall model reproduces the bulk of the observed 
velocities, differences between the data and the model prevent any definitive 
conclusions about the gas undergoing free-fall from being drawn.
Nevertheless, the velocity gradient is broadly consistent with free-fall. 
Since this velocity gradient is generic for free-fall in elliptical galaxies, 
the discovery of additional filaments exhibiting this gradient suggests that
some of them, at the very least, are in approximate free-fall.

A number of additional problems with the model adopted here must also be addressed.
First, the filament is not oriented radially with respect to the BCG nucleus. This
may be caused by a transverse velocity offset between the ICM and the BCG, which 
would be expected given the presence of a cold front.
This model also neglects the gravity of the companion galaxy. Assuming the companion
lies in the plane of the sky and has a mass-to-light ratio comparable to the BCG,
the inner portion of the filament should experience 3$\times$ more gravitational
acceleration from the companion galaxy than from the BCG. This could also affect
the orientation of the infalling gas. Without a reliable measurement of the 
line-of-sight separation between the BCG, companion galaxy, and filament, we
cannot create a more robust gravitational free-fall model.
Finally, the model we have adopted here has a large number of unconstrained parameters, 
notably the inclination angle and bulk velocity of the ICM. As a result, virtually
any linear velocity structure can be reproduced with an appropriate choice of
these parameters.

\subsection{Star Formation}
\label{starformation}

Attributing all of the H$\alpha$ emission to star formation, the luminosity of
$8\e{41}\ergps$ \citep{donahue07} corresponds to a star formation rate of 
$15-20\Msunpyr$, using the $L_{\rm H\alpha}-{\rm SFR}$ scaling relations of 
\citet{kennicutt98}. 
This overestimates the true star formation rate, as starlight alone fails to 
account for the observed spectra of emission-line nebulae \citep{johnstone07}. 
Additional heating sources are required to supplement the ionization of the nebula.
Collisional heating by ionizing particles, such as cosmic rays that penetrate the 
filament, is favoured in the models of \citet{ferland09}. Mixing of the gas in the 
filaments with the hot X-ray emitting plasma was argued to provide the dominant 
source of heat in Virgo \citep{werner13} and other ellipticals \citep{werner14}.
The ionization source in 2A~0335+096 is not currently known, although the spatial 
coincidence between the filaments and 0.5~keV gas suggests that star formation 
alone does not power the nebula.

An analysis of the $B-I$ colour gradient of the central galaxy shows an excess in 
blue emission from $4-30\kpc$ compared to the central galaxies in clusters without 
short cooling times \citep{RH88}. This extends well beyond the companion galaxy,
indicating that active star formation is occurring along the filament.
Within 4~kpc the colour gradient reddens to a level consistent with the control 
sample. \citet{wilman11} argued that this reddening cannot be caused by dust extinction, 
since any attenuation by dust would be negligible in their $K$-band observation.
In the BCG alone, several different measurements place the star formation rate at
a few $\Msunpyr$. Infrared photometry within a $6''$ aperture, which excludes both the
companion galaxy and the filament, derives a SFR of $2.1\Msunpyr$ \citep{odea08}.
Infrared spectroscopy \citep{donahue11} and UV imaging \citep{donahue07} over 
regions that include both galaxies but not the filament measure SFRs of $0.7$ and 
$3-7\Msunpyr$, respectively.

The total molecular gas mass in the BCG is $3.2\e{8}\Msun$. Using the CO(3-2) line to 
trace the emitting area yields a molecular gas surface density of $68\Msun\pc^{-2}$.
Assuming that star formation is distributed over the same area, the maximum SFR
to place the 2A~0335+096 BCG within the scatter of the Kennicutt-Schmidt relation 
\citep{kennicutt98, kennicutt12} is $\sim0.9\Msunpyr$. The SFR derived from IR 
spectroscopy is consistent with this limit.
For the $7.8\e{8}\Msun$ of molecular gas in the filament, the corresponding surface
density of molecular gas is $46\Msun\pc^{-2}$. Placing this on the Kennicutt-Schmidt 
relation requires a SFR of roughly $0.09\Msunpyr$. However, the low confining pressure
of the ICM outside of the cluster core reduces the star formation efficiency in
systems with similar filaments \citep[e.g.][]{verdugo15, salome16a, salome16b}. We
should therefore expect a lower SFR, although the excess blue emission still supports
ongoing star formation within the filament.

\section{Conclusions}
\label{sec:summary}

In this work we have presented ALMA observations of the CO(1-0) and CO(3-2) 
lines of the 2A~0335+096 BCG. We detect $1.13\e{9}\Msun$ of molecular gas that 
is distributed between two distinct structures: a component near the centre of 
the BCG and a 7~kpc long filament beginning 6~kpc from the BCG nucleus and 
extending nearly radially outward. Most of the molecular gas, $7.8\e{8}\Msun$, 
is located in the filament, while $3.2\e{8}\Msun$ is located in the nucleus. 
The nuclear gas is highly asymmetric, with two clumps of comparable mass south 
of the radio source and several smaller clouds extending toward a nearby companion 
galaxy. The filament has a shallow velocity gradient that is slightly
blueshifted at all points, nearing the systemic velocity of the BCG at its innermost
radius. No molecular gas is detected connecting the filament to the BCG nucleus.

Although a companion galaxy is located 5~kpc from the BCG in projection and 
has a small relative velocity ($\sim200\kmps$),
it is unlikely that the molecular gas has been supplied by a merger.
Tidal stripping alone could deposit molecular gas onto the BCG, but the filament 
must then form via completely independent means.
A plume of increasingly redshifted clouds extending toward the companion 
galaxy suggests that the galaxies have already interacted, with the companion galaxy
disrupting a pre-existing reservoir of molecular gas within the BCG as it passed 
through the nucleus at $<20\Myr$ ago. However, no evidence of stellar disruption
is observed in optical imaging, indicating that any tidal interaction must be weak.

Ram pressure stripping is a feasible way to produce the filament, as similar
structures have been observed in other systems (e.g. ESO~137-01). However,
filamentary emission has also been observed in a number of cool core clusters 
that do not have nearby companion galaxies. While we cannot definitively rule
out ram pressure stripping, it is not representative of cool core clusters
in general. Moreover, if the galaxies have already interacted, then ram pressure 
stripping cannot be a viable mechanism for producing the filament.

The filament of molecular gas is coincident with significant dust extinction, 
luminous H$\alpha$ emission, and 0.5~keV X-ray emitting gas. This spatial 
correlation of gas spanning 5--6 decades in temperature implies that the
molecular gas has condensed out of gas cooling from the hot atmosphere. The
hot atmosphere offers an abundant source of gas with which to form the molecular
gas. Condensation out of the hot atmosphere can form the total molecular gas 
supply rapidly enough to sustain cycles of AGN feedback every $\sim10^{8}\yr$.
Condensation can be triggered either in the wake of the BCG or in the uplift
behind an X-ray cavity. We favour the uplift interpretation because of its
similarity to a growing sample of objects observed with ALMA, which may be
representative of cool core clusters in general.

The H$\alpha$ emission from this cooling filament extends toward an X-ray cavity,
with faint emission spreading around its inner edge. This resembles the distribution
of molecular gas in a growing number of objects observed with ALMA (e.g. PKS0745-191 
and A1835), where significant amounts of molecular gas reside in massive outflows 
linked to the buoyantly rising X-ray cavities. Our observations are consistent with
the ``stimulated feedback'' model, where molecular gas condenses out of low entropy 
gas that is lifted out of thermal equilibrium by X-ray cavities. With an enthalpy
of $1.6\e{59}\erg$, the X-ray cavity possesses ample energy to lift enough low entropy
gas to form the observed molecular gas supply.

As clouds of molecular gas condense out of the uplifted low entropy gas in the hot 
atmosphere, they should decouple from the hot atmosphere and fall back onto the BCG 
under the influence of gravity. Initially outflowing gas will eventually decelerate 
and return to the BCG in an inflow. We are unable to distinguish between inflow 
and outflow with our observations. The velocity of the molecular gas in the filament
increases in magnitude from $-30\kmps$ near its base to $-200\kmps$ at its tail. 
This velocity gradient may correspond to gas accelerated by the cavity, but is also 
consistent with simple models of gravitational infall. 
This ambiguity, however, does not affect our interpretation that the molecular gas
has condensed out of low entropy gas uplifted by an X-ray cavity.

\acknowledgements

We thank the anonymous referee for helpful comments that improved this paper.
Support for this work was provided in part by the National Aeronautics and Space Administration through Chandra Award Number G05-16134X issued by the Chandra X-ray Observatory Center, which is operated by the Smithsonian Astrophysical Observatory for and on behalf of the National Aeronautics Space Administration under contract NAS8-03060.
ANV and BRM acknowledge support from the Natural Sciences and Engineering Research Council of Canada.
BRM further acknowledges support from the Canadian Space Agency Space Science Enhancement Program.
ACF and HRR acknowledge support from ERC Advanced Grant Feedback 340442.
This paper makes use of the following ALMA data: ADS/JAO.ALMA 2012.1.00837.S. ALMA is a partnership of the ESO (representing its member states), NSF (USA) and NINS (Japan), together with NRC (Canada), NSC and ASIAA (Taiwan), and KASI (Republic of Korea), in cooperation with the Republic of Chile. The Joint ALMA Observatory is operated by ESO, AUI/NRAO, and NAOJ.
This research made use of APLpy, an open-source plotting package for Python hosted at http://aplpy.github.com.

\bibliographystyle{apj}
\bibliography{paper}

\end{document}

%% file: defn.tex
\newcommand{\Mpc}{\rm\; Mpc}
\newcommand{\kpc}{\rm\; kpc}
\newcommand{\pc}{\rm\; pc}
\newcommand{\km}{\rm\; km}
\newcommand{\m}{\rm\; m}
\newcommand{\cm}{\rm\; cm}

\newcommand{\cmsq}{\hbox{$\cm^2\,$}}

\newcommand{\yr}{\rm\; yr}

\newcommand{\Myr}{\rm\; Myr}
\newcommand{\s}{\rm\; s}

\newcommand{\GHz}{\rm\; GHz}

\newcommand{\K}{\rm\; K}

\newcommand{\Msun}{\hbox{$\rm\thinspace M_{\odot}$}}

\newcommand{\Msunpyr}{\hbox{$\Msun\yr^{-1}\,$}}

\newcommand{\keV}{\rm\; keV}

\newcommand{\erg}{\rm\; erg}
\newcommand{\Jy}{\rm\; Jy}
\newcommand{\mJy}{\rm\; mJy}

\newcommand{\ergps}{\hbox{$\erg\s^{-1}\,$}}

\newcommand{\kmps}{\hbox{$\km\s^{-1}\,$}}

\newcommand{\kmpspMpc}{\hbox{$\kmps\Mpc^{-1}\,$}}

\newcommand{\Lsun}{\hbox{$\rm\thinspace L_{\odot}$}}

\newcommand{\Zsun}{\hbox{$\thinspace \mathrm{Z}_{\odot}$}}

\newcommand{\asec}{\rm\; arcsec}

\newcommand{\psqcm}{\hbox{$\cm^{-2}\,$}}
\newcommand{\pcmsq}{\hbox{$\cm^{-2}\,$}}
\newcommand{\pcmcu}{\hbox{$\cm^{-3}\,$}}

